\documentclass[iop,twocolappendix,numberedappendix,appendixfloats]{emulateapj}

\usepackage{graphicx} 
\usepackage{amssymb,amsmath,multirow,gensymb}
\usepackage{enumitem,multirow,rotating}
\usepackage[caption=false]{subfig} 
\usepackage{natbib}
\newcommand{\MJysr}{\mbox{MJy sr$^{-1}$}}
\newcommand{\vol}{\mbox{cm$^{-3}$}} 
\newcommand{\kms}{\mbox{km s$^{-1}$}}
\newcommand{\um}{\mbox{$\mu$m}}

\newcommand{\Msun}{\mbox{M$_{\odot}$}}

\begin{document}

\title{Intensity-Corrected Herschel\footnotemark[\text{*}] Observations of Nearby Isolated Low-Mass Clouds}

\author{Sarah I. Sadavoy\altaffilmark{1$\dagger$}, Eric Keto\altaffilmark{1}, Tyler L. Bourke\altaffilmark{1,2}, Michael  M. Dunham\altaffilmark{3,1}, Philip C. Myers\altaffilmark{1}, Ian W. Stephens\altaffilmark{1}, James Di Francesco\altaffilmark{4}, Kristi Webb\altaffilmark{5},  Amelia Stutz\altaffilmark{6,7}, Ralf Launhardt\altaffilmark{7}, John Tobin\altaffilmark{8,9}
	 }
	 
\footnotetext[\text{*}]{\emph{Herschel} is an ESA space observatory with science instruments provided by European-led Principal Investigator consortia and with important participation from NASA.}

\footnotetext[$\dagger$]{Hubble Fellow}
	 
\altaffiltext{1}{Harvard-Smithsonian Center for Astrophysics, 60 Garden Street, Cambridge, MA, 02138, USA}
\altaffiltext{2}{Square Kilometre Array Organisation, Jodrell Bank Observatory, Lower Withington, Cheshire SK11 9DL, UK}
\altaffiltext{3}{Department of Physics, State University of New York at Fredonia, 280 Central Ave, Fredonia, NY 14063, USA}
\altaffiltext{4}{National Research Council Canada, 5071 West Saanich Road, Victoria BC Canada, V9E 2E7}
\altaffiltext{5}{Department of Physics and Astronomy, University of Victoria, PO Box 355, STN CSC, Victoria BC Canada, V8W 3P6}
\altaffiltext{6}{Departmento de Astronom\'{i}a, Facultad Ciencias F\'{i}sicas y Matem\'{a}ticas, Universidad de Concepci\'{o}n, Concepci\'{o}n, Chile 0000-0003-2300-8200}
\altaffiltext{7}{Max-Planck-Institut f\"{u}r Astronomie (MPIA), K\"{o}nigstuhl 17, D-69117 Heidelberg, Germany}
\altaffiltext{8}{Homer L. Dodge Department of Physics and Astronomy, University of Oklahoma, 440 W. Brooks Street, Norman, OK 73019, USA}
\altaffiltext{9}{Leiden Observatory, Leiden University, P.O. Box 9513, 2300-RA Leiden, The Netherlands}


\date{Received ; accepted}

\begin{abstract}
We present intensity-corrected \emph{Herschel} maps at 100 \um, 160 \um, 250 \um, 350 \um, and 500 \um\ for 56 isolated low-mass clouds.  We determine the zero-point corrections for \emph{Herschel} PACS and SPIRE maps from the \emph{Herschel} Science Archive (HSA) using \emph{Planck} data.   Since these HSA maps are small, we cannot correct them using typical methods.   Here, we introduce a technique to measure the zero-point corrections for small \emph{Herschel} maps.  We use radial profiles to identify offsets between the observed HSA intensities and the expected intensities from \emph{Planck}.   Most clouds have reliable offset measurements with this technique.   In addition, we find that roughly half of the clouds have underestimated HSA-SPIRE intensities in their outer envelopes relative to \emph{Planck}, even though the HSA-SPIRE maps were previously zero-point corrected.  Using our technique, we produce corrected \emph{Herschel} intensity maps for all 56 clouds and determine their line-of-sight average dust temperatures and optical depths from modified black body fits.  The clouds have typical temperatures of $\sim 14-20$ K and optical depths of $\sim 10^{-5} - 10^{-3}$.  Across the whole sample, we find an anti-correlation between temperature and optical depth.  We also find lower temperatures than what was measured in previous \emph{Herschel} studies, which subtracted out a background level from their intensity maps to circumvent the zero-point correction.  Accurate \emph{Herschel} observations of clouds are key to obtain accurate density and temperature profiles.   To make such future analyses possible, intensity-corrected maps for all 56 clouds are publicly available in the electronic version.   
\end{abstract} 


\section{Introduction\label{Intro}}

Stars form in dense condensations (or cores) within molecular clouds \citep[e.g.,][]{MyersBenson83, williams00}.  Dense cores have typical temperatures of 10 K and densities of $\gtrsim 10^5$ \vol\ \citep{BerginTafalla07, difran07}.  Dense cores are also relatively quiescent, and are considered to be supported by thermal pressure \citep[e.g,][]{Pineda10}.  For such thermally-supported cores, the critical Jeans mass is $\sim 1\ \Msun$ \citep{McKee07}.  Cores beyond the critical Jeans mass are expected to collapse and form one star or a small stellar system.  

Most cores are associated with molecular clouds that span roughly $\sim 10$ pc in scale \citep{BerginTafalla07, Dunham14}.   Numerous surveys have explored the core and young star populations in these clouds, identifying hundreds of objects in each \citep[e.g.,][]{Gutermuth09, Dunham15, Konyves15, Mairs16}.  Cores are also found in smaller ($< 1$ pc), low-mass clouds along the outskirts of these larger cloud complexes \citep{Leung82, LaunhardtHenning97}.  Hereafter called ``globules'' \citep{BokReilly47, Bok48}, these small clouds have typical masses of $\sim 1-10\ \Msun$ and will contain only one or two dense cores \citep{Reipurth08, Launhardt10}.  Some globules have already formed stars  \citep[e.g.,][]{YunClemens92, Stutz10}, whereas others are entirely starless \citep[e.g.,][]{Crapsi07}.

Globules also have relatively simple structures.  They often have round morphologies, with slight deviations due to filaments or cometary features like tails \citep[e.g.,][]{Leung85, Stutz08, Stutz09, Tobin10, Launhardt13}.  Indeed, the density structures of starless globules are often well fit by simple models of hydrostatic equilibrium \citep[e.g.,][]{Alves01, Kandori05}.  By comparison, the density structures of cores in high mass star-forming regions or in more clustered environments are less clear.  These sources have more complicated properties due to the turbulence from the larger cloud and nearby young stars with outflows, or confusion with neighbouring sources \citep{Reipurth08}.    Thus, the relative isolation and simple structures of globules provide the best means to examine core stability \citep[e.g.,][]{KetoField05, Keto06} and the processes that connect the chemistry, kinematics, magnetic fields, and radiative transfer of dense cores with star formation \citep[e.g.,][]{Tafalla04, Marka12, Bertrang14, Keto14, Keto15}.

Thousands of globules have been identified to date, primarily through optical and near-infrared extinction maps \citep[e.g.,][]{ClemensBarvainis88, Bourke95iras, DutraBica02}.  Nevertheless, only a handful have been well studied.  To explore the physical properties of globules, we need good maps of column density to infer their density structures and masses.  Previous assessments with extinction maps \citep[e.g.,][]{Alves01} were mainly limited to nearby globules due to coarse angular resolutions.  Observations from ground-based (sub)millimeter telescopes provided the necessary spatial resolution to probe the density profiles of globules from thermal dust emission, but alone, these data lack the wavelength coverage to constrain even simple models of power-law models of density \citep[e.g.,][]{MotteAndre01, Shirley02}.  

More recently, observations at $100-500$ \um\ from the \emph{Herschel} Space Observatory \citep{Pilbratt10} have provided the necessary resolution and wavelength coverage to reliably map the density structure of globules from thermal dust emission.  These bands are ideal as they trace the peak of the spectral energy distributions of globules, which have with typical dust  temperatures of $10-20$ K \citep{difran07, Andre14}.  The \emph{Herschel} Science Archive (HSA) contains multi-wavelength data for over 60 globules, although only twelve have so far been studied in detail as part of the ``Early Phases of Star Formation'' (EPoS) survey \citep[e.g.,][]{Stutz10, Nielbock12, Launhardt13}.  The initial EPoS globules were  selected because they have relatively weak far-infrared backgrounds.  As such, they may not be representative of a typical globule.  A larger sample of globules is needed to better understand their properties for a range of masses, stages, and environments.  

Using \emph{Herschel} data of globules is not straightforward, however.  \emph{Herschel} observations do not include absolute flux calibrations, meaning the resulting maps give only relative intensities.  Absolute intensities are critical to accurately convert thermal dust emission to mass and density, trace dust temperatures, and compare with complementary observations (e.g., dust extinction and molecular line emission).  Previous studies have developed methods to correct \emph{Herschel} observations using \emph{Planck} data \citep[e.g.,][]{Bernard10, Lombardi14, Abreu17}.   These methods, however, are not easily applicable to the smaller \emph{Herschel} maps of globules.  The globule maps are typically $2-7$ \emph{Planck} beams across, which makes it difficult to reliably convolved them to \emph{Planck} scales \citep{Bernard10, Lombardi14} or to reliably bridge the spatial scales covered by each telescope in a Fourier analysis \citep{Abreu17}.   Zero-point corrections of the globule maps will require a different technique. 

In this paper, we introduce two methods to correct \emph{Herschel} observations of globules.  We apply these techniques to 56 globules from the HSA, producing the largest database of far-infrared maps of globules to date.   These corrected \emph{Herschel} maps will greatly improve models of density and temperature in globules, which are necessary for chemical models and radiative transfer.  In Section \ref{data}, we describe the \emph{Herschel} and \emph{Planck} data used in this analysis.   In Section \ref{method}, we outline our method for correcting the \emph{Herschel} data.  In Section \ref{results} we determine the zero-point corrections for each map over five \emph{Herschel} wavelengths.  In Section \ref{sed_section}, we produce maps of temperature and optical depth for each globule using our corrected \emph{Herschel} data, and in Section \ref{discussion}, we compare these maps to independent measurements in the literature.  Finally, we summarize our results in Section \ref{summary}.


\section{Data}\label{data}

\subsection{Herschel Data}\label{herschel_obs}

We select 56 low-mass, nearby globules from five \emph{Herschel} surveys with observations from $100-500$ \um.  Table \ref{table_loc} lists the 56 globules with their names from their respective surveys.  The second and third columns give the J2000  right ascension and declination coordinates of the globule centers adopted in our work (see Section \ref{method}).  The fourth column identifies the \emph{Herschel} proposal that observed each globule.  The fifth column names the nearest cloud or cluster association.  The sixth column gives the estimated distance for each globule with references in the final column.  With their small sizes and relative isolation, it is difficult to get accurate distances to globules \citep{Yun01}.  As such, most globules do not have direct distance measurements in the literature.  In these cases, we use the distances of their nearest associations (see Table \ref{table_loc}). 

\begin{deluxetable*}{lccllll}
\tablecaption{Nearby Globules in the \emph{Herschel} Archive \label{table_loc}}
\tablehead{
\multirow{1}{*}{Globule }  & \colhead{RA (J2000)} & \colhead{Dec (J2000)} & Proposal ID	& Assocation & Distance (pc) &\colhead{References} 
}
\startdata
CB 4	&	00:39:04.2		&	+52:51:16	&	KPGT\_okrause\_1	&				& $460 \pm 85$	&	1	\\
CB 6	&	00:49:24.7		&	+50:44:50	&	KPGT\_okrause\_1	& CB 4			& $460 \pm 85$	&	1\\
CB 17	&	04:04:35.6	&	+56:56:07	&	KPGT\_okrause\_1	&				& $480 \pm 90$	&	2	\\
L1521F	&	04:28:39.1	&	+26:51:34	&	OT1\_mdunham\_1	& Taurus			& $135 \pm 40$	&	3	\\
CB 26	&	04:59:50.7	&	+52:04:42	&	KPGT\_okrause\_1	& Taurus-Auriga	& $140 \pm 40$	&	4,5,6	\\
L1544	&	05:04:13.1	&	+25:11:05	&	KPGT\_okrause\_1	& Taurus			& $140 \pm 40$	&	4,5,6	\\
CB 27	&	05:05:09.3	&	+32:42:42	&	KPGT\_okrause\_1	& $\alpha$ Persi	& $180 \pm 10$	&	7	\\
L1552	&	05:17:39.2	&	+26:04:50	&	GT2\_astutz\_2		& Taurus			& $140 \pm 40$	&	4,5,6		\\
CB 29	&	05:22:12.6	&	-03:41:35	&	OT2\_tbourke\_3	& Ori OB1a		& $340 \pm 20$	&	7		\\
B 35A	&	05:44:29.4	&	+09:08:53	&	OT1\_mdunham\_1	& Orion Lam		& $400 \pm 30$	&	4	\\
BHR 22	&	07:14:10.2	&	-48:31:25	&	OT1\_mdunham\_1	& Vela OB2		& $410 \pm 10$	&	7	\\
BHR 17	&	07:19:21.7	&	-44:34:54	&	OT2\_tbourke\_3	& Vela OB2		& $410 \pm 10$	&	7	\\
BHR 16	&	08:05:26.0	&	-39:09:07	&	OT1\_mdunham\_1	& Vela OB2		& $250 - 410$	&	7,8	\\
BHR 12	&	08:09:33.0	&	-36:05:11	&	KPGT\_okrause\_1	& Vela OB2 		& $200 - 410$	&	7,9	\\
DC2573-25	&08:17:01.1	&	-39:48:06	&	OT1\_mdunham\_1	& Vela OB2		& $410 \pm 10$	&	7	\\
BHR 31	&	08:18:43.1	&	-49:43:24	&	OT2\_tbourke\_3	& Vela OB2		& $410 \pm 10$	&	7	\\
BHR 42	&	08:26:11.6		&	-51:39:04	&	OT2\_tbourke\_3	& Vela OB2		& $410 \pm 10$	&	7	\\
BHR 34	&	08:26:31.8	&	-50:39:48	&	OT2\_tbourke\_3	& Vela OB2		& $200 - 410$	&	7,8	\\
BHR 41	&	08:27:39.1	&	-51:10:39	&	OT2\_tbourke\_3	& Vela OB2, BHR 34	 & $200 - 410$	&	7,8	\\
BHR 40	&	08:31:58.8	&	-50:32:30	&	OT2\_tbourke\_3	& Vela OB2, BHR 34	 & $200 - 410$	&	7,8	\\
BHR 38/39	&08:34:06.6	&	-50:18:22	&	OT2\_tbourke\_3	& Vela OB2		& $450 \pm 50$	&	10	\\
BHR 56	&	08:44:02.6	&	-59:54:05	&	OT2\_tbourke\_3	&				& $490 \pm 50$	&	11	\\
DC2742-04	&09:28:51.5	&	-51:36:00	&	OT1\_mdunham\_1	& 				& $200 - 500$	& 8,10 \\
BHR 48/49	&09:36:25.8	&	-48:52:16	&	OT2\_tbourke\_3	& BHR 55 		& $300 \pm 50$	&	10	\\ 
BHR 50	&	09:41:36.9	&	-48:41:38	&	OT2\_tbourke\_3	& BHR 55		 	& $300 \pm 50$	&	10	\\
BHR 68	&	11:50:02.0		&	-58:32:18	&	OT2\_tbourke\_3	& Lower Cen-Crux 	& $120 - 350$	& 7,10	\\
BHR 71	& 	12:01:36.1	&	-65:08:49	&	OT1\_jtobin\_1		& Coalsack		& $150 \pm 30$	& 12 \\
BHR 74	&	12:22:14.1	&	-66:28:00	&	OT2\_tbourke\_3	& Coalsack		& $175 \pm 50$	& 8 \\
BHR 79	&	12:37:22.5	&	-69:28:59	&	OT2\_tbourke\_3	& Musca			& $150 \pm 30$	& 12 \\
BHR 81	&	12:39:37.0	&	-65:25:20	&	OT2\_tbourke\_3	& Coalsack		& $150 \pm 30$	& 12 \\
DC3162+51	&14:26:07.1	&	-55:20:27	&	OT2\_tbourke\_3	& Upper Cen-Lup	& $140 \pm 50$	& 7 \\
BHR 95	&	14:53:28.4	&	-61:35:13	&	OT2\_tbourke\_3	& Circinus, BHR 100	& $350 \pm 50$	& 10 \\
BHR 99	&	15:24:57.1	&	-61:01:42	&	OT2\_tbourke\_3	& Circinus, BHR 100  & $350 \pm 50$	& 10 \\
BHR 100	&	15:25:42.1	&	-61:06:58	&	OT2\_tbourke\_3	& Circinus 		& $350 \pm 50$	& 10 \\
BHR 97	&	15:27:14.8	&	-62:22:29	&	OT2\_tbourke\_3	& Circinus, BHR 100  & $350 \pm 50$	& 10 \\
DC3391+117	&15:59:05.2	&	-37:36:22	&	OT1\_mdunham\_1	& Lupus			& $150 \pm 10$	& 13,14 \\
DC3460+78	&16:36:53.2	&	-35:36:52	&	OT1\_mdunham\_1	& Lupus			& $150 \pm 10$	& 13,14 \\
CB 68	&	16:57:19.4	&	-16:09:21	&	KPGT\_okrause\_1	& Ophiuchus		& $120 \pm 20$ 	& 13,15	\\
BHR 147	&	16:58:31.1	&	-36:42:19	&	OT2\_tbourke\_3	& Lupus, HIP 82747	& $150 \pm 40$	& 16 \\
B 68		&	17:22:38.1	&	-23:50:14	&	KPGT\_okrause\_1	& Pipe			& $140 \pm 20$	& 17,18 \\
CB 101	&	17:53:08.7	&	-08:27:10	&	OT2\_tbourke\_3	& Aquila Rift		& $270 \pm 55$	& 19,20 \\
L422	&	18:12:03.7		&	-08:05:21	&	OT2\_tbourke\_3	& Aquila Rift		& $270 \pm 55$	& 19,20 \\
CB 130	&	18:16:16.3	&	-02:32:40	&	KPGT\_okrause\_1	& Aquila Rift		& $270 \pm 55$	& 19,20 \\
L429	&	18:17:05.5		&	-08:14:41	&	GT2\_astutz\_2		& Aquila Rift		& $270 \pm 55$	& 19,20 \\
L483 &	18:17:29.9		&	-04:39:41 &	OT1\_jtobin\_1		& Aquila Rift		& $270 \pm 55$	& 19,20 \\
CB 170	&	19:01:36.2	&	-05:26:23	&	OT2\_tbourke\_3	& 				& $180 \pm 35$		& 21 \\
CB 175	&	19:02:08.5	&	-05:19:19	&	OT2\_tbourke\_3	& 				& $200 \pm 40$		& 21 \\
CB 176\tablenotemark{a}	&	19:02:15.1	&	-04:22:52	&	OT2\_tbourke\_3	&  CB 175			& $200 \pm 40$		& 21 \\
L723	&	19:17:53.6		&	+19:12:16	&	OT1\_mdunham\_1	& 				& $300 \pm 150$		& 22 \\
L673	&	19:20:25.3		&	+11:22:14	&	OT1\_mdunham\_1	& CB 188			& $260 \pm 50$		& 2,3 \\
B 335	&	19:37:00.8		&	+07:34:07	&	KPGT\_okrause\_1	& 				& $105 \pm 15$		& 23 \\
CB 230	&	21:17:38.3	&	+68:17:26	&	KPGT\_okrause\_1	& Cepheus		& $295 \pm 55$		& 2 \\
L1014	&	21:24:06.9	&	+49:59:00	&	OT1\_mdunham\_1	& Northern Coalsack & $260 \pm 50$		& 3 \\
L1165	&	22:06:50.6	&	+59:02:43	&	OT1\_mdunham\_1	& HD 209811		& $300 \pm 50$		& 24 \\
L1221	&	22:28:07.0	&	+69:00:39	&	OT1\_mdunham\_1	& L1219			& $400 \pm 50$		& 25 \\
CB 244	&	23:25:44.8	&	+74:17:36	&	KPGT\_okrause\_1	& Cepheus		& $180 \pm 40$		& 25 \\
\enddata
\tablenotetext{a}{We assume the same distance as CB 175, but caution that CB 175 and CB 176 have different gas velocities and may not be related.  CB 176 has a velocity of $\approx 16$ \kms, whereas CB 175 is at $\approx 10$ \kms\ \citep{Clemens91}.}
\tablecomments{We adopt errors of 50 pc for distances measurements without reported uncertainties.  References for distances also indicate the measurement method and if the distance was not measured for the globule directly. (1) \citealt{Barman15} (reddening), (2) \citealt{Das15} (reddening), (3) \citealt{Maheswar11} (reddening), (4) \citealt{Kenyon94} (reddening for Taurus), (5) \citealt{Schlafly14} (reddening for Taurus), (6) \citealt{Torres07} (stellar parallax for Taurus), (7) \citealt{deZeeuw99} (stellar parallax for $\alpha$ Persi, Vela, Lower Cen-Crux, Upper Cen-Lup), (8) \citealt{Racca09} (reddening), (9) \citealt{Knude99} (reddening), (10) \citealt{Bourke95} (reddening), (11) \citealt{Vieira03} (stellar photometry to Herbig Ae star GSC 8581-2002), (12) \citealt{Corradi97} (reddening for Coalsack and Musca), (13) \citealt{Lombardi08} (parallax for Lupus), (14) \citealt{Crawford00} (sodium absorption in Lupus), (15) \citealt{Loinard08} (parallax for Ophiuchus), (16) \citealt{vandenAncker98} (Hipparcos parallax to Herbig Ae star, HIP 82747), (17) \citealt{Lombardi06} (reddening+parallex for Pipe), (18) \citealt{AlvesFranco07} (polarization+parallax for Aquila), (19) \citealt{Lallement14} (reddening for Aquila), (20) \citealt{Straizys03} (reddening for Aquila), (21) \citealt{MaheswarBhatt06} (reddening), (22) \citealt{Goldsmith84} (reddening), (23) \citealt{Olofsson09} (extinction to background stars), (24) \citealt{Gyulbudagyan85} (association with star HD 209811 with the parallax distance from \citealt{Gaia1}), (25) \citealt{Kun98} (stellar photometry). }
\end{deluxetable*}


We use photometry maps from the HSA from the Photodetector Array Camera and Spectrometer \citep[PACS;][]{Poglitsch10} and the Spectral Photometric Imaging Receiver \citep[SPIRE;][]{Griffin10}.   For the PACS data, we use the Level 2.5 data products at 100 \um\ and 160 \um.  Since only a few globules have 70 \um\ observations with PACS, we do not include this band in our analysis.  The PACS 100 \um\ and 160 \um\ maps were made with the PACS-only small map observing mode and have typical sizes of $\lesssim 10$\arcmin.  The HSA PACS data were reduced using version 14.2.0 of the pipeline.  For our analysis, we assume effective beam sizes of 7\arcsec.1 and 11\arcsec.2 for the 100 \um\ and 160 \um\ bands, respectively \citep[e.g.,][]{Aniano11}. 

For the SPIRE data, we use the Level 2 data products from the HSA at 250 \um, 350 \um, and 500 \um.   These maps were observed with the SPIRE-only large scan observing mode and cover areas of typically $30-50$\arcmin.  L1521F and L1544 did not have dedicated SPIRE-only observations, and as such, we use the Level 2.5 data products from the larger PACS/SPIRE parallel mode observations taken by the \emph{Herschel} Gould Belt Survey \citep[HGBS,][]{Andre10}.  In general, the HGBS clouds cover a much larger area, but the globule-specific maps have better sensitivities by a factor up to a factor of $\sim 2$.   The HSA SPIRE data were reduced using version 14.1.0 of the pipeline.  For our analysis, we assume effective beam sizes of 18\arcsec.2, 24\arcsec.9, and 36\arcsec.3 for the 250 \um, 350 \um, and 500 \um\ bands, respectively \citep[e.g.,][]{Griffin10}.  

\subsection{Planck Data}\label{planck_obs}

We use the \emph{Planck} data products from the \emph{Planck} 2013 all-sky model of thermal dust emission \citep{Planck_allsky_2014}\footnote{The 2013 all-sky \emph{Planck} data products were taken from https://wiki.cosmos.esa.int/planckpla/index.php/CMB\_and\_astro-physical\_component\_maps.}.  These data products give the parameters from modified blackbody fits to IRAS and \emph{Planck} spectral energy distributions (SEDs) from 100 \um\ to 2 mm.  The fitted parameters are dust temperature, $T_d$, dust emissivity index, $\beta$, and dust optical depth at 353 GHz, $\tau_{353}$ (see also Section \ref{sed_section} for explanations of SED fitting).  The temperature and optical depth maps have 5\arcmin\ resolution, whereas the dust emissivity index map has 30\arcmin\ resolution.  For each globule, we extracted smaller 3$\degree$ maps of $T_d$, $\tau_{353}$, and $\beta$, using barycentric interpolation to convert the all-sky data from a HEALPix system \citep[e.g.,][]{Gorski05} to standard Cartesian coordinates.    This method is a simple, first-order linear interpolation similar to bilinear interpolation, but instead interpolates using a triangulation of the three nearest neighbors.Ê This routine is useful in cases where the input data are not on Cartesian grids. 

The \emph{Planck} 2013 all-sky maps of temperature, optical depth, and dust emissivity index provide the best measurements of the SED parameters for our globules.   There are more recent \emph{Planck} data products that include two temperature components \citep[e.g.,][]{MeisnerFinkbeiner15} or more sophisticated methods to subtract the cosmic infrared background \citep[e.g.,][]{Planck_gnilc_2016}.  These products, however, subtracted out point sources from the \emph{Planck} and IRAS data to avoid artifacts when bright, compact objects in the higher resolution maps are convolved to lower resolution.  Since our globules generally appear as point sources with Planck, these products are unsuitable for our analysis. 

\section{Zero-Point Corrections} \label{method}

\emph{Herschel} does not measure absolute fluxes due to an unknown instrumental thermal background.  In practice, one can measure a zero-point correction using a clean background level (e.g., a clean background should have zero emission).  \emph{Herschel} maps, however, do not generally include locations without emission due to widespread emission at far-infrared and (sub)millimeter wavelengths throughout the Galaxy.  Hence, they require zero-point corrections that are estimated from comparisons to calibrated data from other facilities.  

Zero-point corrections are applied to SPIRE maps at Level 2 or higher from the HSA.  These corrections are based on \emph{Planck} observations at 545 GHz and 857 GHz.  In brief, the HSA calculates ``color corrections'' to determine the emission that \emph{Herschel} would detect from the observed \emph{Planck} data.  These color corrections are calculated from the shapes of the \emph{Planck} filters relative to the SPIRE filters, assuming a typical SED for the dust emission.  The corrections are most reliable for the 350 \um\ and 500 \um\ SPIRE bands because their filters overlap well with the 857 GHz ($\approx$ 350 \um)  and 545 GHz  ($\approx$ 550 \um) bands, respectively.  For the SPIRE 250 \um\ band, the color corrections are extrapolated from the 857 GHz data, and are more sensitive to the assumed SED parameters\footnote{See the SPIRE Data Reduction Guide and SPIRE Handbook for more details.}.  

\begin{figure}[h!]
\includegraphics[width=0.475\textwidth]{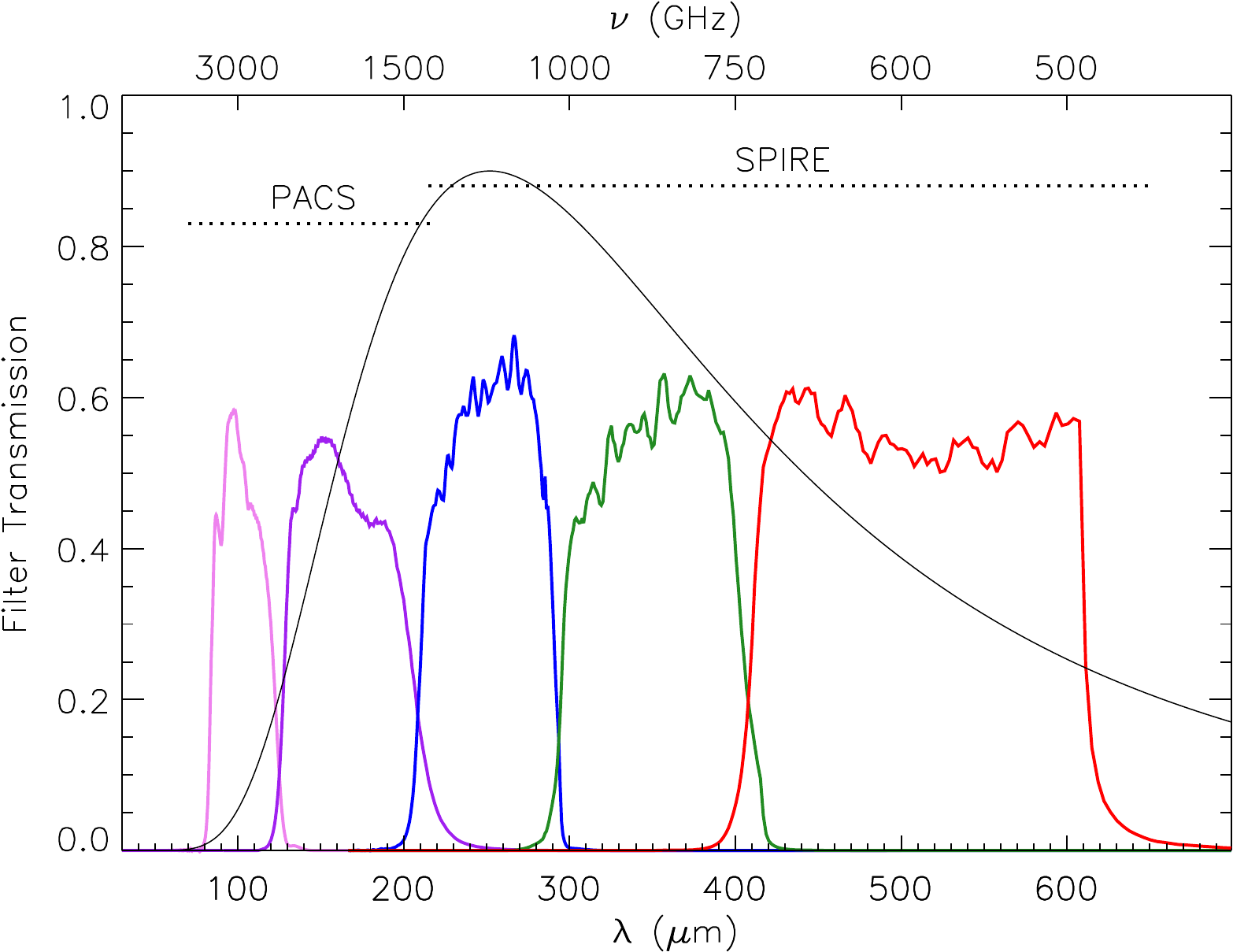}
\caption{\emph{Herschel} filters at 100 \um, 160 \um, 250 \um, 350 \um, and 500 \um\ (from left to right).  The black curve shows a modified blackbody function at a temperature of 10 K, which is representative of cold dust seen in the globules. The filter functions are available from the instrument calibration context within HIPE.  \label{filters}}
\end{figure}

In contrast, the PACS Level 2.5 maps in the HSA are \emph{not} zero-point corrected.   \citet{Lombardi14} outlines the methodology for such calculations for large \emph{Herschel} maps, which we also follow in this paper.  First, we use the all-sky \emph{Planck} maps of temperature, optical depth, and dust emissivity index to reconstruct the modified blackbody function for each pixel at 5\arcmin\ resolution.  Second, we integrate these blackbody functions over the \emph{Herschel} filter functions to determine the expected emission that would be detected by \emph{Herschel}.  Figure \ref{filters} compares a sample modified blackbody function with the PACS and SPIRE filter functions at 100 \um, 160 \um, 250 \um, 350 \um, and 500 \um.  For simplicity, we use the point-source filters for the SPIRE bands and apply an extended source correction (typically less than 1\%) to account for extended emission.   

\begin{figure}[h!]
\includegraphics[width=0.475\textwidth]{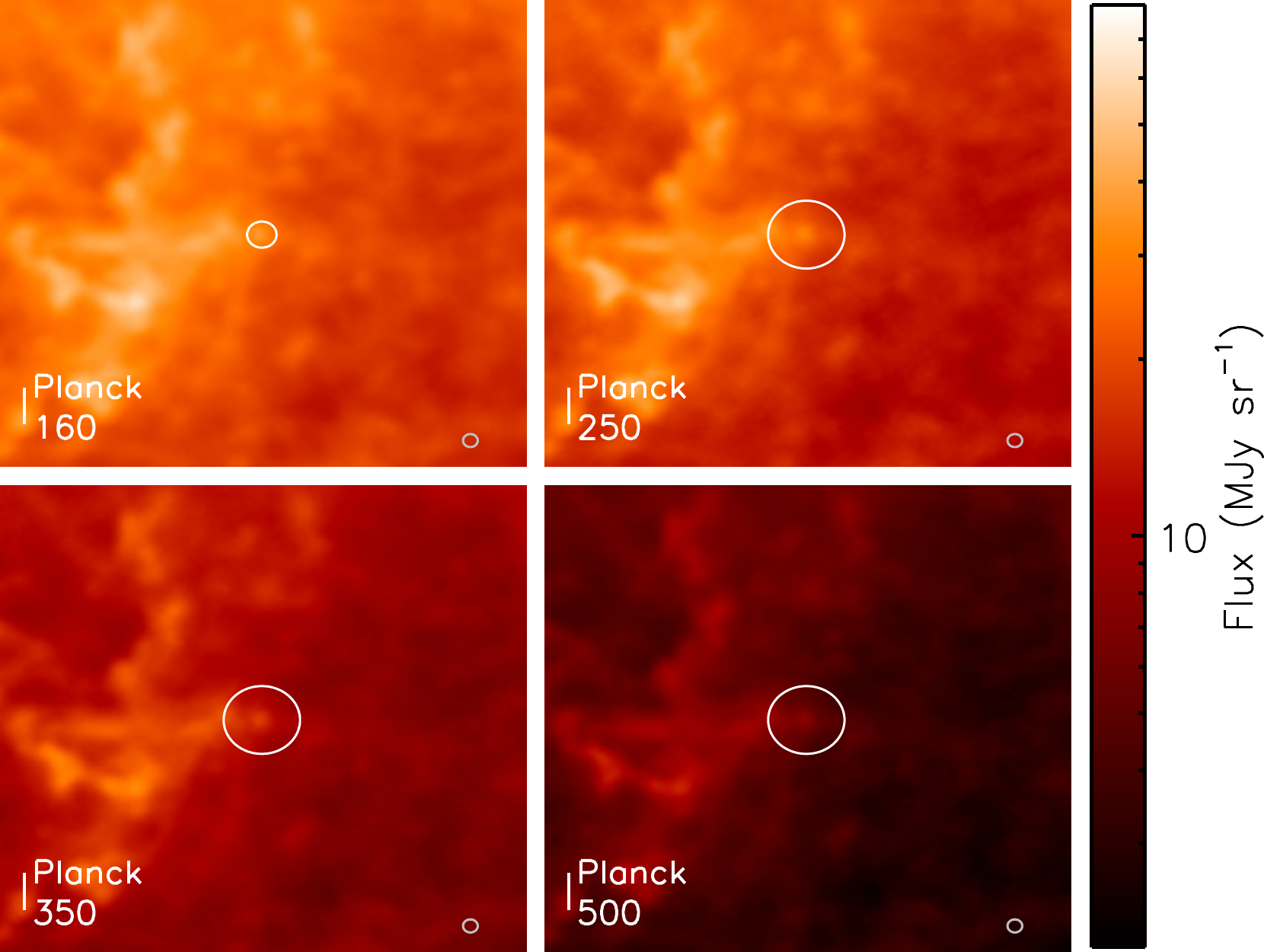}
\caption{\emph{Planck}-determined intensity maps at 160 \um, 250 \um, 350 \um, and 500 \um\ for a 3\degree\ field around CB 4.  These maps are made using the \emph{Planck} all-sky SED model parameters to produce modified blackbody functions, which are then integrated over the \emph{Herschel} filter functions.   The corresponding map at 100 \um\ is not shown.  All maps are on the same logarithmic color scale and at a common resolution of 5\arcmin.  The larger white circle shows the approximate size of corresponding \emph{Herschel} observations of CB 4 and the smaller grey circle shows the 5\arcmin\ beam resolution.  \label{cb4_expected_fluxes}}
\end{figure}

Figure \ref{cb4_expected_fluxes} shows an example of the expected \emph{Herschel} intensity maps at 160 \um, 250 \um, 350 \um, and 500 \um\ for CB 4.  The maps span 3\degree\ and have a resolution of 5\arcmin.  Hereafter, we call these results \emph{Planck}-determined intensity maps, and we represent them by the symbol $I_{\lambda}^{Planck}$, where $\lambda$ indicates the wavelength of the corresponding \emph{Herschel} band.  Figure \ref{cb4_expected_fluxes} also shows the approximate size of the PACS and SPIRE observations for CB 4.  These map sizes are comparable in size to that of the other globules in our study.

\citet{Lombardi14} found the zero-point corrections to their \emph{Herschel} maps of Orion by comparing the \emph{Herschel} intensity maps with their \emph{Planck}-determined intensity maps pixel-by-pixel, where the \emph{Herschel} observations were convolved to the same resolution and pixel scale as the \emph{Planck}-determined intensity maps \citep[see also,][]{Bernard10}.  From this comparison, they obtain an average zero-point correction for the entire map.  In an alternative approach, \citet{Abreu17} corrected \emph{Herschel} maps with \emph{Planck} data in Fourier space.  This approach uses the spatial information from \emph{Planck} to apply the zero-point correction locally rather than adopting a single value for the whole map as in the pixel-by-pixel case.  Local variations in the zero-point correction can be significant (up to roughly 50\%), particularly in the PACS bands.

Both techniques outlined above utilized large maps that span several degrees.    By contrast, the HSA maps of globules are much smaller.  Figure \ref{cb4_expected_fluxes} shows the typical map size of the globules compared to a \emph{Planck} beam size of 5\arcmin.  The PACS fields are only $\sim 10$\arcmin\ across for most globules.  Even the larger SPIRE maps are only $7-8$ beams across at 5\arcmin\ resolution.  With such small maps, we cannot reliably convolve them to \emph{Planck} resolutions for the pixel-by-pixel approach from \citet{Lombardi14}, nor can we reliably trace their emission over all spatial scales for the Fourier space approach from \citet{Abreu17}.  Instead, we propose an alternative measurement technique to determine the zero-point corrections local to each globule, which we describe below.

\subsection{The Radial Profile Method}\label{radial_profiles}

We use radial profiles of both the HSA maps and the \emph{Planck}-determined intensity maps to identify any offsets between the observed emission and the expected emission.  The radial profiles are constructed from azimuthally-averaged annuli from the center of the globule (as given in Table \ref{table_loc}).   For the PACS data, we initially mask out pixels at the edge of the map which tend to be noisy due to reduced coverage.  We use the PACS coverage maps to define the masks by excluding regions with low coverage relative to the center of the map.  Since the coverage maps vary with \emph{Herschel} project, time on source, and map size, we define the masks for each globule by eye with typical coverage levels that are 0.25-0.5 times lower than the  value in the map center. The resulting profiles are generally insensitive to the limit used to define the mask.  

\begin{figure}[h!]
\includegraphics[width=0.475\textwidth]{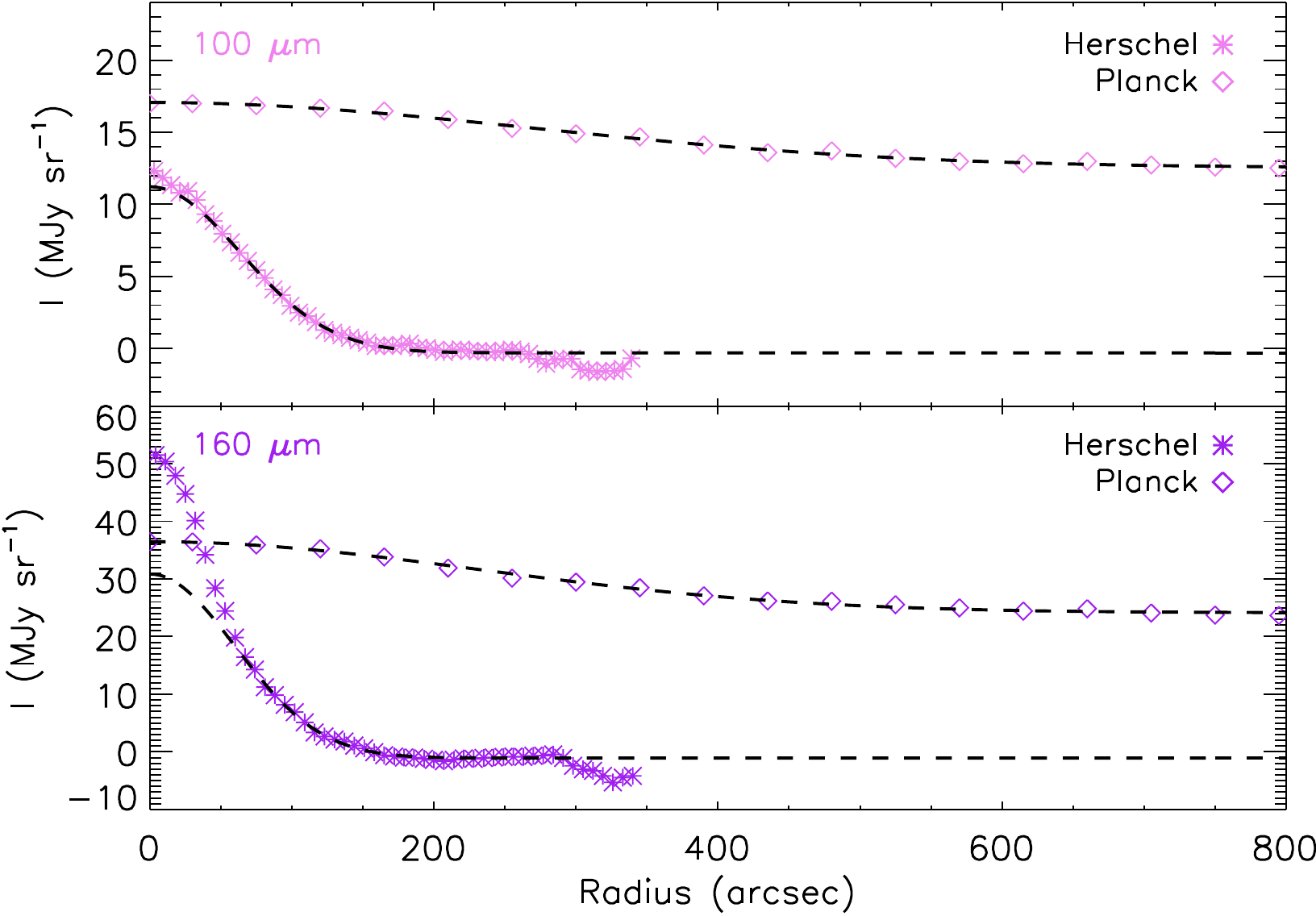}
\caption{Azimuthally-averaged radial profiles of observed intensity at 100$\ \um$\ (top) and 160 \um\ (bottom) for CB 4.  The profiles with star symbols correspond to the HSA-PACS intensities, whereas the open diamonds correspond to the expected emission from the \emph{Planck}-determined intensity maps (see the previous Section).  Black dashed curves show the best-fit Gaussians for each profile, excluding the emission peaks (see text). \label{profiles_cb4}}
\end{figure}

Figure \ref{profiles_cb4} shows the 100 \um\ and 160 \um\ radial profiles for CB 4 as an example.   Both the HSA-PACS and \emph{Planck}-determined profiles are centrally peaked due to emission from the globule, although the \emph{Planck}-determined profiles are much broader because of their lower resolution.  CB 4 has a semi-major axis of $\sim 2$\arcmin\ \citep{ClemensBarvainis88}, and is subsequently resolved by PACS and unresolved by \emph{Planck}.  The radial profiles also flatten out at large angular extents of $\gtrsim 200$\arcsec\ for PACS and $\gtrsim 500$\arcsec\ for \emph{Planck}.  The emission at large angular distances from the globule should primarily trace the large-scale, diffuse background material.  In the absence of small-scale structure, the HSA-PACS and \emph{Planck}-determined intensities should match at these large angular extents.   In contrast, Figure \ref{profiles_cb4} shows a large intensity offset between the HSA-PACS and \emph{Planck}-determined profiles.  We attribute these offsets to the missing zero-point corrections.

To measure the intensity offset between the HSA-PACS and the \emph{Planck}-determined intensity maps, we fit their radial profiles with Gaussian functions to identify their respective intensities at large angular extents.  The dashed curves in Figure \ref{profiles_cb4} show the corresponding best-fit Gaussian functions.  For the \emph{Planck}-determined profiles, we typically fit Gaussians for angular extents $<800$\arcsec. For the HSA-PACS profiles, we exclude the emission peak to ensure a good fit at large angular extents. (Note that we are not interested in fitting the emission peaks.)  Since some globules in our sample have a range of profiles from those with sharp intensity peaks to those that are very flat, we cannot use a fixed radial limit to measure the intensity offsets.  Instead, we require at least 300 pixels in the annuli at 160 \um\ and at least 500 pixels in the annuli at 100 \um\ for the Gaussian fits.  For 14 globules (labeled in Table \ref{table_pacs}), we also truncate the upper radius used in the Gaussian fits to $300-400$\arcsec\ to exclude sudden changes in emission at the edge of the profiles that deviate from the general trend.  These jumps are not seen in the \emph{Planck} profiles. 

\begin{figure}[h!]
\includegraphics[width=0.475\textwidth]{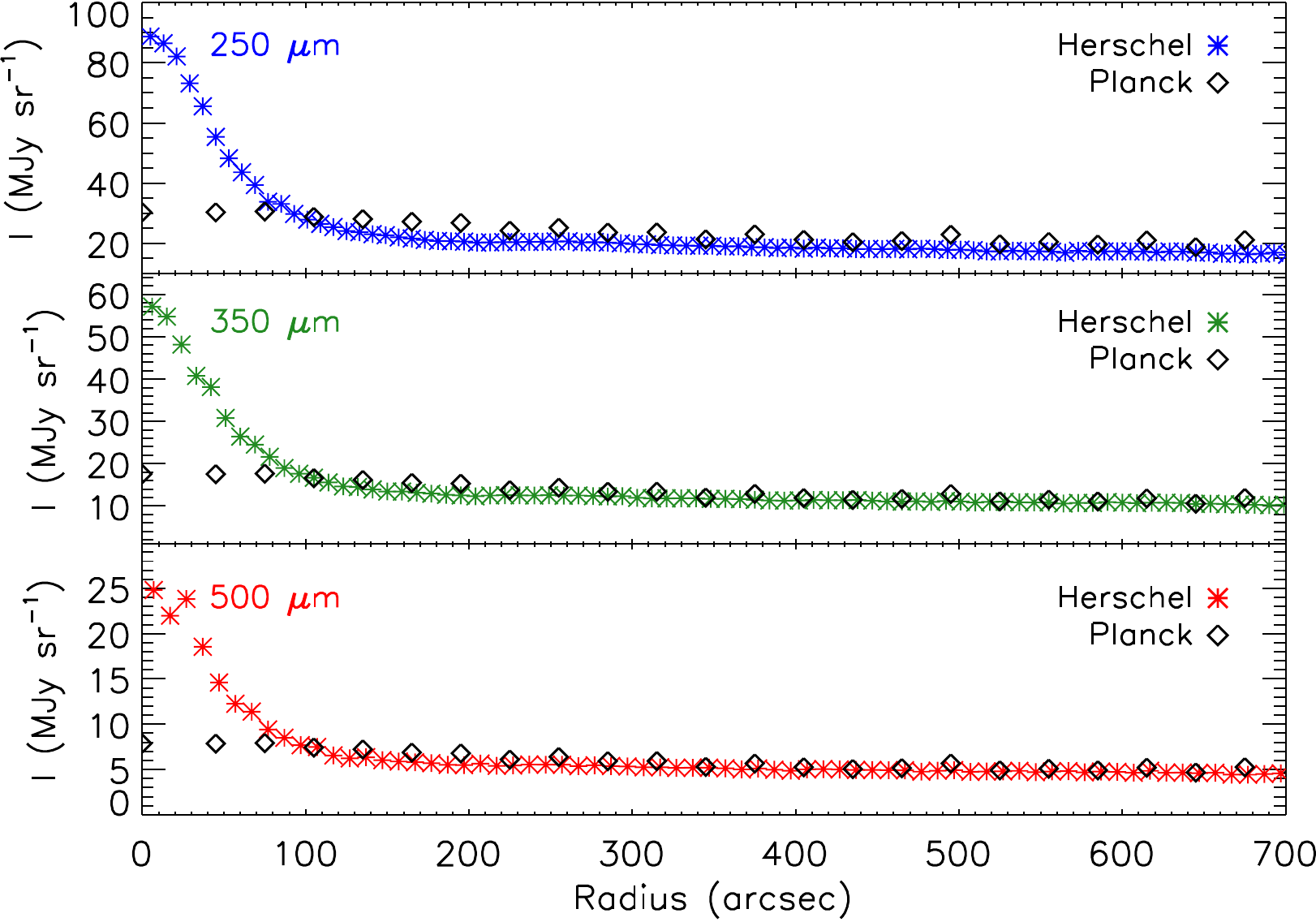}
\caption{Same as Figure \ref{profiles_cb4} except for profiles at 250 \um, $350\ \um$, and 500 \um. Gaussian fits to the \emph{Herschel} and reconstructed \emph{Planck} profiles (not shown) suggest the two profiles agree at angular extents $\gtrsim 300$\arcsec\ within 10\%.  \label{profiles_cb4_spire}}
\end{figure}

The radial profile method assumes that the HSA-PACS and \emph{Planck}-determined intensities should be equal at large angular extents from the cloud centers.  Indeed, we find better agreement between the HSA-SPIRE intensities, which were previously zero-point corrected (see Section \ref{herschel_obs}), and their corresponding \emph{Planck}-determined intensities at large angular extents.  Figure \ref{profiles_cb4_spire} shows the radial profiles from the HSA-SPIRE and \emph{Planck}-determined maps of CB 4 at 250 \um, 350 \um, and 500 \um.  At angular extents $> 300$\arcsec, the HSA-SPIRE and \emph{Planck}-determined intensities agree within 10\%.  Thus, it is reasonable to assume that the HSA-PACS intensities should also agree with \emph{Planck} at large angular distances.   

\subsection{Offset Groups}\label{slices}

\begin{figure}[h!]
\includegraphics[width=0.475\textwidth]{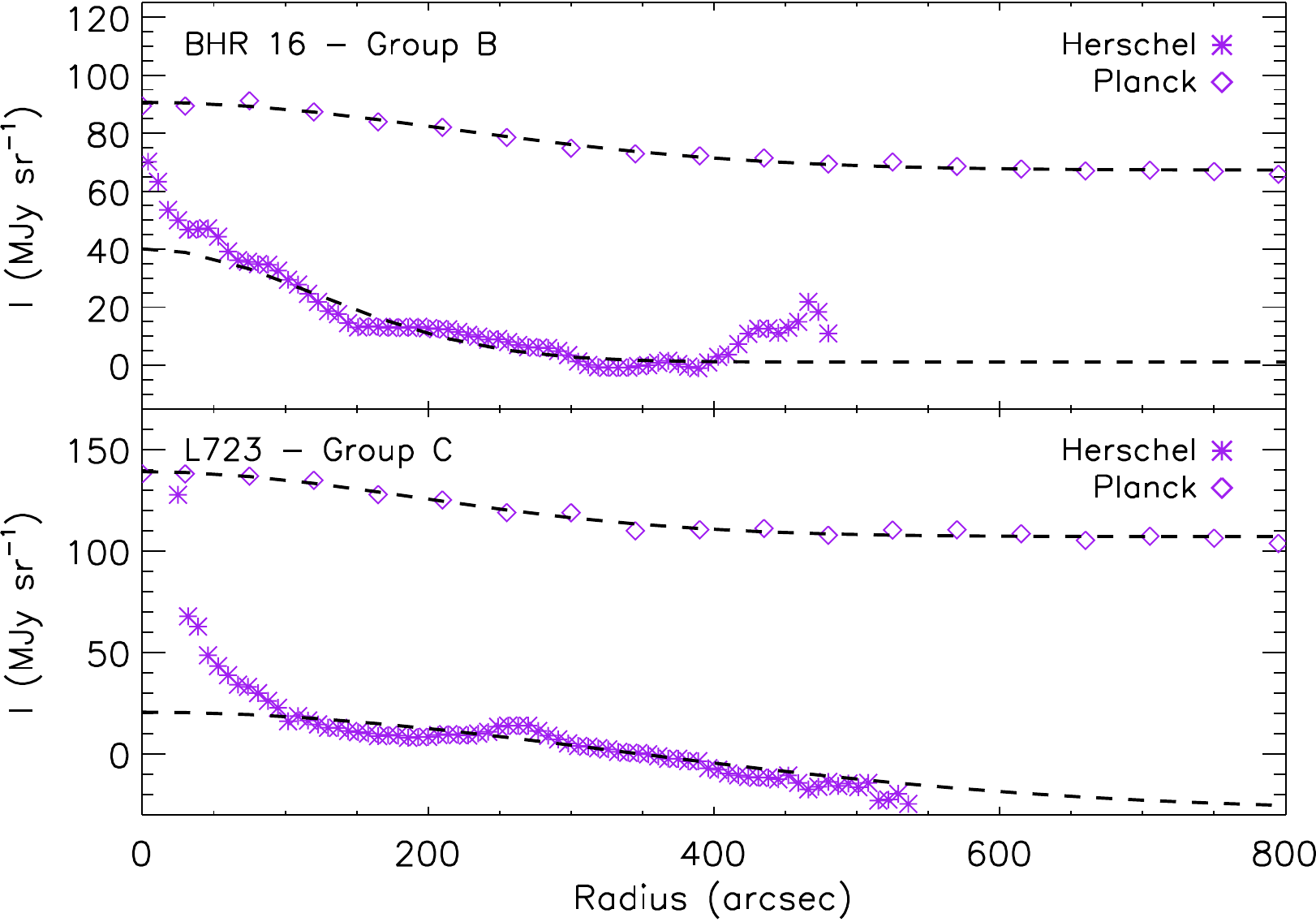}
\caption{Radial profiles at 160 \um\ of BHR 16 (top) and L723 (bottom).  The symbols are the same as in Figure \ref{profiles_cb4}. BHR 16 is considered Group B, whereas L723 is Group C (see text for group definitions). \label{profiles_group}}
\end{figure}

Figure \ref{profiles_cb4} shows the typical offset fit results for a cloud with radial profiles that are well-fit by Gaussians.  Not all globules have such clean radial profiles, however.  Figure \ref{profiles_group} shows the HSA and \emph{Planck}-determine radial profiles at 160 \um\ with their best-fit Gaussians for BHR 16 (top) and L723 (bottom) as examples of more complicated clouds.   BHR 16 and L723 do not have flat HSA radial profiles at large angular extents ($> 200\arcsec$) from their centers, which makes their best-fit Gaussians less reliable.  In the case of BHR 16, we can still fit a Gaussian function to its HSA 160 \um\ radial profile, although there is a larger margin of uncertainty due to its wavy structure.  For L723, the HSA 160 \um\ radial profile continuously decreases for angular extents $>300$\arcsec, and as such, we cannot get a reasonable measure of its background level using Gaussian fits (see Figure \ref{profiles_cb4}).  We need an additional measure of the offsets to test the reliability of the radial profile method for clouds with substructure like BHR 16 or to estimate the offset for clouds with radial profiles that are not well characterized like L723.

For our secondary measurements, we use intensity slices through the HSA and \emph{Planck}-determined maps.  We use the median values from the HSA and \emph{Planck} slices over the same angular extents to estimate the intensity offsets between them.  For simplicity, we take slices through the centers of the globules (see Table \ref{table_loc}) along right ascension and declination, excluding the central 400\arcsec\ to avoid any biases from a bright, central source.   (For BHR 71 and L483, which are smaller maps, we exclude the central 200\arcsec\ from the HSA-PACS slices to have a large enough sample of pixels.) Some clouds have no bright central peak, especially at 100 \um.  For these clouds, we use the median values across the entire slice for better statistics.  

\begin{figure}[h!]
\includegraphics[width=0.475\textwidth]{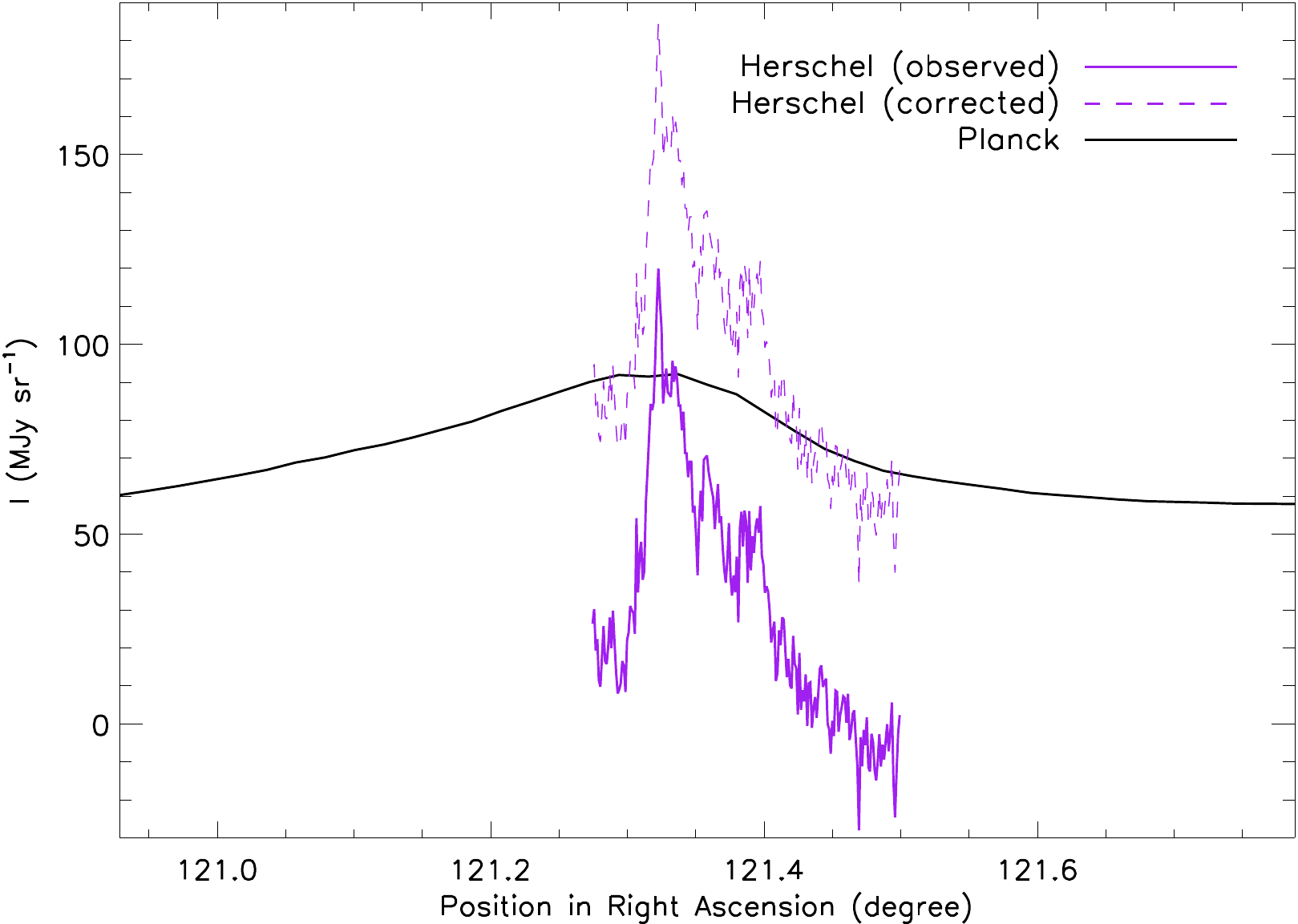}
\caption{Intensity slices along right ascension at 160 \um\ through the center of BHR 16.  The black curve shows the slice from the \emph{Planck}-determined 160 \um\ map and the solid purple curve shows the slice from the HSA-PACS 160 \um\ map.  The dashed purple curve shows the ``corrected'' PACS 160 \um\ distribution after applying an offset correction of 64.5 \MJysr.   \label{slice_bhr16}}
\end{figure}

Figure \ref{slice_bhr16} shows the HSA-PACS and \emph{Planck}-determined intensity slices at 160 \um\ through BHR 16 in right ascension.  The black curve shows the slice through the corresponding \emph{Planck}-determined map, whereas the purple solid curve shows the profile through the same slice from the HSA data.  We use the same mask as the radial profile method to exclude noisy edge pixels for cleaner slices.  The intensity slices show a clear offset between the two profiles.  The dashed purple curve shows the ``corrected'' PACS 160 \um\ slice using the median intensities as described above.  

The intensity slices are harder to constrain than the radial profiles because the measured offset can vary with different position angles.  Indeed, the radial profile method represents a global average of all possible position angles through the core, whereas the slices represent individual position angles.  Therefore, we only use the intensity slices as a check for those globules with questionable fits to their radial profiles (e.g., BHR 16) or for those globules with radial profiles that do not appear to flatten at large angular extents (e.g., L723).  In the case of BHR 16, the radial profile method gives an offset of $66.2 \pm 0.6$ \MJysr\ at 160 \um, whereas the intensity slices give offsets of $\approx$ 64.5-70 \MJysr.  The two methods are therefore consistent, which  gives confidence to the radial profile value even if the HSA profile itself is not smooth.

We visually inspect the radial profiles and fits of all globules, and group them into the three categories that represent the reliability of their measured zero-point offsets.  These groups are defined as:

\begin{enumerate}
\item{\textbf{Group A:} The most reliable measurements.  These globules have clean radial profiles that are well fit with Gaussians based on visual inspection.}
\item{\textbf{Group B:} Somewhat reliable measurements.  These globules have questionable fits to their radial profiles (e.g., due to structure), but the offsets from the radial profile method are consistent with the values from the intensity slices.   }
\item{\textbf{Group C:} The least reliable measurements.  These globules also have questionable fits to their radial profiles, but in these cases, the radial profile offsets are inconsistent with the values from the intensity slices.  } 
\end{enumerate}

For Group A and B clouds, we adopt the intensity offsets from the radial profiles.  Errors in these offsets are determined by adding in quadrature the uncertainties in the vertical shifts from the corresponding Gaussian fits to the \emph{Herschel} and \emph{Planck}-determined radial profiles.  For Group C clouds, we consider two cases.   Clouds with questionable fits to their radial profiles (e.g., there is some structure, but the profile flattens out at large angular extents) have their radial profile offsets estimates, whereas clouds with poorly constrained radial profiles (e.g., the profiles do not flatten at large angular extents; see L723 in Figure \ref{profiles_group}) have the average offset value from the intensity slices alone.  For all Group C clouds, we use a larger, fixed error of 5 \MJysr\ at 100 \um\ and 160 \um.  For most globules, the right ascension and declination PACS slices differ by $\lesssim 10$ \MJysr, so an error of 5 \MJysr\ represents the typical uncertainty. 

\section{Results}\label{results}

\subsection{PACS Zero-Point Corrections} \label{pacs_corr} 

We measure the zero-point corrections for all 56 globules using the radial profile method as outlined in the previous section.   Table \ref{table_pacs} lists the measured zero-point corrections for the HSA-PACS maps.  The first column gives the globule name.  The second and third columns give the offset and uncertainty for the PACS 100 \um\ data.  The fourth column identifies the Group (reliability, see below) for the 100 \um\ correction.  The fifth, sixth, and seventh columns give the corresponding offset, uncertainty, and Group for the 160 \um\ data, respectively.  We rank the HSA-PACS $100\ \um$\ and $160\ \um$\ zero-point corrections separately because their radial profiles often have different shapes.  Most of the globules are cold ($< 20$ K), and therefore only weakly detected at 100 \um.  We note that these offsets must be added to the HSA-PACS maps so they match with predictions from \emph{Planck}.

\begin{deluxetable*}{lcccccccc}
\tablecaption{Measured Offsets in the 100 \um\ and 160 \um\ PACS Bands \label{table_pacs}}
\tablehead{
\colhead{ } & \multicolumn{4}{c}{Zero-Point Corrections at 100 \um} & \multicolumn{4}{c}{Zero-Point Corrections at 160 \um} \\
\multirow{2}{*}{Globule } & \colhead{Offset} & \colhead{Error} & \multirow{2}{*}{Method\tablenotemark{a}} & \multirow{2}{*}{Group\tablenotemark{b}}  & \colhead{Offset} & \colhead{Error} &\multirow{2}{*}{Method\tablenotemark{a}} & \multirow{2}{*}{Group\tablenotemark{b}}  \\
\colhead{ } & \colhead{(\MJysr)} & \colhead{(\MJysr)}& \colhead{ } & \colhead{ } & \colhead{(\MJysr)}  &\colhead{(\MJysr)} & \colhead{ } & \colhead{ } 
}
\startdata
CB 4		&	12.9	&	0.1	&	RP	&	A	&	25.2	&	0.1	&	RP	&	A	\\
CB 6		&	10.0	&	0.1	&	RP	&	A	&	20.1	&	0.2	&	RP	&	A	\\
CB 17	&	17.7	&	0.05	&	RP	&	A	&	43.0	&	0.2	&	RP	&	A	\\
L1521F	&	-2.7	&	5.0	&	RP+S	&	C	&	42.3	&	5.0	&	RP+S	&	C	\\
CB 26\tablenotemark{c}	&	18.4	&	0.1	&	RP	&	A	&	48.2	&	5.0	&	RP+S	&	C	\\
L1544	&	22.1	&	0.2	&	RP	&	A	&	60.9	&	0.6	&	RP+S	&	B	\\
CB 27\tablenotemark{c}	&	18.1	&	5.0	&	RP+S	&	C	&	50.0	&	5.0	&	RP+S	&	C	\\
L1552	&	10.9	&	0.2	&	RP+S	&	B	&	69.2	&	0.6	&	RP+S	&	B	\\
CB 29\tablenotemark{d}	&	11.2	&	5.0	&	RP+S	&	C	&	40.1	&	5.0	&	RP+S	&	C	\\
B 35A	&	-2.2	&	5.0	&	RP+S	&	C	&	16.5	&	5.0	&	S	&	C	\\
BHR 22\tablenotemark{d}	&	-1.8	&	0.3	&	RP+S	&	B	&	25.4	&	0.6	&	RP+S	&	B	\\
BHR 17	&	0.0	&	5.0	&	RP+S	&	C	&	14.8	&	0.3	&	RP	&	A	\\
BHR 16	&	4.5	&	5.0	&	S	&	C	&	66.2	&	0.6	&	RP+S	&	B	\\
BHR 12	&	43.4	&	0.4	&	RP+S	&	B	&	97.0	&	1.0	&	RP+S	&	B	\\
DC2573-25\tablenotemark{d}	&	33.3	&	5.0	&	S	&	C	&	140.9	&	0.8	&	RP+S	&	B	\\
BHR 31	&	33.1	&	0.4	&	RP	&	A	&	64.0	&	0.4	&	RP	&	A	\\
BHR 42\tablenotemark{e}	&	18.4	&	0.3	&	RP+S	&	B	&	51.8	&	0.6	&	RP+S	&	B	\\
BHR 34\tablenotemark{e}	&	15.1	&	0.2	&	RP+s	&	B	&	52.0	&	0.2	&	RP+S	&	B	\\
BHR 41	&	24.2	&	0.6	&	RP	&	A	&	69.8	&	0.9	&	RP+S	&	B	\\
BHR 40\tablenotemark{c,e}	&	21.7	&	5.0	&	RP+S	&	C	&	66.4	&	5.0	&	RP+S	&	C	\\
BHR 38/39	&	18.4	&	5.0	&	RP+S	&	C	&	63.6	&	5.0	&	RP+S	&	C	\\
BHR 56	&	2.7	&	5.0	&	S	&	C	&	24.1	&	0.4	&	RP+S	&	B	\\
DC2742-04\tablenotemark{e}	&	45.7	&	0.3	&	RP+S	&	B	&	91.7	&	0.5	&	RP+S	&	B	\\
BHR 48/49\tablenotemark{e}	&	8.1	&	0.5	&	RP+S	&	B	&	69.1	&	1.4	&	RP+S	&	B	\\
BHR 50	&	13.2	&	0.9	&	RP	&	A	&	50.0	&	1.5	&	RP+S	&	B	\\
BHR 68\tablenotemark{c}	&	32.2	&	0.3	&	RP	&	A	&	78.0	&	1.0	&	RP+S	&	B	\\
BHR 71\tablenotemark{f}	&	63.9	&	0.6	&	RP+S	&	B	&	136.1	&	5.0	&	RP+S	&	C	\\
BHR 74	&	18.7	&	0.2	&	RP	&	A	&	51.3	&	0.2	&	RP	&	A	\\
BHR 79\tablenotemark{c,e}	&	6.6	&	0.3	&	RP+S	&	B	&	-25.0	&	0.9	&	RP+S	&	B	\\
BHR 81\tablenotemark{e}	&	38.7	&	0.2	&	RP+S	&	B	&	93.4	&	1.2	&	RP+S	&	B	\\
DC3162+51	&	43.2	&	5.0	&	RP+S	&	C	&	94.5	&	5.0	&	RP+S	&	C	\\
BHR 95\tablenotemark{c}	&	70.6	&	0.5	&	RP+S	&	B	&	140.9	&	0.8	&	RP	&	A	\\
BHR 99	&	47.4	&	0.7	&	RP+S	&	B	&	86.2	&	0.5	&	RP+S	&	B	\\
BHR 100\tablenotemark{e}	&	38.0	&	0.9	&	RP	&	A	&	84.7	&	5.0	&	RP+S	&	C	\\
BHR 97	&	35.8	&	0.5	&	RP+S	&	B	&	69.3	&	0.4	&	RP+S	&	B	\\
DC3391+117\tablenotemark{c}	&	15.9	&	0.2	&	RP+S	&	B	&	34.3	&	0.3	&	RP+S	&	B	\\
DC3460+78\tablenotemark{c}	&	35.6	&	5.0	&	S	&	C	&	84.1	&	5.0	&	RP+S	&	C	\\
CB 68	&	16.4	&	0.3	&	RP	&	A	&	41.6	&	0.4	&	RP	&	A	\\
BHR 147\tablenotemark{d}	&	55.8	&	0.5	&	RP+S	&	B	&	132.3	&	1.2	&	RP+S	&	B	\\
B 68\tablenotemark{c}		&	44.8	&	0.3	&	RP+S	&	B	&	79.0	&	5.0	&	S	&	C	\\
CB 101\tablenotemark{d}	&	17.4	&	5.0	&	RP+S	&	C	&	81.4	&	5.0	&	RP+S	&	C	\\
L422\tablenotemark{d}		&	40.0	&	0.4	&	RP	&	A	&	129.4	&	0.7	&	RP+S	&	B	\\
CB 130	&	43.0	&	0.2	&	RP	&	A	&	105.1	&	0.4	&	RP	&	A	\\
L429\tablenotemark{e}		&	65.8	&	5.0	&	RP+S	&	C	&	162.2	&	0.9	&	RP+S	&	B	\\
L483\tablenotemark{f}		&	39.5	&	5.0	&	RP+S	&	C	&	85.0	&	5.0	&	RP+S	&	C	\\
CB 170\tablenotemark{c,e}	&	39.3	&	5.0	&	S	&	C	&	70.3	&	0.3	&	RP+S	&	B	\\
CB 175\tablenotemark{c}	&	50.9	&	0.3	&	RP+S	&	B	&	68.3	&	0.5	&	RP+S	&	B	\\
CB 176\tablenotemark{e}	&	33.9	&	0.3	&	RP+S	&	B	&	83.7	&	0.2	&	RP	&	A	\\
L723\tablenotemark{c,e}		&	32.2	&	0.3	&	RP+S	&	B	&	110	&	5.0	&	S	&	C	\\
L673	\tablenotemark{e}	&	129.0	&	5.0	&	S	&	C	&	301.1	&	1.7	&	RP+S	&	B	\\
B 335\tablenotemark{c,e}	&	20.6	&	0.2	&	RP	&	A	&	39.2	&	0.4	&	RP	&	A	\\
CB 230	&	23.5	&	0.2	&	RP	&	A	&	58.7	&	0.6	&	RP+S	&	B	\\
L1014	&	50.7	&	0.3	&	RP	&	A	&	146.7	&	5.0	&	RP+S	&	C	\\
L1165\tablenotemark{c}	&	40.2	&	0.6	&	RP+S	&	B	&	113.8	&	0.6	&	RP+S	&	B	\\
L1221	&	12.6	&	0.2	&	RP	&	A	&	44.7	&	0.7	&	RP	&	A	\\
CB 244	&	12.2	&	0.2	&	RP	&	A	&	45.2	&	1.2	&	RP+S	&	B	
\enddata
\tablenotetext{a}{Offsets as measured by the radial profiles (RP) method or intensity slices (S). When both methods are used (RP+S), the listed offsets are from the radial profile method and the measurements are compared against intensity slices.}
\tablenotetext{b}{Confidence in the zero-point corrections.  Group A are most reliable, Group B are somewhat reliable, and Group C are unreliable.  For Group C sources, we assume fixed offset errors of 5 \MJysr.  See text for the definitions of each group.}
\tablenotetext{c}{These globule are highly offset from the center of their map, which will affect the reliability of their radial profile measurements.}
\tablenotetext{d}{These globules have irregular coverage maps at 160 \um\ maps due to missing scans.}
\tablenotetext{e}{These globules have truncated radial profiles due to a jump in emission at large angular extents.}
\tablenotetext{f}{These PACS maps are smaller ($\sim 6$\arcmin\ across) than the other fields, which makes it harder to measure their offsets.}
\end{deluxetable*} 

We rank the reliability of the HSA 100 \um\ and 160 \um\ offsets separately into the three Groups defined in Section \ref{slices}. For the 100 \um\ data, there are 19 globules in Group A (most reliable), 20 in Group B (somewhat reliable), and 17 in Group C (unreliable), and for the 160 \um\ data, there are 12 globules are in Group A, 28 in Group B, and 16 in Group C.  Group A globules are typically more isolated and have high emission contrast relative to their local background.  Groups B and C globules are more confused due to the presence of tails, secondary cores, or a bright, complicated background.

\subsection{Additional SPIRE Corrections} \label{spire_corr}

\begin{figure}[h!]
\includegraphics[width=0.475\textwidth]{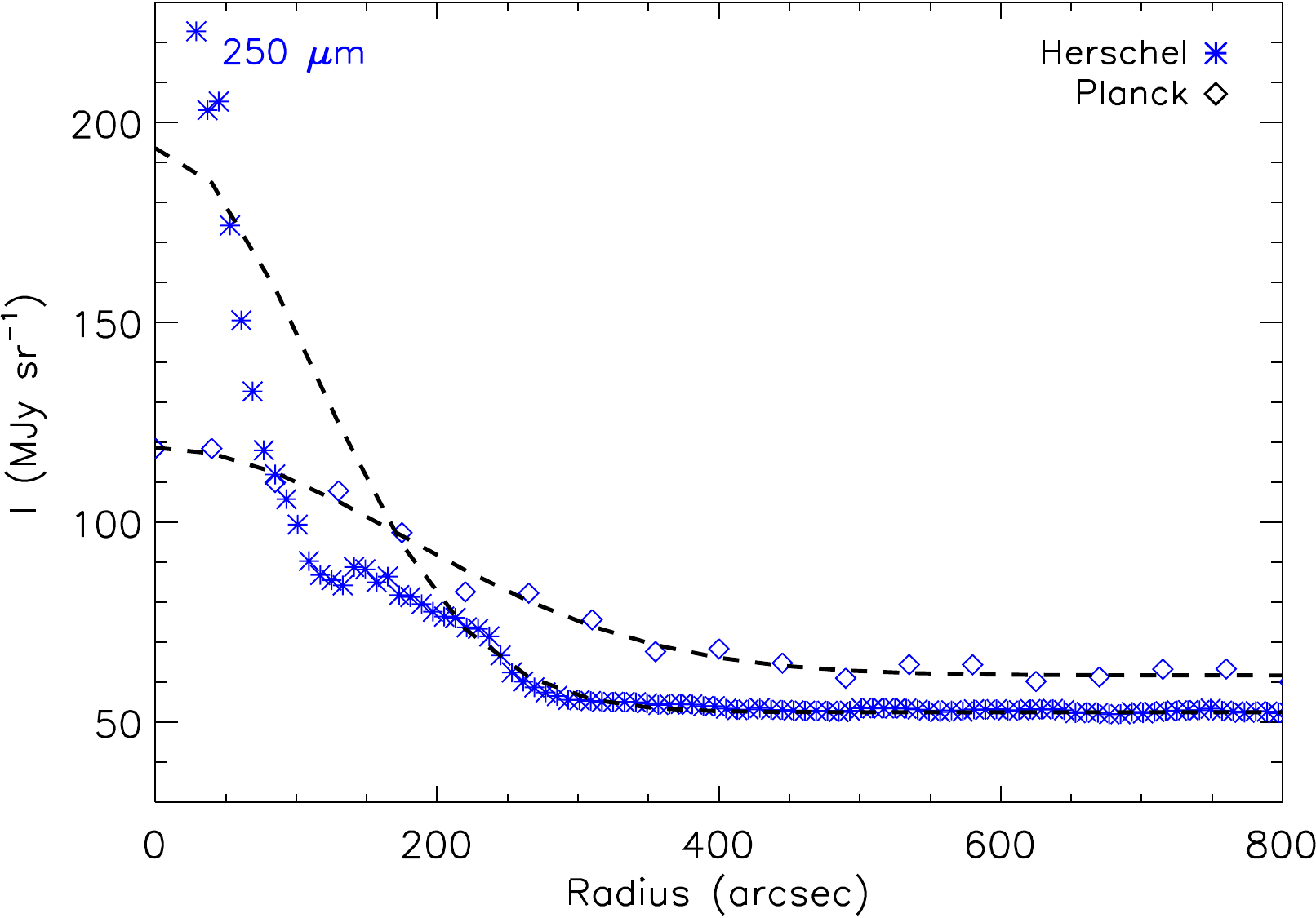}
\caption{Radial profiles of BHR 68 at 250 \um.  The SPIRE-HSA profile is shown with stars and the \emph{Planck}-determined profile is shown with diamonds.   Gaussian fits to both profiles differ by $\sim 10\ \MJysr$\ at angular extents $> 400$\arcsec.  The 350 \um\ and $500\ \um$\ profiles (not shown) also have deviations.\label{profiles_bhr68_spire}}
\end{figure}

Figure \ref{profiles_cb4_spire} shows decent agreement between the HSA-SPIRE intensities and their corresponding \emph{Planck}-determined intensities at angular extents $> 400$\arcsec\ for CB 4, but we still find deviations of $\lesssim 10$\%.   Other globules show even more significant deviations.   Figure \ref{profiles_bhr68_spire} shows the radial profiles of BHR 68 at 250 \um\ as an example.  At angular extents $> 400\arcsec$, which is well off the central globule, the HSA-SPIRE profiles are lower in intensity than the \emph{Planck}-determined profile by roughly 10 \MJysr.  This deviation suggests the HSA 250 \um\ intensities of BHR 68 are underestimated by $\sim 17$\%\ relative to the \emph{Planck}-determined intensities over angular extents of $r > 400$\arcsec. 

Table \ref{table_spire} gives the intensity deviation between the HSA-SPIRE maps and the \emph{Planck}-determined maps following the radial profile technique outlined in Section \ref{radial_profiles}.  We list the deviations and errors at 250 \um\ in columns two and three, at 350 \um\ in columns four and five, and at 500 \um\ in columns six and seven.  Column eight gives the method(s) used to measure the deviations, and column nine gives the radii used to fit the HSA-profiles with Gaussians.  For simplicity, we use the same radii for all three SPIRE bands.  For the \emph{Planck}-determined profiles, we fit Gaussians out to the upper radius used for the HSA-SPIRE profiles.  

\begin{deluxetable*}{lcccccccc}
\tablecaption{Deviations between the HSA SPIRE maps and \emph{Planck}-determined maps \label{table_spire}}
\tablehead{
\colhead{ }		& \multicolumn{2}{c}{SPIRE 250 \um} & \multicolumn{2}{c}{SPIRE 350 \um} & \multicolumn{2}{c}{SPIRE 500 \um} & \colhead{ } & \colhead{ } \\
\multirow{2}{*}{Globule } & \colhead{Deviation} & \colhead{Error} & \colhead{Deviation} & \colhead{Error} & \colhead{Deviation} & \colhead{Error}   & \multirow{2}{*}{Method\tablenotemark{a}} & \colhead{Radii} \\
\colhead{ } & \colhead{(\MJysr)} & \colhead{(\MJysr)}& \colhead{(\MJysr)} & \colhead{(\MJysr)}  &\colhead{(\MJysr)} & \colhead{(\MJysr)} & \colhead{ } & \colhead{arcsec}
}
\startdata
\multicolumn{8}{c}{Group A\tablenotemark{b}} \\
\hline
CB 4		&	2.5	&	0.2	&	0.3	&	0.1	&	0.2	&	0.05	&	RP	& $100-800$ \\
CB 6		&	1.6	&	0.1	&	0.3	&	0.1	&	0.04	&	0.04	&	RP	& $100-800$\\
CB 17	&	3.0	&	0.1	&	0.9	&	0.1	&	0.6	&	0.05	&	RP	& $100-800$\\
CB 26	&	5.6	&	0.4	&	2.9	&	0.3	&	1.8	&	0.1	&	RP	& $100-800$\\
L1544	&	1.7	&	0.6	&	0.6	&	0.4	&	0.7	&	0.2	&	RP	& $200-1000$\\
CB 27	&	2.8	&	0.4	&	0.6	&	0.3	&	0.5	&	0.1	&	RP	& $100-800$\\
L1552	&	4.7	&	0.6	&	2.2	&	0.4	&	1.8	&	0.2	&	RP	& $200-800$\\
CB 29	&	2.0	&	0.5	&	0.7	&	0.3	&	0.3	&	0.1	&	RP	& $200-800$\\
B 35A	&	4.1	&	1.9	&	1.0	&	1.2	&	0.5	&	0.5	&	RP	& $200-1000$\\
BHR 22	&	1.4	&	0.4	&	0.05	&	0.2	&	-0.03	&	0.1	&	RP	& $200-1000$\\
BHR 17	&	1.0	&	0.2	&	-0.2	&	0.1	&	-0.2	&	0.05	&	RP	& $200-1000$\\
BHR 31	&	4.4	&	0.4	&	1.5	&	0.2	&	1.1	&	0.1	&	RP	& $100-800$\\
BHR 42	&	3.2	&	0.3	&	0.8	&	0.2	&	0.6	&	0.1	&	RP	& $100-800$\\
BHR 34	&	5.9	&	0.5	&	1.8	&	0.2	&	1.0	&	0.1	&	RP	& $100-1000$\\
BHR 41	&	7.4	&	1.1	&	3.1	&	0.6	&	1.9	&	0.3	&	RP	& $100-800$\\
BHR 38/39	&	6.7	&	0.3	&	2.4	&	0.2	&	1.5	&	0.1	&	RP	& $100-800$\\
DC2742-04	&	9.1	&	0.4	&	2.6	&	0.3	&	1.6	&	0.1	&	RP	& $100-800$\\
BHR 68	&	9.2	&	0.4	&	2.8	&	0.3	&	1.5	&	0.1	&	RP	& $200-1000$\\
BHR 74	&	6.5	&	0.1	&	1.7	&	0.1	&	0.8	&	0.03	&	RP	& $100-800$\\
BHR 79	&	2.4	&	0.4	&	0.2	&	0.2	&	0.03	&	0.1	&	RP	& $100-800$\\
BHR 99	&	5.2	&	0.4	&	1.8	&	0.2	&	1.2	&	0.1	&	RP	& $100-800$\\
BHR 100	&	6.8	&	0.4	&	2.6	&	0.2	&	1.5	&	0.1	&	RP	& $100-800$\\
DC3391+117	&	1.0	&	0.3	&	0.1	&	0.2	&	0.2	&	0.1	&	RP	& $200-1000$\\
CB 68	&	1.5	&	0.2	&	0.2	&	0.1	&	0.2	&	0.1	&	RP	& $200-1000$\\
B 68		&	4.2	&	0.3	&	1.2	&	0.2	&	0.8	&	0.1	&	RP	& $100-600$\\
CB 101	&	2.5	&	0.2	&	0.9	&	0.1	&	0.8	&	0.1	&	RP	& $100-800$\\
CB 130	&	7.2	&	0.3	&	4.1	&	0.2	&	3.0	&	0.1	&	RP	& $100-800$\\
L483		&	10.6	&	0.9	&	5.9	&	0.5	&	3.2	&	0.2	&	RP	& $200-800$\\
CB 170	&	7.2	&	0.2	&	2.2	&	0.1	&	1.2	&	0.05	&	RP	& $100-800$\\
CB 175	&	5.6	&	0.3	&	2.3	&	0.2	&	1.3	&	0.1	&	RP	& $200-800$\\
CB 176	&	5.4	&	0.2	&	1.4	&	0.1	&	0.9	&	0.05	&	RP	& $100-800$\\
L723		&	7.3	&	0.6	&	3.1	&	0.3	&	2.1	&	0.1	&	RP	& $100-1000$\\
B 335	&	2.4	&	0.2	&	0.6	&	0.1	&	0.25	&	0.05	&	RP	& $200-800$\\
CB 230	&	3.6	&	0.8	&	1.6	&	0.5	&	1.1	&	0.2	&	RP	& $100-600$\\
L1165	&	11.4	&	0.3	&	3.7	&	0.2	&	2.1	&	0.1	&	RP	& $100-1000$\\
L1221	&	6.4	&	0.8	&	2.2	&	0.5	&	1.2	&	0.3	&	RP	& $200-1000$\\
\hline
\multicolumn{8}{c}{Group B\tablenotemark{b}} \\
\hline
L1521F	&	1.8	&	0.5	&	1.0	&	0.3	&	1.0	&	0.2	&	RP+S	& $200-1500$\\
BHR 16	&	10.3	&	0.4	&	3.7	&	0.3	&	2.2	&	0.1	&	RP+S	& $200-1000$\\
DC2573-25	&	14.0	&	0.7	&	5.4	&	0.5	&	3.5	&	0.2	&	RP+S	& $100-800$\\
BHR 40	&	4.5	&	0.3	&	1.6	&	0.2	&	1.1	&	0.1	&	RP+S	& $50-800$\\
BHR 56	&	1.7	&	0.3	&	0.1	&	0.2	&	0.1	&	0.1	&	RP+S	& $200-800$\\
BHR 71	&	12.7	&	1.1	&	3.7	&	0.7	&	2.2	&	0.3	&	RP+S	& $100-1000$\\
BHR 81	&	8.2	&	0.4	&	2.6	&	0.2	&	1.7	&	0.1	&	RP+S	& $100-1000$\\
BHR 95	&	10.6	&	0.6	&	4.1	&	0.3	&	2.7	&	0.1	&	RP+S	& $200-1000$\\
DC3460+78	&	6.7	&	0.8	&	3.3	&	0.6	&	2.2	&	0.3	&	RP+S	& $400-800$\\
BHR 147	&	11.5	&	0.6	&	4.7	&	0.3	&	2.7	&	0.1	&	RP+S	& $100-800$\\
L422		&	6.8	&	0.4	&	2.2	&	0.3	&	2.1	&	0.1	&	RP+S	& $100-1200$\\
L429		&	13.7	&	0.5	&	6.0	&	0.4	&	3.7	&	0.2	&	RP+S	& $200-1000$\\
L673		&	8.5	&	1.0	&	8.7	&	0.5	&	7.3	&	0.2	&	RP+S	& $100-800$\\
L1014	&	11.1	&	2.1	&	5.5	&	0.4	&	3.6	&	0.2	&	RP+S	& $100-1000$\\
CB 244	&	4.2	&	0.5	&	2.3	&	0.3	&	1.3	&	0.2	&	RP+S	& $200-1000$\\
\hline
\multicolumn{8}{c}{Group C\tablenotemark{b}} \\
\hline
BHR 12		&	12.6	&	2.0	&	4.1	&	1.0	&	2.4	&	0.5	&	RP+S	& $100-600$\\
BHR 48/49	&	4.0	&	2.0	&	0.8	&	1.0	&	0.7	&	0.5	&	RP+S	& $100-1000$\\
BHR 50		&	2.9	&	2.0	&	0.2	&	1.0	&	0.3	&	0.5	&	RP+S	& $200-1000$\\
DC3162+51	&	8.9	&	2.0	&	2.7	&	1.0	&	1.6	&	0.5	&	RP+S	& $200-1000$\\
BHR 97		&	5.8	&	2.0	&	1.9	&	1.0	&	1.1	&	0.5	&	RP+S	& $100-800$
\enddata
\tablenotetext{a}{All deviations are determined using the radial profiles (RP) method. Those with less reliable profiles are compared against intensity slices (RP+S).}
\tablenotetext{b}{Confidence in the intensity deviations.  Group A are most reliable, Group B are somewhat reliable, and Group C are unreliable. For Group C sources, we assume fixed errors of 2 \MJysr, 1 \MJysr, and 0.5 \MJysr\ at 250 \um, 350 \um, and 500 \um, respectively.  See text for the definitions of each group.}
\end{deluxetable*} 

We rank the HSA-SPIRE deviations using the same classifications as the HSA-PACS offsets (see Section \ref{slices}).  Unlike the PACS data, the HSA-SPIRE profiles tend to follow a similar shape, and therefore we can assign the same Group to all three SPIRE bands.  Thus, Table \ref{table_spire} is ordered by Group.  Most (36) of the globules have reliable deviation measurements and are in Group A.  For the remaining globules, 15 are in Group B and 5 are in Group C.  Since the HSA-SPIRE fields are larger than the PACS fields, we are better able to trace their background emission.  For the Group C globules, we adopt fixed errors of 2 \MJysr\ at 250 \um, 1 \MJysr\ at 350 \um, and 0.5 \MJysr\ at 500 \um.  

\begin{figure}[h!]
\includegraphics[width=0.475\textwidth]{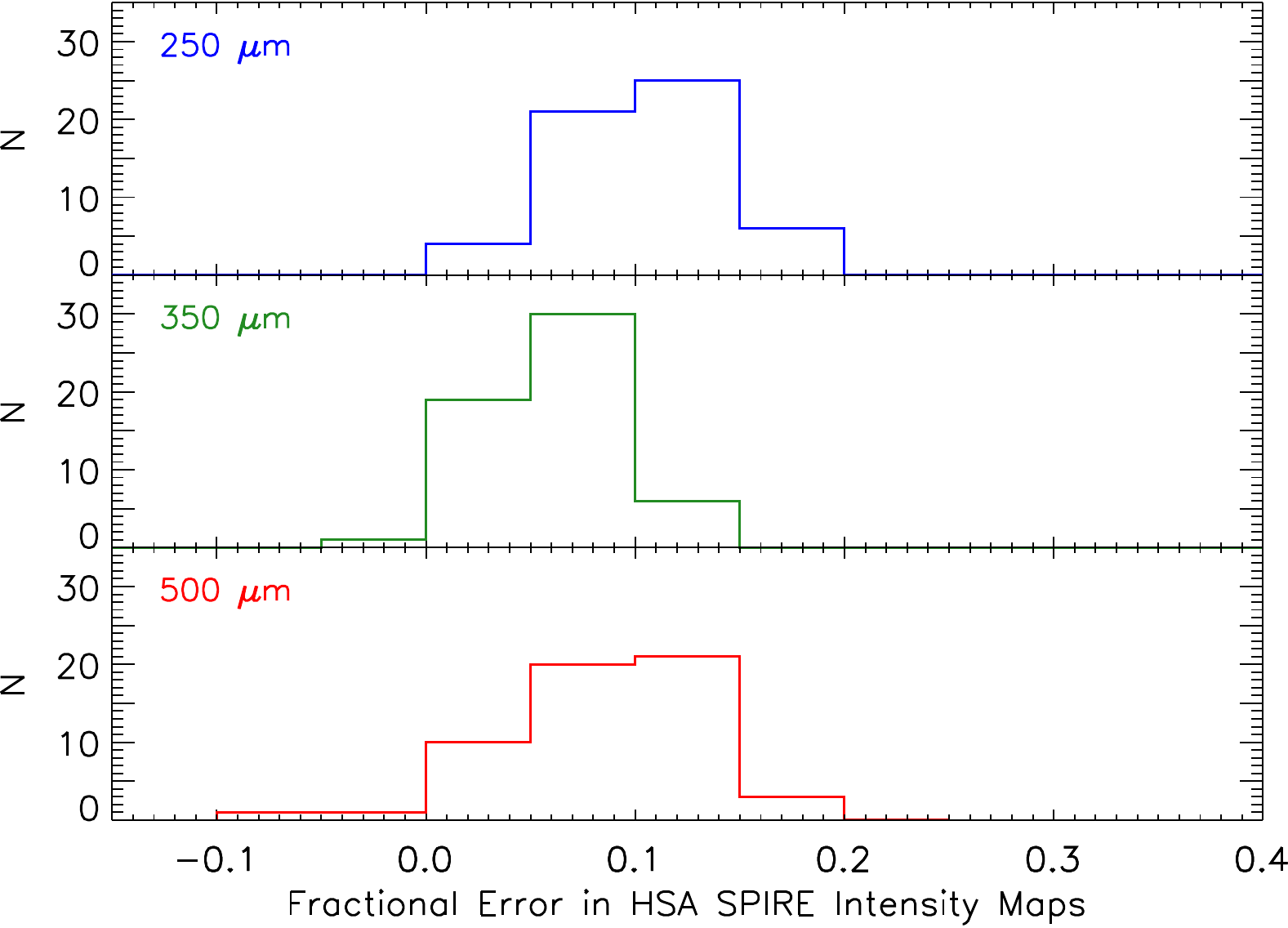}
\caption{Histograms of median fractional error between the HSA-SPIRE intensities relative to the \emph{Planck}-determined intensities for all 56 globules.  The fractional errors are measured at angular extents between $\gtrsim 600-1000$\arcsec\ to exclude emission from the globules themselves.  A value of zero indicates that the HSA-SPIRE intensities perfectly match the \emph{Planck}-determined intensities.  The three panels show the results at 250 \um\ (top), 350 \um\ (middle) and 500 \um\ (bottom).  \label{histo_spire}}
\end{figure}

The HSA-SPIRE deviations are also generally highest at 250 \um\ and lowest at 500 \um, but these values and their significance vary from source to source.  Figure \ref{histo_spire} shows the median fractional error in the HSA-SPIRE intensity maps relative to the \emph{Planck}-determined intensities between radial extents of $600-1000\arcsec$, e.g., $(I^{Planck}_{\lambda} - I^{HSA}_{\lambda})/I^{Planck}_{\lambda}$.  A value of zero indicates that the SPIRE emission perfectly matches the \emph{Planck}-determined maps.    Instead, many of the globules have deviations that are $> 10$\%, and these large values are seen across all Groups, and in globules with or without stars.   The deviations also tend to be positive, indicating that the HSA-SPIRE maps generally underestimate the thermal dust emission relative to \emph{Planck}.   We note that these deviations will be insignificant ($< 1\%$) toward the bright centers of the clouds. They will affect the intensities of the background emission and outer, diffuse envelopes, however.

Deviations between the HSA-SPIRE intensities and the corresponding \emph{Planck}-determined intensities suggest that the HSA-SPIRE data were not properly corrected.   As mentioned in Section \ref{method}, the Level 2 HSA-SPIRE maps were zero-point corrected using color corrections from \emph{Planck} 857 GHz and 545 GHz observations, assuming a specific temperature and dust emissivity.  Uncertainties in the assumed SED parameters will affect the color corrections.  The largest error will be at 250 \um, since the pipeline must extrapolate the \emph{Planck} 857 GHz data to 250 \um\ (see Section 5.10 in the SPIRE Handbook\footnote{http://herschel.esac.esa.int/Docs/SPIRE/spire\_handbook.pdf}).   Figure \ref{histo_spire} shows more globules with significant ($>10$\%) deviations at 250 \um\ compared to the 350 \um\ and 500 \um\ bands.   Nevertheless, if the HSA-SPIRE deviations were solely due to the  \emph{Planck} color corrections, then the 350 \um\ and $500\ \um$\ profiles should have only minor differences as these bands have excellent overlap with the \emph{Planck} 857 GHz and 545 GHz filters.  Figure \ref{histo_spire} shows that almost half of the globules have large ($>10$\%) deviations at $500\ \um$\ and most globules have moderate ($>5$\%) deviations at 350 \um.   So the improper zero-point corrections applied by the HSA are unlikely to be caused by the \emph{Planck} color corrections.

We attribute the HSA-SPIRE deviations to the relatively small map sizes of globules.  The SPIRE fields are larger than the PACS maps, but are still only $7-8$ beams across at 5\arcmin\ resolution (see Figure \ref{cb4_expected_fluxes}).  Such small areas may not sample well the background emission, especially if the globule is located in a bright, highly structured environment  (see Section \ref{gbs} for a comparison with larger clouds).   The  globules with the best matching profiles (e.g., $(I^{Planck}_{\lambda} - I^{HSA}_{\lambda})/I^{Planck}_{\lambda} < 5\%$) tend to have simple, compact structures and relatively weak, diffuse backgrounds, whereas the globules with the most significant deviations (e.g., $(I^{Planck}_{\lambda} - I^{HSA}_{\lambda})/I^{Planck}_{\lambda} >10\%$) tend to be more extended with brighter, highly structured backgrounds.   When the latter are convolved to 5\arcmin\ resolution, the bright structured emission may overestimate the background and subsequently, underestimate the zero-point correction that must be applied.   In this explanation, the HSA-SPIRE maps should preferentially underestimate the emission relative to the \emph{Planck}-determined maps, which is the case for most of the globules (e.g., see Figure \ref{profiles_bhr68_spire} and Table \ref{table_spire}).

To illustrate this scenario more clearly, we perform a case study on NGC 7538 in Appendix \ref{case_study}.  NGC 7538 is a high-mass star-forming region with extended, bright structured emission.  It was observed entirely in a $2.7\degree\ \times 2.7\degree$ field as part of the \emph{Herschel} infrared Galactic Plane Survey \citep[Hi-Gal; Field 112,][]{Molinari10}.  A small subregion around NGC 7538 IRS1-3 was also observed in an $\sim 11\arcmin$ scan map with SPIRE as follow-up observations due to saturation in the \emph{Herschel} OB Young Stars survey \citep[HOBYS,][]{Motte10}.  Both HSA maps were processed with the same pipeline (version 14.1.0).  We use the Level 2 data for these observations which include their nominal zero-point corrections.  Although both HSA-SPIRE maps should give the same result, we instead find that the smaller NGC 7538 IRS1-3 field greatly underestimates the emission seen in the larger NGC 7538 field.  Moreover, the intensities from the larger HSA map matches the intensities from \emph{Planck}-determined maps of NGC 7538 at large angular distances, suggesting that it has more reliable zero-point corrections.  Thus, we see a disconnect between large and small HSA-SPIRE maps, where the smaller maps generally underestimate the true intensities.  These deviations should be accounted for as they may impact the derived temperatures and densities at the edges of these globules.

\section{Globule Properties from SED Fitting} \label{sed_section}

To fit he SEDs, we use the PACS 160 \um\ and SPIRE 250 \um, 350 \um, and 500 \um\ data.  The 100 \um\ band is excluded hereafter because most globules are only weakly detected at 100 \um.   We correct the PACS 160 \um\ intensities by adding the offsets in Table \ref{table_pacs} to the HSA-PACS maps.  We also add the deviations in Table \ref{table_spire} with fractional errors $>5$\% to the HSA-SPIRE maps to better match the SPIRE intensities with \emph{Planck}. 

\begin{figure*}
\centering
\includegraphics[scale=0.6]{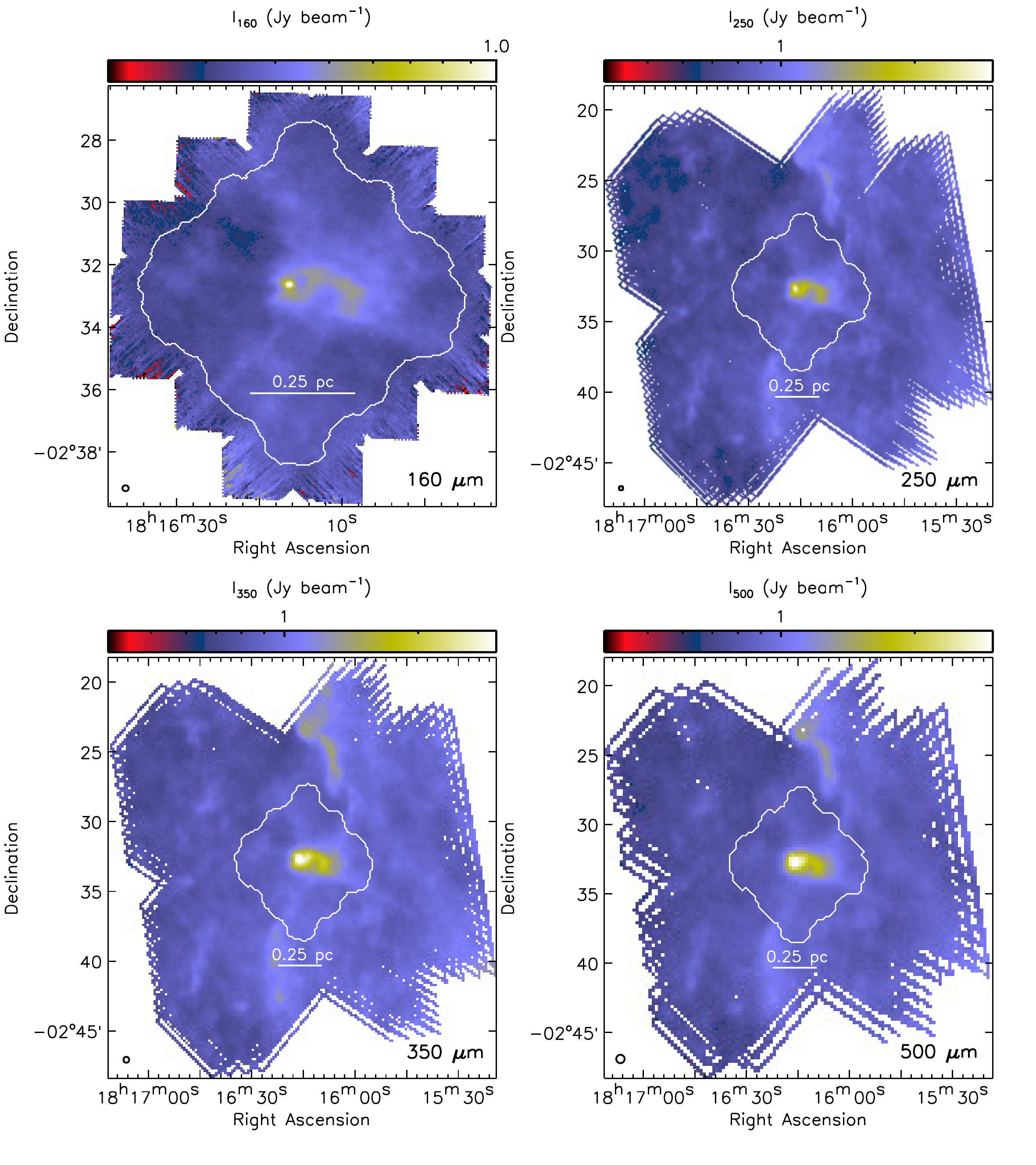}
\vspace{-5mm}
\caption{Corrected \emph{Herschel} intensity maps of CB 130 at 160 \um, 250 \um, 350 \um, and 500 \um.  Each panel shows a 0.25 pc scale bar.  Note that the 160 \um\ map spans $\sim 10$\arcmin, whereas the three SPIRE maps span $\sim 25$\arcmin.   The beam sizes of the maps are in the lower-right corners.   The \emph{Herschel} maps include the flux offsets from Tables \ref{table_pacs} and \ref{table_spire}.  Each panel also shows the mask used for the 160 \um\ radial profiles and intensity slices as a reference.   Note that the color scale the PACS 160 \um\ map differs from the color scale used for the SPIRE maps.  \label{cb130_fluxes}}
\end{figure*}

Figure \ref{cb130_fluxes} shows as an example the corrected intensity maps of CB 130 at 160 \um, 250 \um, 350 \um, and 500 \um.   The 160 \um\ panel also shows the outline of the mask that was used for the radial profiles and intensity slices in our analysis at 160 \um\ (see Section \ref{radial_profiles}).  We show CB 130 because it highlights well the extended, diffuse emission that is ubiquitous in the maps of many globules.  The corrected intensity maps for all globules are available as FITS images online.

Here we describe our SED-fitting procedure to obtain dust temperature and optical depth for all 56 globules.  We use the same procedure for all sources.  Images of both parameters are shown in Figure \ref{app_fig}.

\subsection{Modified Blackbody Model}

To determine dust temperature and optical depth at each pixel for all globules, we fit their observed SEDs at a common resolution with a single temperature, modified blackbody function,

\begin{equation}
S_{\nu} = B_{\nu}(T)(1 - e^{-\tau_{\nu}})\Omega \label{sed_eq},
\end{equation}
where $S_{\nu}$ is the flux per beam, $B_{\nu}$ is the black body equation at a dust temperature, $T$, $\tau_{\nu}$ is the optical depth, and $\Omega$ is the solid angle of the observations.  In the case of optically thin dust emission, Equation \ref{sed_eq} becomes $S_{\nu} = \tau_{\nu}B_{\nu}(T)\Omega$.

We assume the optical depth is a power-law at these far-infrared wavelengths of the form, $\tau_{\nu} \sim \nu^{\beta}$.  Since we lack sufficient long wavelength data to constrain the dust emissivity index, $\beta$ \citep[e.g.,][]{Kelly12, Sadavoy13, MChen16}, we use the values of $\beta$ from the \emph{Planck} models, which appear to measure broadly the global dust properties of both large and small clouds.  For example, the \emph{Planck}-determined indices are generally $\beta \approx 1.8$  across both molecular clouds  \citep{Planck_beta2} and the cold cores \citep{Juvela15}.  In keeping with the \emph{Planck} SED parameters used in this analysis, we fit for $\tau$ at 353 GHz,  

\begin{equation}
\tau_{\nu} = \tau_{353} \left(\frac{\nu}{353\ \mbox{GHz}}\right)^{\beta} \label{tau_eq}.
\end{equation}

Our modified black body function differs from that used in the previous \emph{Herschel} studies of globules \citep[e.g.,][]{Nielbock12, Launhardt13}.  These studies primarily used dust opacities from the grain growth models of \citet{Ossenkopf94}.  These dust grain models are most appropriate for dense ($n > 10^5$ \vol) protostellar cores, whereas globules may have much lower densities, particularly in their extended envelopes \citep{Launhardt10}.  Therefore, it is possible that the globules have different dust properties.  \citet{Webb17} combined near-infrared extinction data with \emph{Herschel} thermal dust emission maps of CB 68, L1552, and L429 to compare the consistency of eight different dust models from \citet{Ossenkopf94} and \citet{Ormel11}.  They found broad agreement for several models, suggesting that the globules may have a range of grain properties.  Our sample also includes both starless globules and globules with embedded stars, and dust properties may differ between these evolutionary states.  Thus, we prefer to use the \emph{Planck} SED parameters to infer the dust properties rather than assume a fixed dust model for all the globules.

\citet{Launhardt13} also excluded an additional background term in their SED analysis that was used to account for diffuse Galactic emission and the Cosmic Infrared Background \citep[CIB; see also, ][]{Nielbock12}.  The average CIB contribution was removed from the \emph{Planck} dust parameter maps \citep{Planck_allsky_2014}, and therefore will not contribute to the \emph{Planck}-determined intensity maps used to correct the PACS and SPIRE data.  Diffuse Galactic emission, both foreground and background to our globules, is more difficult to remove as these features can be both bright and highly structured.   For example, Figure \ref{cb130_fluxes} shows extensive, structured emission throughout the PACS and SPIRE maps of CB 130.  Detailed modeling of these globules may be required to remove intervening Galactic emission.  Such models are beyond the scope of the current paper.  Thus, we focus on line-of-sight average SED fits at this time.

\subsection{SED Fitting}

We fit the above modified black body function to the observed SEDs at $160-500$ \um\ using 500 Monte Carlo trials for each pixel (see below).   We first convolve the zero-point corrected $160-500$ \um\ maps to a common resolution of 36.3\arcsec\ corresponding the the lowest resolution 500 \um\ data and then regrid them to a common pixel scale.  Since the 160 \um\ maps are smaller than the SPIRE maps, we restrict the area for the SED fits by selecting only those pixels with good coverage at $160\ \um$\ to mask out the noisy edge pixels (e.g., see the first panel of Figure \ref{cb130_fluxes}).   We then fit the $160-500$ \um\ SEDs at their reference wavelengths in a pixel-by-pixel manner using the IDL program, \emph{mpfitfun} \citep{Markwardt09}.  This program uses an iterative $\chi^2$ minimization technique to compare observed data to a user-defined model within errors.

To estimate the errors, we use a two-step process.  First, we adopt the median value in the \emph{Herschel} error maps at 36.3\arcsec\ resolution as the 1 $\sigma$ noise uncertainty for each band.  We then add in quadrature the offset uncertainties to obtain an overall intensity map error.  These errors are additive and are used directly in the SED fitting.  Second, we account for flux calibration errors and \emph{Herschel} color correction errors using a Monte Carlo analysis in a similar manner as described in \citet{Sadavoy13}.  Both of these errors are multiplicative.  We represent them as Gaussian functions centered on unity and then generate 500 random correction factors for the PACS and SPIRE instruments.  The Gaussian width is set by the magnitude of the uncertainty.  For the calibration errors, we assume a conservative value of 10\%\ for both the PACS and SPIRE bands \citep[e.g., the estimated calibration error for SPIRE is 4\%,][]{Bendo13}.  We use the same correction factor for the three SPIRE bands because their calibrations are all correlated.  For the \emph{Herschel} color corrections, we calculate a mean value ($\lesssim$ 3\%) and error ($\lesssim$ 5\%) assuming a range of SED profiles with temperatures between $10-25$ K and dust emissivity indices between 1.5-2.5 \citep[e.g., see][]{Pezzuto12, Sadavoy13}.  The color corrections are not correlated, so we apply the appropriate value to each of the \emph{Herschel} bands separately.

Since we are essentially fitting 500 SEDs, we produce a broad distribution of best-fit temperatures and optical depths for every pixel.  From these distributions, we obtain a mean value and 1 $\sigma$ standard deviation of each parameter.  Figure \ref{app_fig} shows maps of mean dust temperature and optical depth at 353 GHz for all 56 globules.   The typical 1 $\sigma$ errors are $\lesssim$ 5\%\ for temperature and range between $10-20$\%\ for the optical depth. 

Dust temperature maps such as those derived here represent line-of-sight averages, whereas these globules are expected to have temperature gradients due to shielding from the interstellar radiation field (ISRF) and internal heating sources.   Average temperatures will diminish these gradients.  For example, we find typical temperatures of $12-14$ K in the centers of dense, starless cores,  whereas predicted dust temperatures from radiative transfer models \citep[e.g.,][]{Evans01, Nielbock12, Roy14} and observed gas temperatures from high-density gas tracers \citep[e.g.,][]{Tafalla02, Crapsi07,Pagani07} often find temperatures $< 10$ K in the centers of low-mass starless clouds.   Similarly, while many of our temperature maps show compact, warm spots toward protostars, the absence of a warm spot does not mean the globule is starless.  In particular, we do not see a warm spot in our line-of-sight average temperature maps corresponding to the well-known, low-luminosity protostar in L1014 \citep{Young04L1014}. 

From the optical depth maps, the globules show a wide range of morphologies on 36.3\arcsec\ scales.  This resolution is 8 times better than Planck at 353 GHz.  More than half of the globules have substructure, including multiple compact objects, pillar-like features, or elongated tails extending from the main globule.   The more complex globules include both starless sources (e.g., L1544, CB$\ 175$) and protostellar objects (e.g., BHR 17, L723), as classified from the literature \citep[e.g.,][]{MaheswarBhatt06, Crapsi07, Haikala10, Lopez15}.  Therefore, globules on average do not have simple structures \citep[e.g., see also][]{Leung85, LaunhardtHenning97, Launhardt10}.      

\section{Discussion}\label{discussion}

\subsection{Temperature and Optical Depth}

We have the largest sample of temperature and optical depth maps of globules that were produced from thermal dust emission to date (see  Figure \ref{app_fig}).   As mentioned above, these maps correspond to line-of-sight averages, and they will include contributions from background dust emission in the Galaxy.  This background must be removed to determine the density profiles and masses of the globules.   Such an analysis is beyond the scope of the current paper.  Nevertheless, with a large sample of globules, we can identify trends in temperature and optical depth.  

\begin{figure}[h!]
\includegraphics[width=0.475\textwidth]{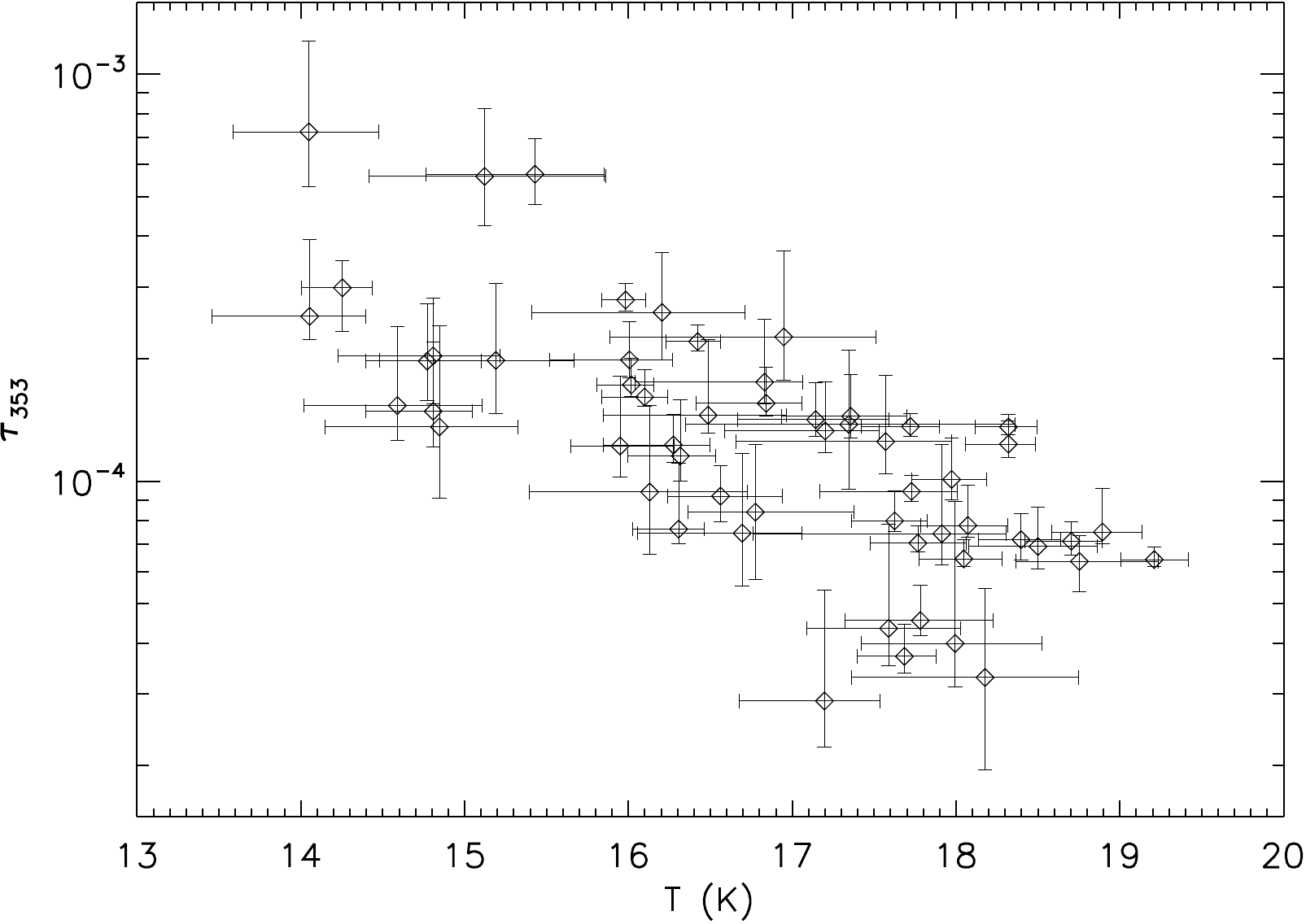}
\caption{Comparison of median temperature and median optical depth for all 56 globules.  Error bars show the first and third quartile values for each cloud.  The median and quartile values are determined only for those pixels with $\tau_{353} > \tau_{med}^{map}$ to avoid substantial background emission.  \label{temp_tau}}
\end{figure}

Figure \ref{temp_tau} compares the median optical depth and median temperature for each globule above an optical depth threshold, $\tau_{med}^{map}$.  We define $\tau_{med}^{map}$ as the median optical depth for the entire map shown in Figure \ref{app_fig}.  This threshold is used to exclude background material from the analysis and ensure that we are measuring the temperature and optical depth properties of the globules alone.  We visually inspected contours of $\tau_{med}^{map}$ to ensure that this level enclosed the higher density material associated with the globules.   The error bars in Figure \ref{temp_tau} represent the first and third quartile ranges of each parameter for material with $\tau > \tau_{med}^{map}$.

Figure \ref{temp_tau} shows a clear anti-correlation between optical depth and temperature, although there is a fair amount of scatter.  In general, globules with higher optical depths at submillimeter and far-infrared wavelengths should also have higher optical depths at optical and near-infrared wavelengths.  Hence, these globules should be well shielded from ionizing photons from the interstellar radiation field, which makes their interior gas temperatures lower \citep{Evans01, Stamatellos07}.  Similar trends have been seen in dense cores embedded in large molecular clouds \citep[e.g.,][]{Rathborne08, Friesen09, Marsh16} and are also interpreted as evidence of self-shielding.

\begin{figure*}
\centering
\includegraphics[trim=9mm 2mm 1mm 1mm,clip=true,scale=0.63]{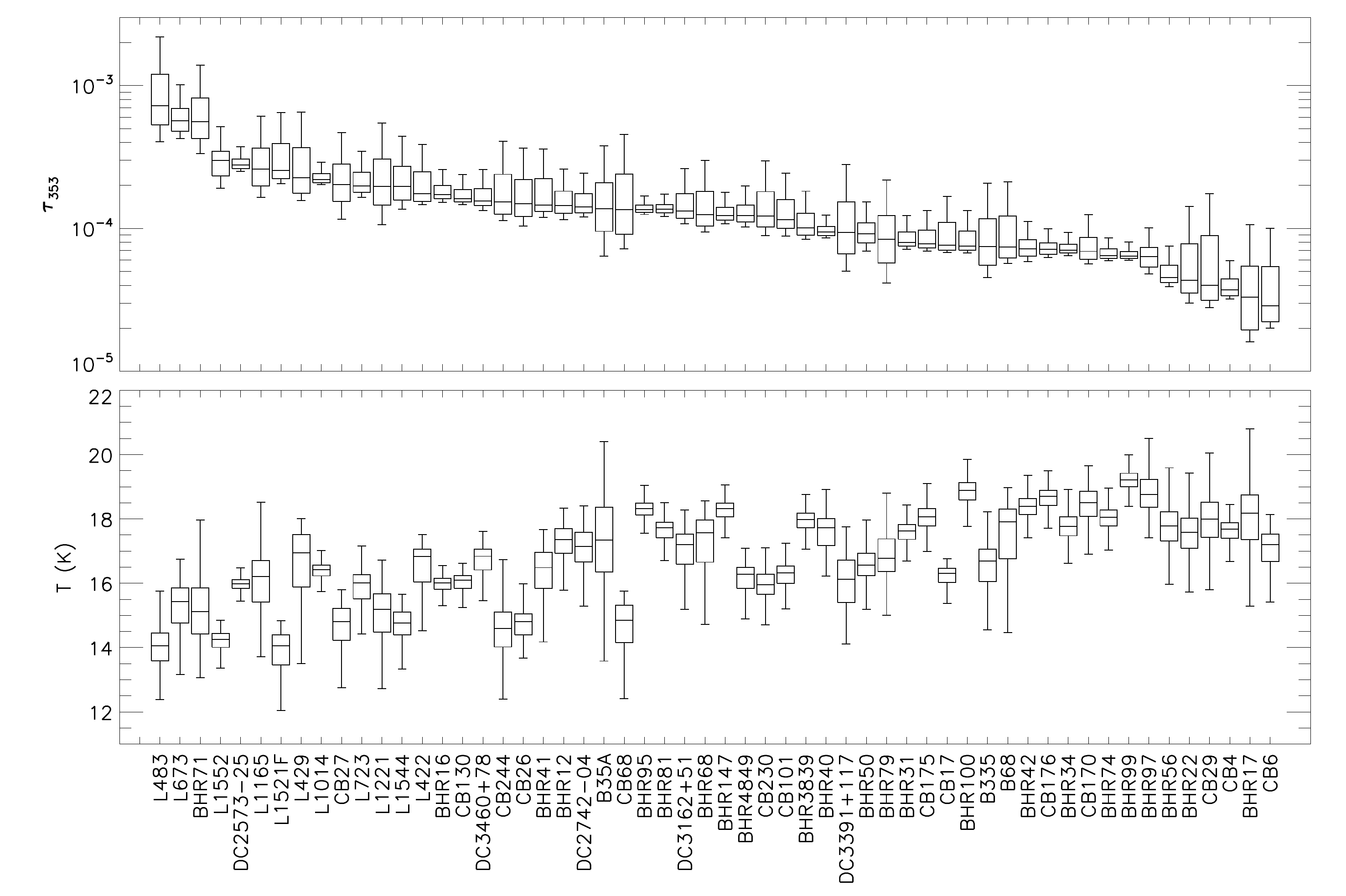}
\caption{Box and whisker plots of optical depth (top) and temperature (bottom) for all 56 globules.  The globules are arranged in order of decreasing median optical depth.  As in Figure \ref{temp_tau}, only those pixels with $\tau_{353} > \tau_{med}^{map}$ are included.  The upper and lower bounds of the boxes show the first and third quartile ranges with horizontal lines depicting the median value.  The error bars (whiskers) represent 1.5 times the quartile values, and correspond to the upper and lower limits of the data.     \label{boxplot}}
\end{figure*}

Figure \ref{boxplot} shows box and whisker plots of both temperature and optical depth to compare the individual globules.  For this figure, we again only include those pixels with $\tau > \tau_{med}^{map}$.  The globules in both panels are ordered by decreasing median optical depth.  Median values are shown by the horizontal lines, whereas the boxes give the first and third quartiles.  The whiskers are shown as error bars and indicate values that are 1.5 times below and above the quartiles.  Thus, the error bars are representative of statistical upper and lower limits in the distribution of optical depth and temperature (for $\tau > \tau_{med}^{map}$).  

Similar to Figure \ref{temp_tau}, Figure \ref{boxplot} shows a clear trend with optical depth, where the globule with higher optical depths are generally cooler than those with lower optical depths.   The temperature distribution is not smooth, however, indicating that there are other factors that determine the temperature and optical depth structures of globules.  For example, globule properties can be greatly affected by the interstellar radiation field, star formation activity, or internal and external turbulence \citep{Leung82, Myers83, Dunham06, Launhardt10}. 

Regarding the optical depths, Figure \ref{temp_tau} shows a very smooth downward trend across all globules, with a typical median value around $\tau \sim 10^{-4}$.  L483, BHR 71, and L673 show the highest median optical depths, although the median values of L483 and BHR 71 are likely biased by the smaller sizes of their \emph{Herschel} dust maps.  L673 is very extended and its \emph{Herschel} temperature and optical depth maps may be too small to cover it fully.  Nevertheless, L673 is also larger ($\sim$ 2 pc $\times$ 1 pc) and more massive ($\sim 90$ \Msun) than most globules in our sample \citep{Visser02}.  Therefore, its higher optical depths could be indicative of genuinely higher densities.  On the other end of the distribution, CB 4, BHR 17, and CB 6 have the lowest median optical depths.  If low optical depths are indicative of lower densities, then we would naively expect these globules to be starless.  BHR 17 and CB 6, however, have young, embedded stars, whereas only CB 4 appears to be starless \citep{Launhardt10, Bourke95iras}.  Therefore, the median optical depths of these globules are not necessarily representative of their star formation activity.

\subsection{Comparison to the EPoS Survey}\label{robust}

\emph{Herschel} observations of twelve  EPoS globules are reported in the literature.  The remaining 44 globules presented here have not been included in a study previously.  The EPoS studies explored individual globules \citep[CB 17, B 68, and CB 244][]{Stutz10, Nielbock12, Schmalzl14}, samples of six \citep{Lippok16}, or a small survey of all twelve \citep{Launhardt13}.  These papers also included complementary  near-infrared extinction maps, continuum observations in the mid-infrared and the (sub)millimeter, and CO line emission.  With these extra datasets, the authors performed more detailed modeling, including radiative transfer models to produce 3-D temperature and density profiles of the cores.  We will compute comparable density profiles in a future study.  Here, we focus only on the pixel-by-pixel SED fitting that is common to our analyses.  We also focus on our temperature maps to avoid converting our optical depth maps to column density ($N(H) \propto \tau_{\nu} / \kappa_{\nu,dust}$).

The aforementioned EPoS studies were primarily conducted before maps of the \emph{Planck}-derived SED parameters became  available.   As such, they avoided the \emph{Herschel} zero-point corrections by subtracting out a background level based on relatively clean, emission-free sections of the intensity maps.  They used the same clean region for all \emph{Herschel} bands for consistency.   For truly isolated globules, this background subtraction will not affect greatly the emission from the globule itself.  In practice, it is difficult to get a genuinely ``clean region'' in the small globule maps (e.g., see Figure \ref{cb130_fluxes}).  If the selected ``clean region'' still has diffuse emission from the globule, then the background subtraction will have an effect similar to how ground-based (sub)millimeter telescopes filter out the atmospheric foreground \citep{Nielbock12}.  Indeed, these studies tended to produce smaller, and more irregularly shaped temperature maps  than the temperature maps presented here. 

This background subtraction will be most pronounced in the outer envelopes of the globules where the emission is lower and the contributions from the ``background'' are more significant, whereas the effect will be least significant in the bright interiors of the globules.  Indeed, we find consistent temperatures with the EPoS studies in the dense centers and different temperatures in the outer envelopes of the globules we examine in common.  Globules that show the most significant deviations in their outer envelopes are CB 4, CB 17, CB 26, CB 27, B 68, and CB 230 \citep[see][]{Nielbock12, Launhardt13, Schmalzl14, Lippok16}.  For example, we find uniform dust temperatures of $\approx$ 17.5 K at large angular extents from CB 4, whereas \citet{Launhardt13} find temperatures of $\approx$ 22 K.  We have a consistent temperature of $\approx$ 14.5 K in the dense core center, suggesting that the difference seen at large angular extents is not systematic throughout the entire map.   

We attribute the higher temperatures in the outer envelopes of the globules from the previous studies entirely to the background subtraction rather than differences in SED models.  As mentioned in Section \ref{sed_section}, the EPoS studies generally adopted the OH5a dust opacities \citep{Ossenkopf94} for their SED fitting.  OH5a dust opacities have an approximate power-law shape with a slope of $\beta \approx 1.8$ at the \emph{Herschel} wavelengths, which is slightly steeper than typical $\beta$ values of $1.6-1.7$ from the \emph{Planck} dust emissivity maps.    A steeper value of $\beta$ will decrease the line-of-sight dust temperature due to a degeneracy between these two parameters \citep[e.g., see][]{Doty94, Shetty09}.  If we instead adopt the OH5a dust opacities for CB 4, the dust temperatures in the outer envelope decrease slightly to $\sim 16.5$ K and deviate further from the temperatures in \citet{Launhardt13}.  Thus, the differences in temperature cannot be attributed to a different dust opacity law in the SED fits.

Higher dust temperatures at large angular extents can have significant consequences for the radiative transfer models of these globules.  Globules are generally modeled as simple spherical structures that are heated externally by the ISRF and internally by young embedded stars.  Higher temperatures of only a few Kelvin in their outer envelopes can result in model fits that suggest much higher local ISRFs \citep[e.g.,][]{Nielbock12, Schmalzl14, Lippok16}.   We find that most globules have background temperatures of $17-20\ $K, despite our sample covering a wide range of positions in the Galaxy.  A few globules  have cooler background temperatures of $\sim 15$ K (e.g., see L1521F in Figure \ref{app_fig}), which suggests a weaker ISRF than the average globule.    In the case of L1521F,  the globule is located near the Taurus molecular cloud, and may be better shielded from the ISRF.   While we defer radiative transfer models with our maps to a future study, our observations suggest that the relatively high local ISRF intensities inferred in previous studies may be overestimated.  

We also note that several globules with embedded protostars (e.g., CB 17, CB 26, BHR 12, CB 68) show only weak indications of internal heating from their young stars compared to previous maps from EPoS \citep{Launhardt13}.  Since the dust emission is brightest toward the protostars, the background subtraction will be less significant in these cases.  Instead, these differences may reflect our respective SED fitting techniques, such as the choice of dust opacity law or how the data around the unresolved protostars were convolved.  Nevertheless, emission associated with embedded stars is very localized and will not be relevant for determining the zero-point corrections.

\subsection{Comparison to the Herschel Gould Belt Survey}\label{gbs}

Fifteen nearby molecular clouds were observed with both PACS and SPIRE as part of the HGBS.  These maps cover several degrees on the sky and have uniformly applied zero-point corrections to the PACS and SPIRE data using \emph{Planck} data following \citet{Bernard10}.  For such large maps, the PACS and SPIRE data can be reliably convolved to the \emph{Planck} resolution and compared pixel-by-pixel to get the average offset across the entire cloud.   Four of our globules are covered in larger molecular cloud maps in the HGBS.  These globules are L1544 and L1521F in Taurus, CB 68 in Ophiuchus, and B 68 in the Pipe Nebula.   Thus, we can directly test the zero-point corrections between these globules and their coincident larger HGBS maps. 

For this comparison, we use the publicly available temperature map of B 68 from \citet{Roy14}.  The \emph{Herschel} observations of B 68 are a subset of larger maps of the Pipe Nebula.  The Pipe was observed using the parallel PACS/SPIRE observing mode, covering an area of roughly $1.5\degree\ \times 1.5\degree$.  \citet{Roy14} produced line-of-sight averaged temperature maps and column density maps over a $20\arcmin \times 20\arcmin$ field centered on B 68 by fitting the \emph{Planck}-corrected \emph{Herschel} SEDs to a modified blackbody function with a fixed dust emissivity index of $\beta = 2$.  They also produce 3-D temperature and density profiles assuming an isotropic ISRF, which we do not consider here.

\begin{figure}[h!]
\centering
\includegraphics[width=0.475\textwidth]{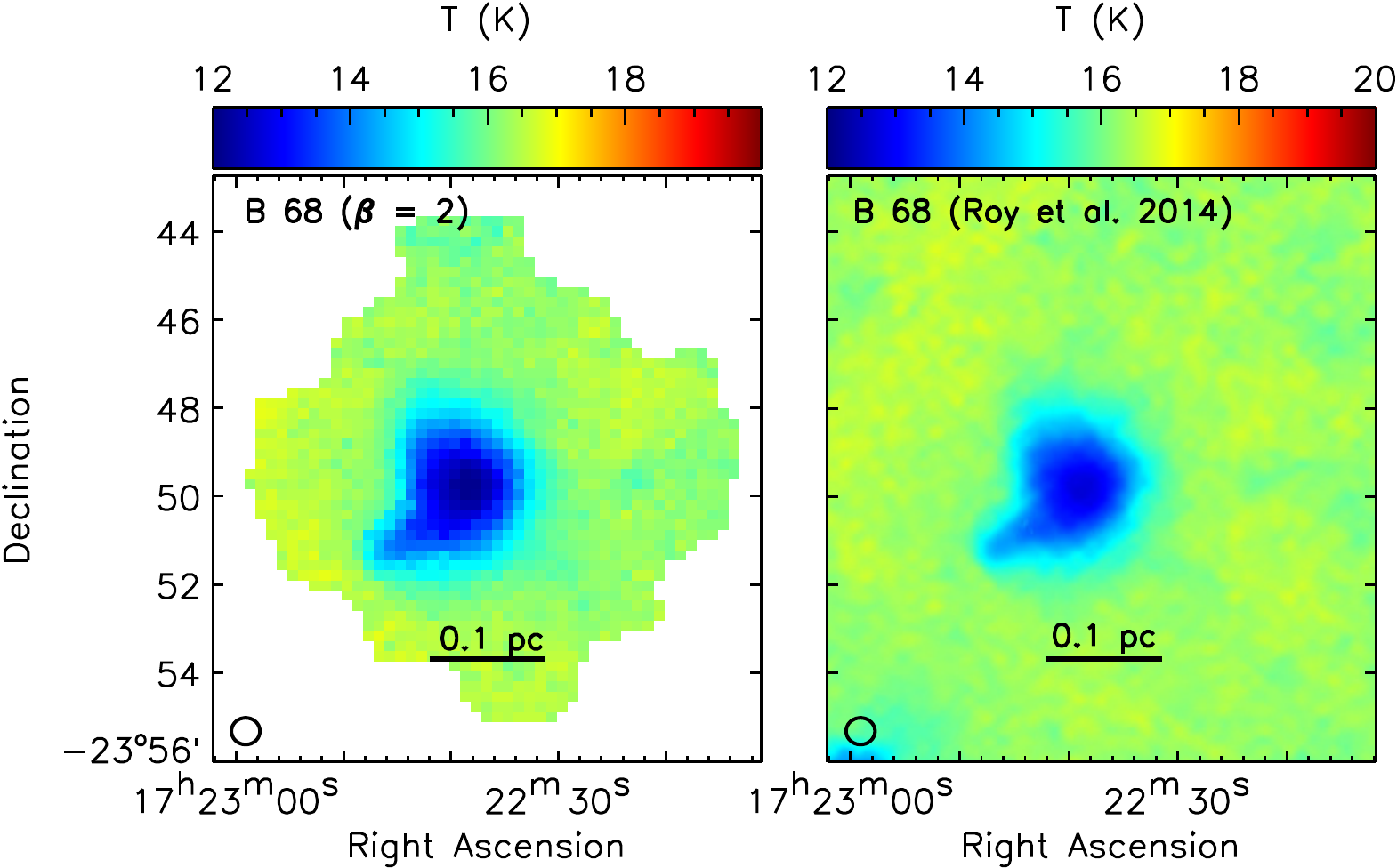}
\caption{Comparison of temperatures in B 68 from our work (left) and \citet[][right]{Roy14}.  We re-fitted our SEDs assuming $\beta = 2$ to match the SED model of \citet{Roy14}.  The two maps are on the same color scale as the B 68 map in Figure \ref{app_fig} for comparison.  Note that the steeper value of $\beta$ decreases the dust temperature by $\lesssim 2$ K. Our main analysis for B 68 assumes $\beta \approx 1.7$ from the \emph{Planck}-derived dust emissivity index map (see Figure \ref{app_fig}). \label{b68_temp}}
\end{figure}

Figure \ref{b68_temp} compares our temperature map of B 68 with that from \citet{Roy14}.  In this case, we use a fixed value of $\beta = 2$ to match the SED model of \citet{Roy14}.  This change in dust emissivity decreases the temperature at large angular extents by $\sim 2$ K compared to our primary analysis that uses $\beta \approx 1.7$ from the \emph{Planck} models (see Figure \ref{app_fig}).  With $\beta = 2$ for both maps, Figure \ref{b68_temp} shows excellent agreement between our zero-point corrected temperature map and the Gould Belt data.  The temperatures generally agree within a few percent across the entire field, suggesting they are consistent well within the measurement uncertainties.

The consistency between our analysis and the independently corrected temperatures from the Gould Belt survey indicates that our radial profile and intensity slice methods robustly measure the zero-point corrections for both PACS and SPIRE.  For B 68, the PACS corrections are in Group C (least reliable) and the SPIRE corrections are in Group A (most reliable).  Even with a Group C measure of the 160 \um\ offset, we were still able to match the results from \citet{Roy14}.  Thus, we can expect other globules with zero-point offsets in Groups A or B to be accurately corrected for future analyses (see Table \ref{table_sum} for a summary of the different groups at each band for all the globules).

\section{Summary}\label{summary}

In this paper, we produce corrected \emph{Herschel} maps for 56 low-mass globules using data products from the HSA.  The HSA maps are corrected against \emph{Planck}-determined emission maps using radial profiles to identify offsets between the HSA intensities and the \emph{Planck}-determined intensities at large angular extents off the globules.  These corrected \emph{Herschel} maps are available online.  We then fit their SEDs with a modified black body function to obtain line-of-sight average dust temperatures and optical depths.  Our main conclusions are:

\begin{enumerate}

\item Most (70-75\%) of the globules have reliable (Group A) or somewhat reliable (Group B) offset measurements for the HSA-PACS maps at 100 \um\ and 160 \um.  The least reliable (Group C) globules tend to have very small maps, incomplete coverage, or a large offset from the center of field.

\item Half of the globules have significant deviations ($>10$\%) in the HSA-SPIRE maps, even though these data were already corrected against \emph{Planck}.  We suggest that the HSA-SPIRE maps of globules may have inaccurate zero-point corrections because of the small sizes of these maps.  When convolved to \emph{Planck} resolutions, bright, diffuse background emission may contribute too strongly in the smaller maps and results in underestimated zero-point corrections.  It is better to correct these small maps at their native resolution, e.g., using the radial profile method discussed here.  

\item Using our techniques, we measure the intensity offsets for SPIRE.  Most (86\%) of the globules have reliable (Groups A or B) offsets.  The least reliable (Group C) globules generally have highly structured background emission.  

\item We compare our work to the EPoS survey, which background-subtracted the \emph{Herschel} data in the absence of available zero-point corrections \citep[e.g.,][]{Launhardt13}.  We tend to find consistent temperatures in the dense interiors of the globules but lower temperatures (by $2-5$ K) in their outer envelopes.  Since the background subtraction level will be most significant in the outer envelopes, our lower temperatures are likely more realistic.  

\item We also compared our dust temperatures for B 68 with independent measurements of the core that were included in a larger map of the Pipe Nebula from the \emph{Herschel} Gould Belt Survey \citep{Roy14}.   The  \emph{Herschel} intensities for the Pipe Nebula were also corrected using intensity maps based on \emph{Planck}-derived SED parameters.   We find excellent agreement between this study and that of \citet{Roy14}.  This agreement demonstrates the reliability of our techniques to correct small \emph{Herschel} maps. 

\item With a large sample of globules, we see a clear anti-correlation between their dust temperatures and optical depths.  We attribute this anti-correlation to the effect of self-shielding from the interstellar radiation field, where those globules with higher optical depths are more efficiently shielded and therefore cooler on average.  Most globules have a median optical depth of $\sim 1 \times 10^{-4}$ at 353 GHz.

\end{enumerate}

This paper contains the largest sample of globules examined with \emph{Herschel} to date.  The \emph{Herschel} data are key to observing and modeling low-mass globules and accurate corrections are necessary to produce reliable density profiles and connect their structure and state to the physical processes regulating star formation.  One key benefit to our technique is that we do not subtract a background level or convolve the \emph{Herschel} maps to the \emph{Planck} resolution.  Both background subtraction and convolving to the \emph{Planck} resolution appear to be problematic for small \emph{Herschel} fields with bright extended emission.   We further propose that the radial profiles may help recover diffuse emission that is filtered out by ground-based (sub)millimeter telescopes.  

Our analysis is meant to be a first look at these data.  With this large sample, future projects can compare their physical properties in subsamples, such as evolutionary state, size and mass, distance, or local environment.  The 56 globules presented in this study span a range of locations in the Galaxy, from highly isolated objects to sources on the outskirts of high-mass star-forming clouds.    The data published here represent the strongest constraints to date for the dust temperature, (column) density, and mass for a large sample of globules.

\vspace{1cm}
\begin{acknowledgements}
We thank the anonymous referee for comments that improved the clarity of this paper and its discussion.  SIS acknowledges the support for this work provided by NASA through Hubble Fellowship grant HST-HF2-51381.001-A awarded by the Space Telescope Science Institute, which is operated by the Association of Universities for Research in Astronomy, Inc., for NASA, under contract NAS 5-26555.  AS is thankful for funding from the ``Concurso Proyec- tos Internacionales de Investigaci\'{o}n, Convocatoria 2015'' (project code PII20150171) and the BASAL Centro de Astrof\'{i}sica y Tecnolog\'{i}as Afines (CATA) PFB-06/2007.  The authors thank Douglas Finkbeiner for discussions on the methodology and Ivan Valtchanov for discussions on the SPIRE deviations.   This project used data taken by the \emph{Herschel} PACS and SPIRE instruments.  This project also used data based on observations obtained with \emph{Planck} (http://www.esa.int/Planck), an ESA science mission with instruments and contributions directly funded by ESA Member States, NASA, and Canada.  This research made use of the SIMBAD database, operated at CDS, Strasbourg, France  and the VizieR catalogue access tool, CDS, Strasbourg, France.

PACS has been developed by a consortium of institutes led by MPE (Germany) and including UVIE (Austria); KU Leuven, CSL, IMEC (Belgium); CEA, LAM (France); MPIA (Germany); INAF-IFSI/OAA/OAP/OAT, LENS, SISSA (Italy); IAC (Spain). This development has been supported by the funding agencies BMVIT (Austria), ESA-PRODEX (Belgium), CEA/CNES (France), DLR (Germany), ASI/INAF (Italy), and CICYT/MCYT (Spain).  SPIRE has been developed by a consortium of institutes led by Cardiff University (UK) and including Univ. Lethbridge (Canada); NAOC (China); CEA, LAM (France); IFSI, Univ. Padua (Italy); IAC (Spain); Stockholm Observatory (Sweden); Imperial College London, RAL, UCL-MSSL, UKATC, Univ. Sussex (UK); and Caltech, JPL, NHSC, Univ. Colorado (USA). This development has been supported by national funding agencies: CSA (Canada); NAOC (China); CEA, CNES, CNRS (France); ASI (Italy); MCINN (Spain); SNSB (Sweden); STFC, UKSA (UK); and NASA (USA).  

\end{acknowledgements}

\bibliographystyle{apj}
\bibliography{references}

\clearpage
\newpage

\begin{appendix}

\section{HSA-SPIRE Deviations in NGC 7538} \label{case_study}

NGC 7538 is a bright, high-mass star-forming region roughly 2.7 kpc away \citep{Moscadelli09}.  We selected this region because (1) it has extended emission, which is common to the globules with large intensity deviations in the SPIRE bands, and (2) it was observed multiple times with SPIRE in maps that cover different areas of the cloud.  NGC$\ 7538$ was mapped fully in a $\sim 2.7\degree \times 2.7$\degree\ chunk of the Galactic Plane as part of the Hi-Gal survey \citep{Molinari10} and in a $\sim 1\degree \times 1$\degree\ map for the HOBYS survey \citep[see][]{Fallscheer13}.  A smaller 11\arcmin\ section around NGC 7538 IRS 1-3 was also observed with SPIRE due to saturation.  For this case study, we use the smallest 11\arcmin\ map and largest 2.7\degree\ map of NGC 7538 to illustrate how map size affects the zero-point corrections.  

\begin{figure}[h!]
\centering
\includegraphics[width=0.475\textwidth]{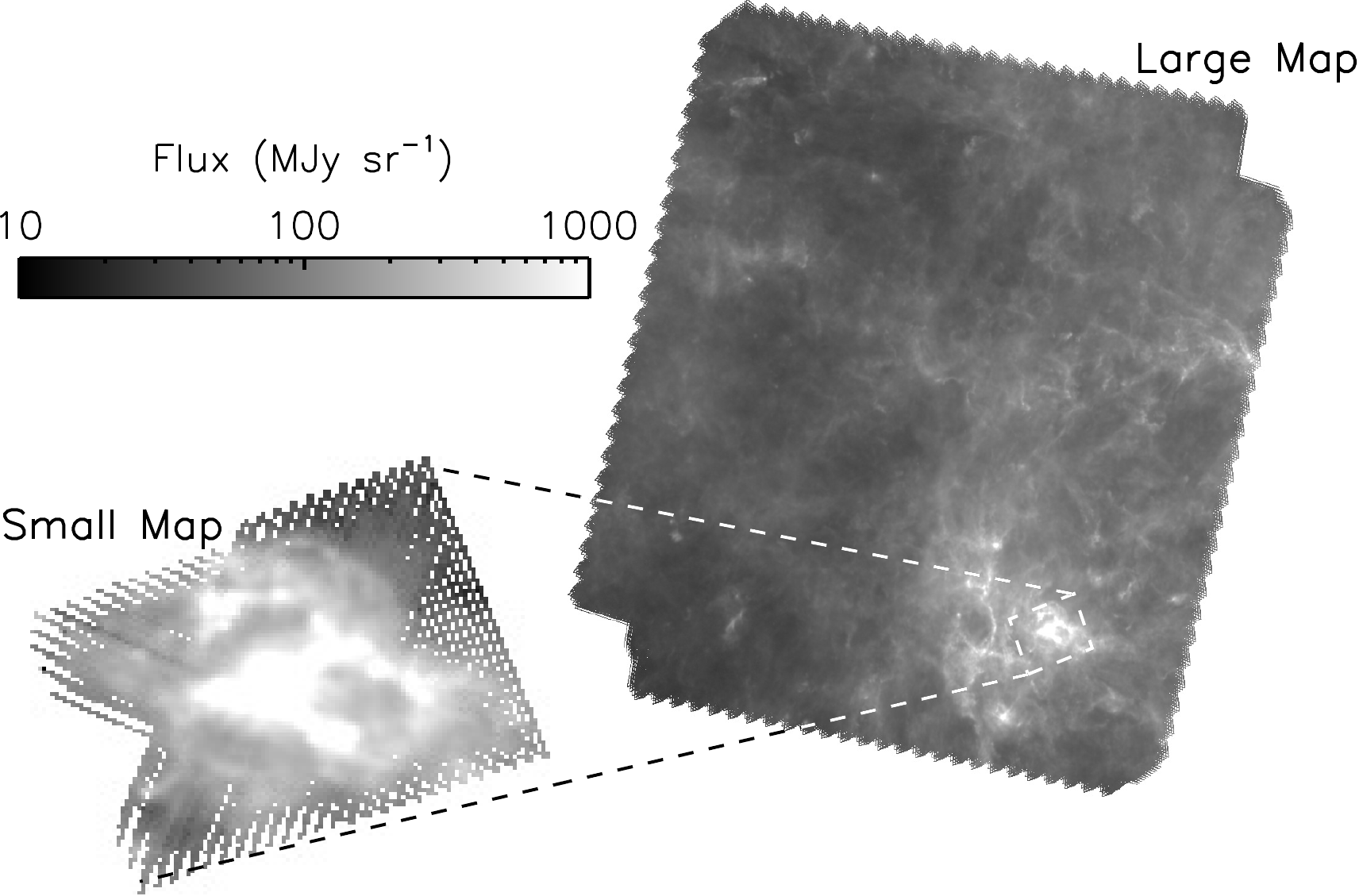}
\caption{Observations of NGC 7538 at 350 \um\ from the HSA.  The larger map spans $2.7\degree \times 2.7$\degree, whereas the smaller map spans $\sim 11$\arcmin\ and covers only NGC 7538 IRS1-3.  The approximate area covered by the small map relative to the large map is shown by the dashed rectangle.  Both figures use the same logarithmic color scaling. These data were reduced with the same version of the HSA pipeline and include a zero-point correction.  \label{ngc7538}}
\end{figure}

Figure \ref{ngc7538} shows the large and small maps of NGC 7538 at 350 \um.  We use the Level 2.5 HSA data products for the large map (to combine the nominal and orthogonal scans) and the Level 2 HSA data products for the small map.   The SPIRE data were reduced using version 14 of the HSA pipeline and both maps are zero-point corrected by the pipeline as described in Section \ref{method}.  Thus, the SPIRE maps are treated equally for both fields.

\begin{figure}[h!]
\includegraphics[width=0.475\textwidth]{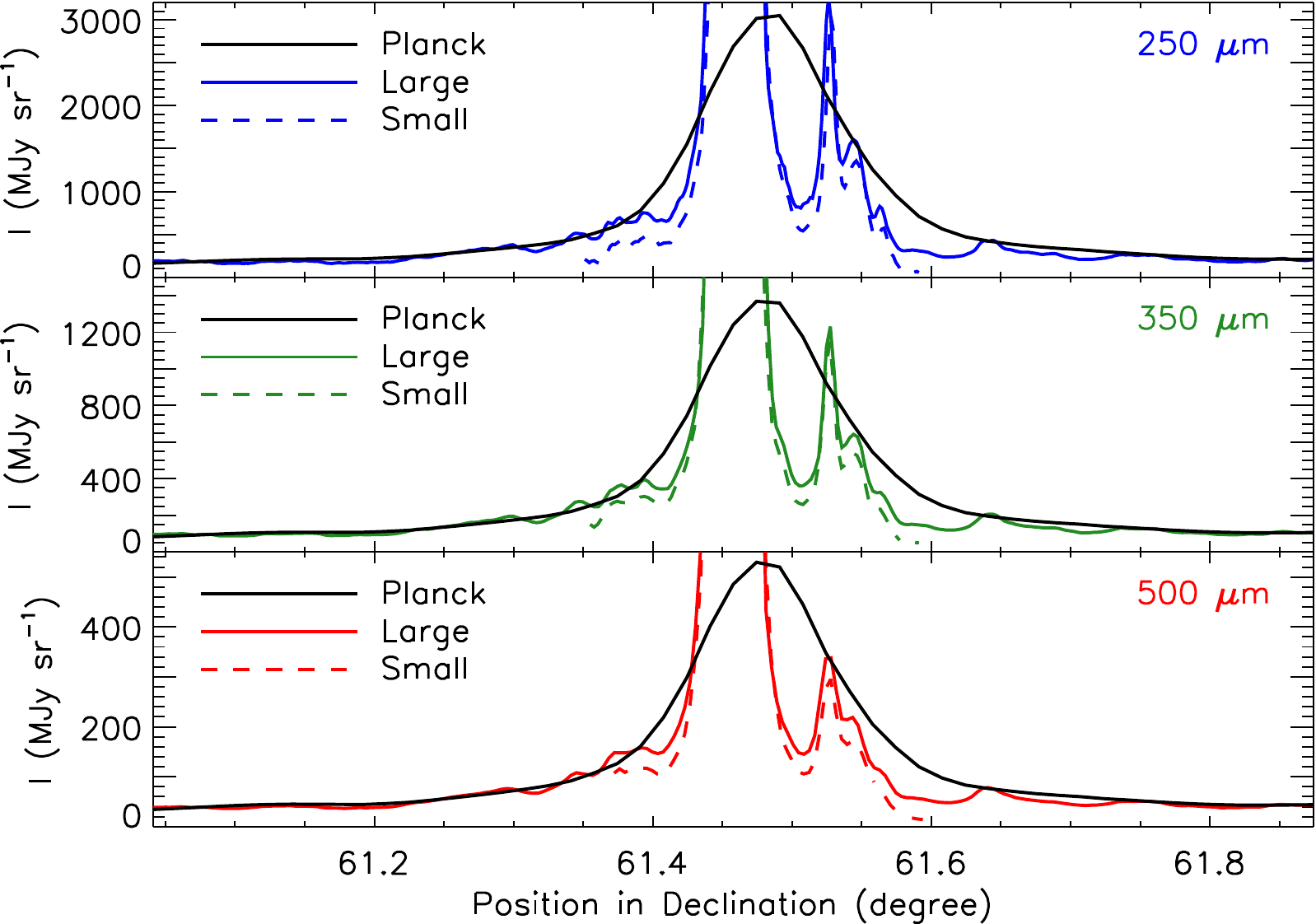} 
\caption{Intensity slices through NGC 7538 in declination.  The solid black curve correspond to the \emph{Planck}-determined slices, the solid colored curves give the SPIRE slices from the large maps, and the dashed colored curves are the SPIRE slices from the small maps.  The slices from the larger map agree well with the \emph{Planck}-determined slices at large angular extents off the main cluster.  The slices from the smaller map, however, greatly underestimate the intensity.  The plots are both truncated in the vertical direction to highlight the intensity offset.     \label{slice_ngc7538}}
\end{figure}

A quick analysis of the maps in Figure \ref{ngc7538} shows nearly identical features.  Note that the two maps in Figure \ref{ngc7538} use the same logarithmic color scale.  Nevertheless, in a more careful analysis, we find substantial deviations in intensity between the small and large map of NGC 7538 for all three SPIRE bands.  Figure \ref{slice_ngc7538} compares intensity slices through both the large and small maps (solid and dashed colored curves) and the corresponding \emph{Planck}-determined data (black solid curves).  We use intensity slices because the radial profiles contain too much structure to be well fit by a Gaussian function.  The slices from the \emph{Planck}-determined maps and the large HSA-SPIRE maps agree relatively well (within 10\%) at angular distances $> 0.2$\degree\ from the center of the cluster.   The small map slices, however, are systematically lower in intensity by $\sim 270$ \MJysr\ at 250 \um, $\sim 100$ \MJysr\ at 350 \um, and $\sim$ 30 \MJysr\ at 500 \um. These offsets can account for $\sim$ 50\%\ of the intensity at the edge of the small map.

The agreement between the larger HSA-SPIRE maps and the \emph{Planck}-determined maps suggests that the former recovered their zero-point corrections.   In contrast, the smaller HSA-SPIRE maps shows substantial deviations in intensity, even though they were produced from data obtained with the same instrument.  The HSA-SPIRE slices from both the small and large maps show similar structures, so the underestimated intensities are unlikely from errors in the observations themselves.  Furthermore, the SPIRE data from both surveys were reduced using the same version of the HSA pipeline, so they will have the same flux calibrations.  That means the large deivations cannot be explained by differences in the data reduction.

The smaller map of NGC 7538 IRS1-3 shares the same issue as the globule fields with bright, extended emission.  Namely, the HSA-SPIRE intensities are underestimated relative to the \emph{Planck}-determined emission.  Since the larger map of NGC 7538 does not share this deviation, we propose that small, HSA-SPIRE scan maps with bright extended emission have inaccurate zero-point corrections.   Bright emission can bias the HSA data when convolved to 5\arcmin\ resolution, which in turn causes the zero-point corrections to be underestimated.  This particular problem is amplified in the small map of NGC 7538 IRS 1-3 relative to the globules, as this field is only 11\arcmin\ in size and has very bright emission across the entire SPIRE field.  The HSA-SPIRE maps for the globules are typically $\sim 40$\arcmin\ in size and have less extended emission, so we can expect any deviations relative to \emph{Planck} to be smaller than for NGC 7538 (see Table \ref{table_spire}).  Nevertheless, about half of the globules have non-negligible ($>10$\%) deviations at 250 \um\ and 500 \um\ (see Figure \ref{histo_spire}), which should be accounted for prior to further scientific analysis.

\section{Temperature and Optical Depth Maps} \label{app_figures}

Figure \ref{app_fig} shows the temperature and optical depth maps for all 56 globules from the SED fitting outlined in Section \ref{sed_section}.  The maps all have a common resolution of 36.3\arcsec, as shown by the black circles in the lower-left corners.  The globule names are given in the top-left corners.  Note that the color scales are selected to highlight the globules, and the optical depth maps have a logarithmic color scale.  The maps also have a physical scale bar corresponding to 0.1 pc, 0.2 pc, or 0.25 pc depending on the distance of the globule.   For those globules with a range of distance estimates in the literature, we select the upper limit distance for the scale bar.  

Two of the globules, B 35A and BHR 79, have excess 160 \um\ emission at the edge of their maps but within their masks that is not detected in the SPIRE bands.  This excess emission extends to a larger area when convolved to 36.3\arcsec\ resolution and skews the SEDs.  As a result, we find unusually high temperatures of $\gtrsim 30$ K and unusually low optical depths of $< 10^{-5}$ in the regions associated with this excess emission.  Hence, these ``hot spots'' may be unreliable, as there is no corresponding emission in the SPIRE bands.  B 35A is a bright rimmed globule associated with the $\lambda$ Ori system, and as such, its high temperatures could be real.  In contrast, BHR$\ 79$ is in the more quiescent Musca cloud and unlikely to have temperatures $\gtrsim 30$ K at such large angular extents.  In both cases, this emission has been masked out of the temperature maps to keep the focus on the cooler temperatures associated with the globules.

The data in Figure \ref{app_fig} were intensity-corrected prior to SED fitting using the offsets in Table \ref{table_pacs} and deviations in Table \ref{table_spire}.  Table \ref{table_sum} summarizes the zero-point correction Groups for each globule.  See Section \ref{slices} for an explanation of these Groups.

\newpage


\begin{figure*}
\centering
\includegraphics[scale=0.57]{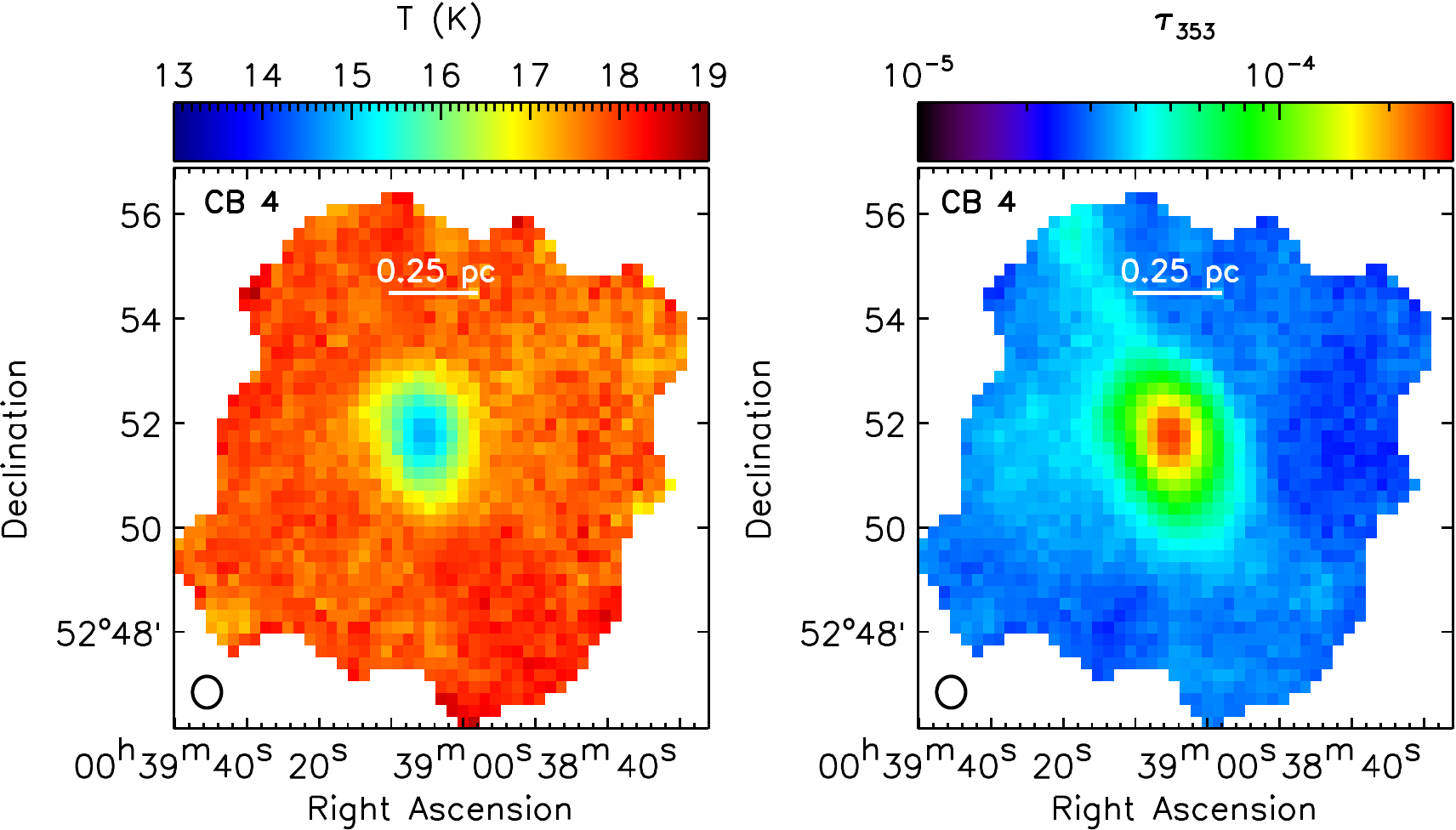}
\qquad
\includegraphics[scale=0.57]{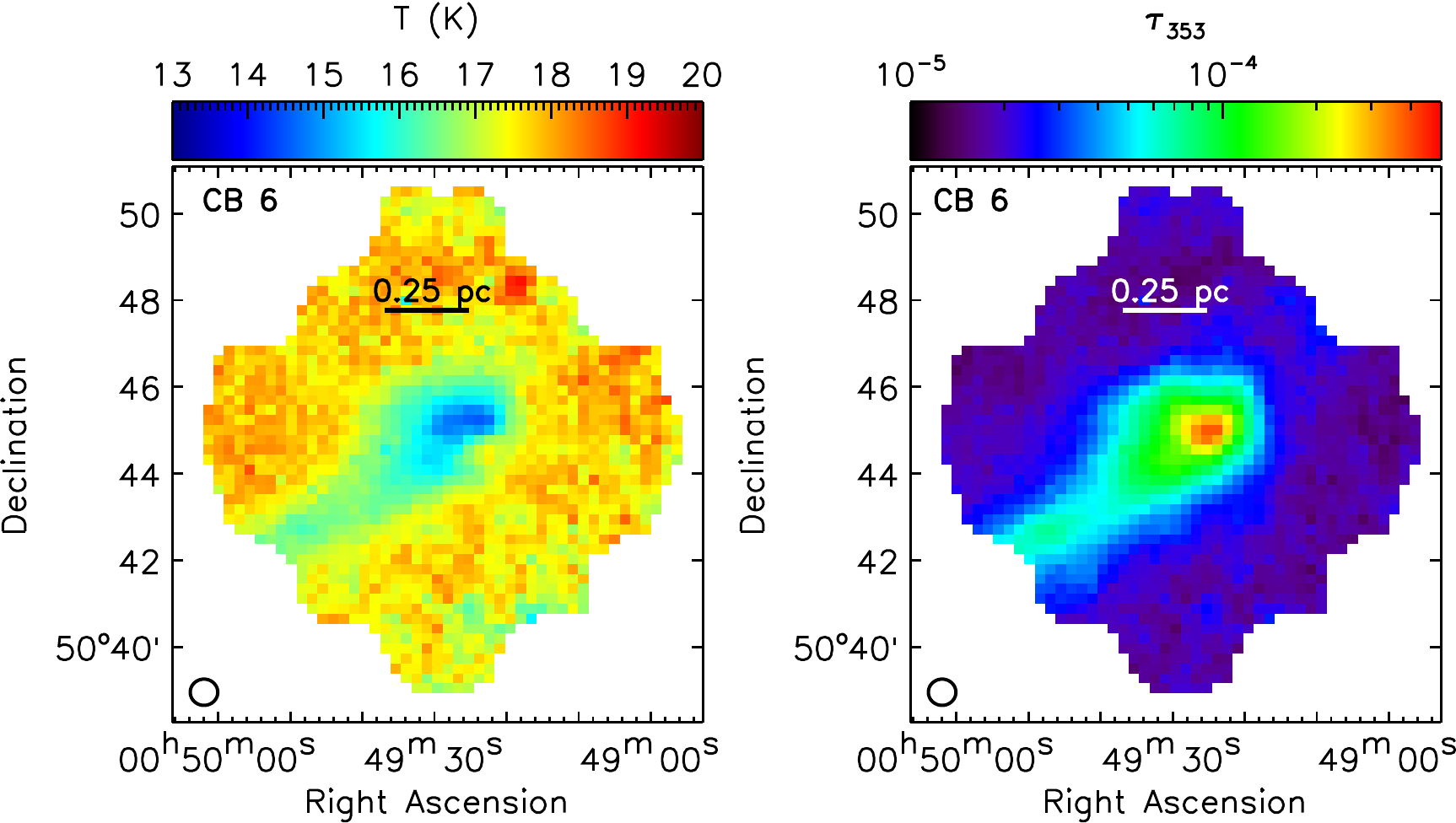}
\qquad
\includegraphics[scale=0.57]{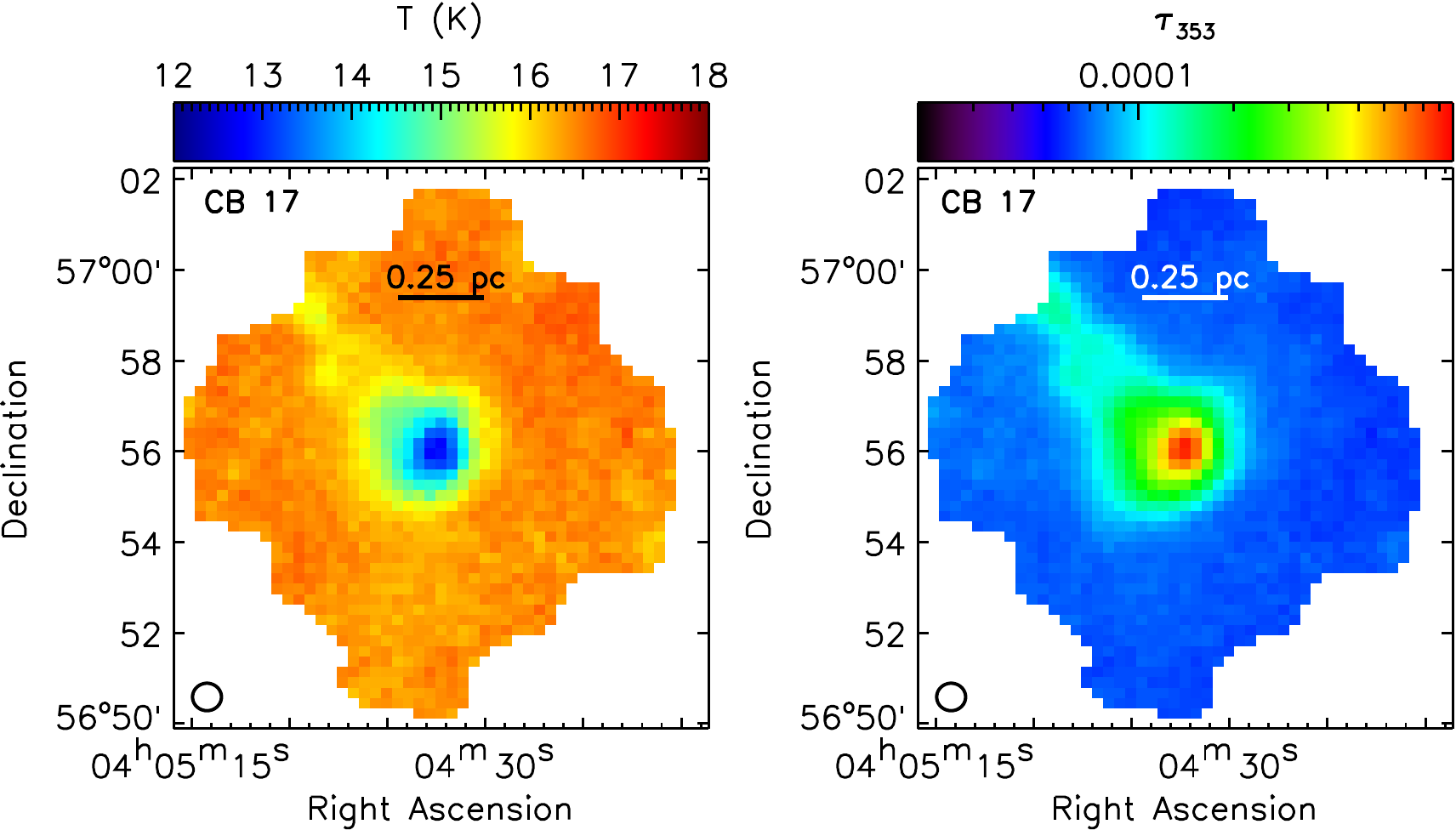}
\caption{Maps of dust temperature (left) and optical depth at 353 GHz (right) for all globules.  These maps are produced following the SED fitting technique outlined in Section \ref{sed_section}.  The PACS and SPIRE data have been zero-point corrected following our estimated offsets.  Each panel shows the map resolution (36.3\arcsec) in the bottom-left corner and the cloud name in the top-left corner.  We also include scale bar assuming the distances listed in Table \ref{table_loc}.  For those globules with multiple distances listed, we select the largest distance value.  These panels correspond to CB 4, CB 6, and CB 17.
\label{app_fig}}
\end{figure*}
\begin{figure*}
\ContinuedFloat
\centering
\includegraphics[scale=0.575]{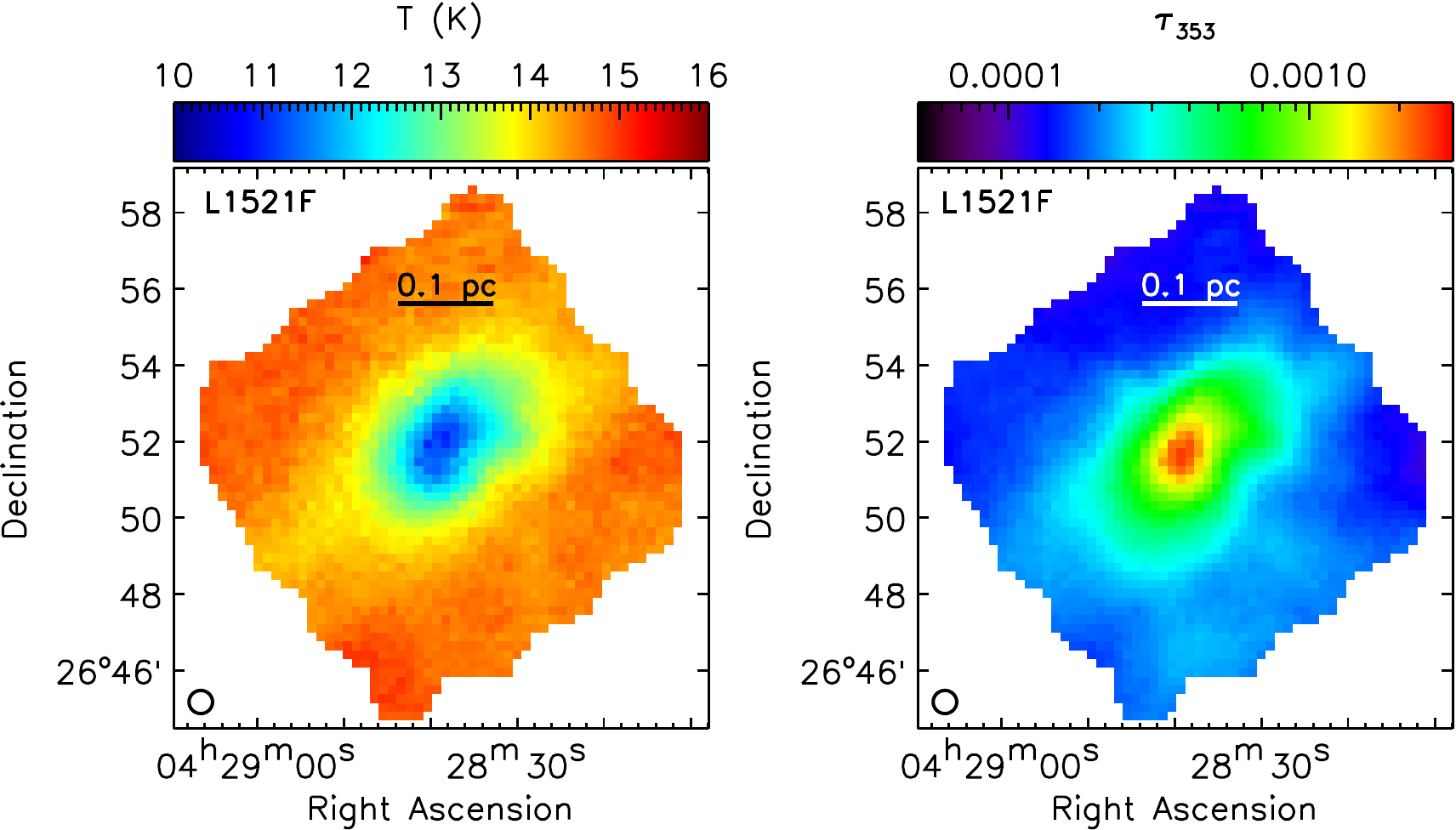}
\qquad
\includegraphics[scale=0.575]{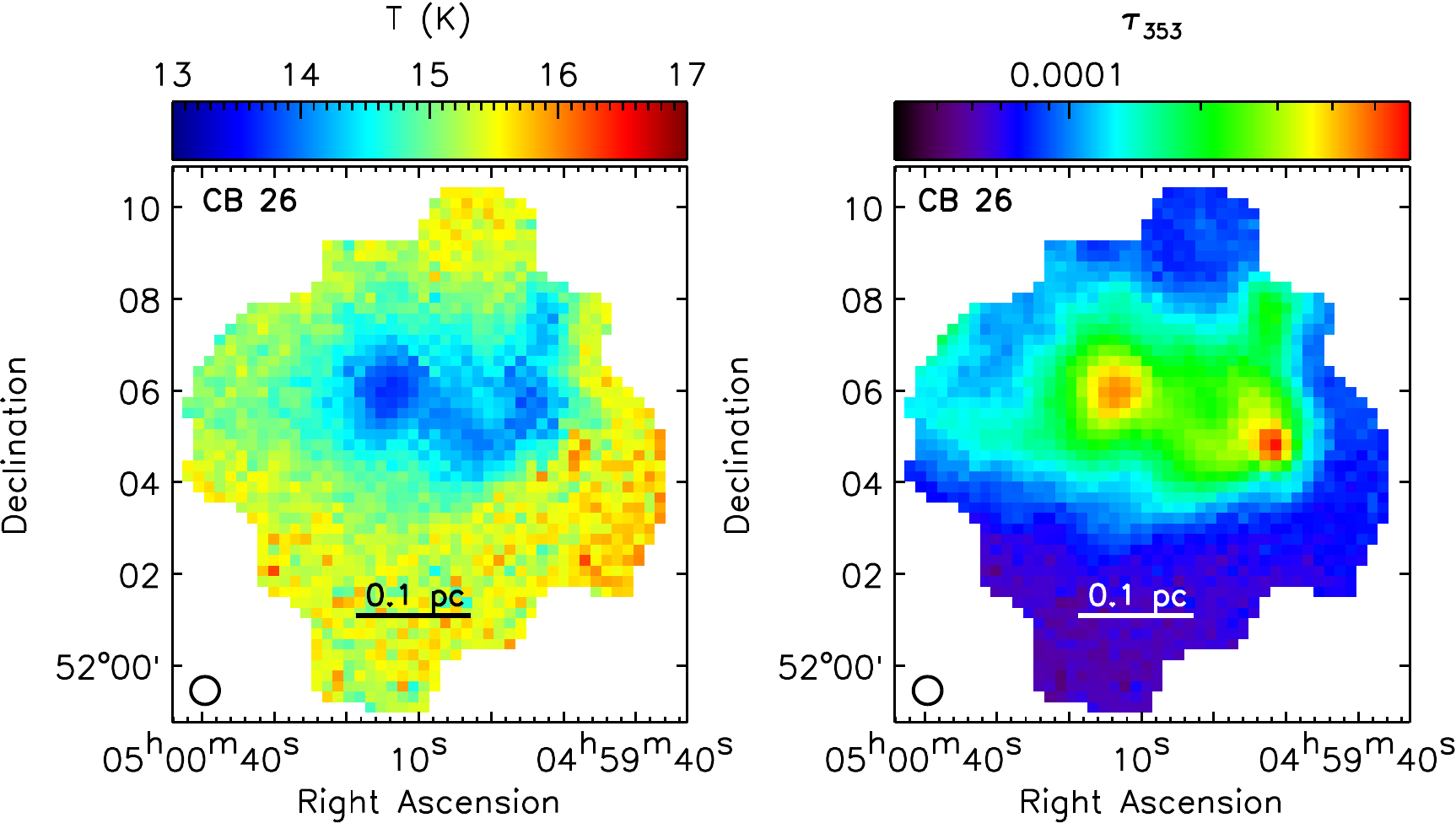}
\qquad
\includegraphics[scale=0.575]{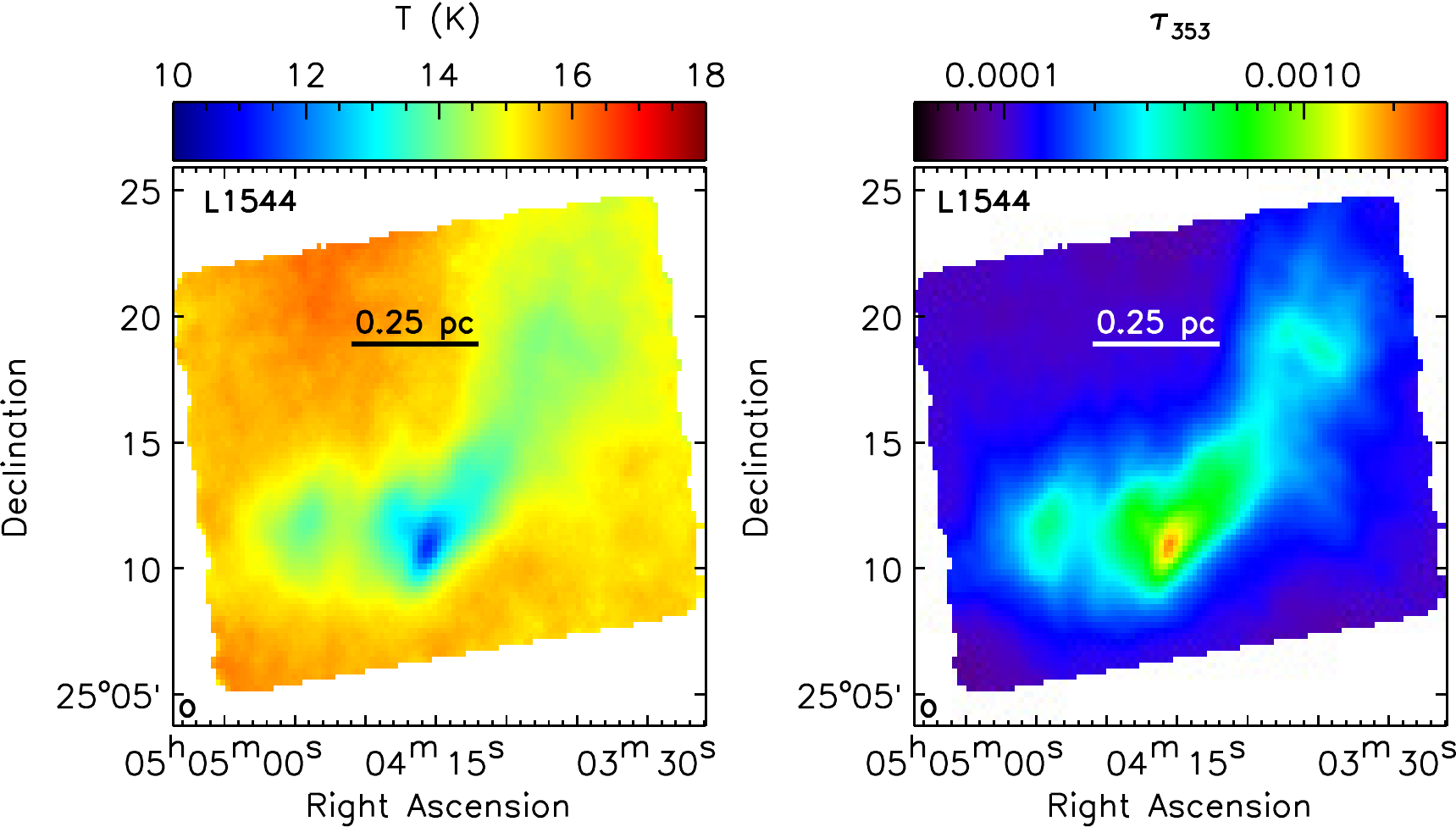}
\qquad
\includegraphics[scale=0.575]{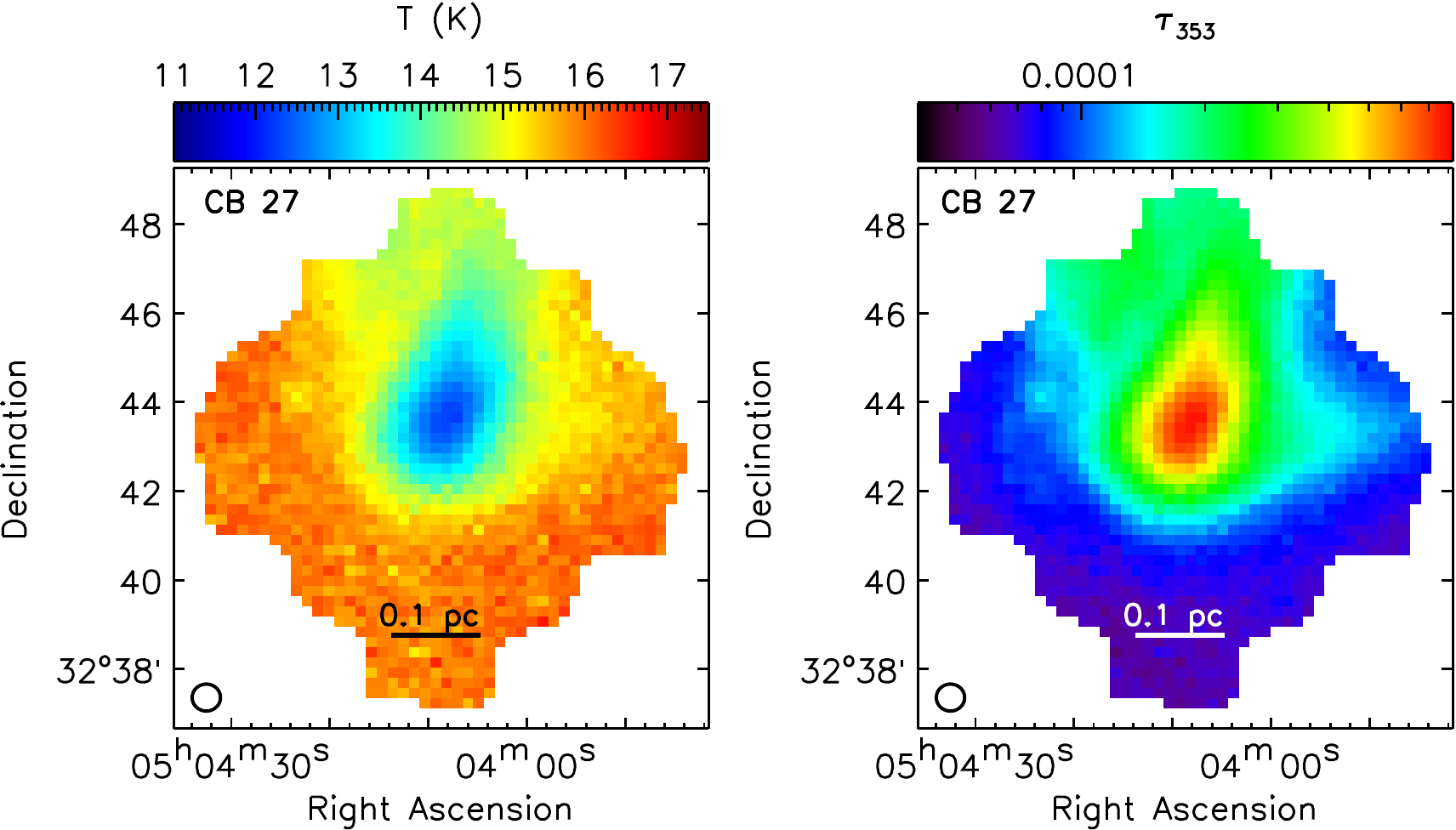}
\caption{Continued -  For L1521F, CB 26, L1544, and CB 27.}
\end{figure*}
\begin{figure*}
\ContinuedFloat
\centering
\includegraphics[scale=0.575]{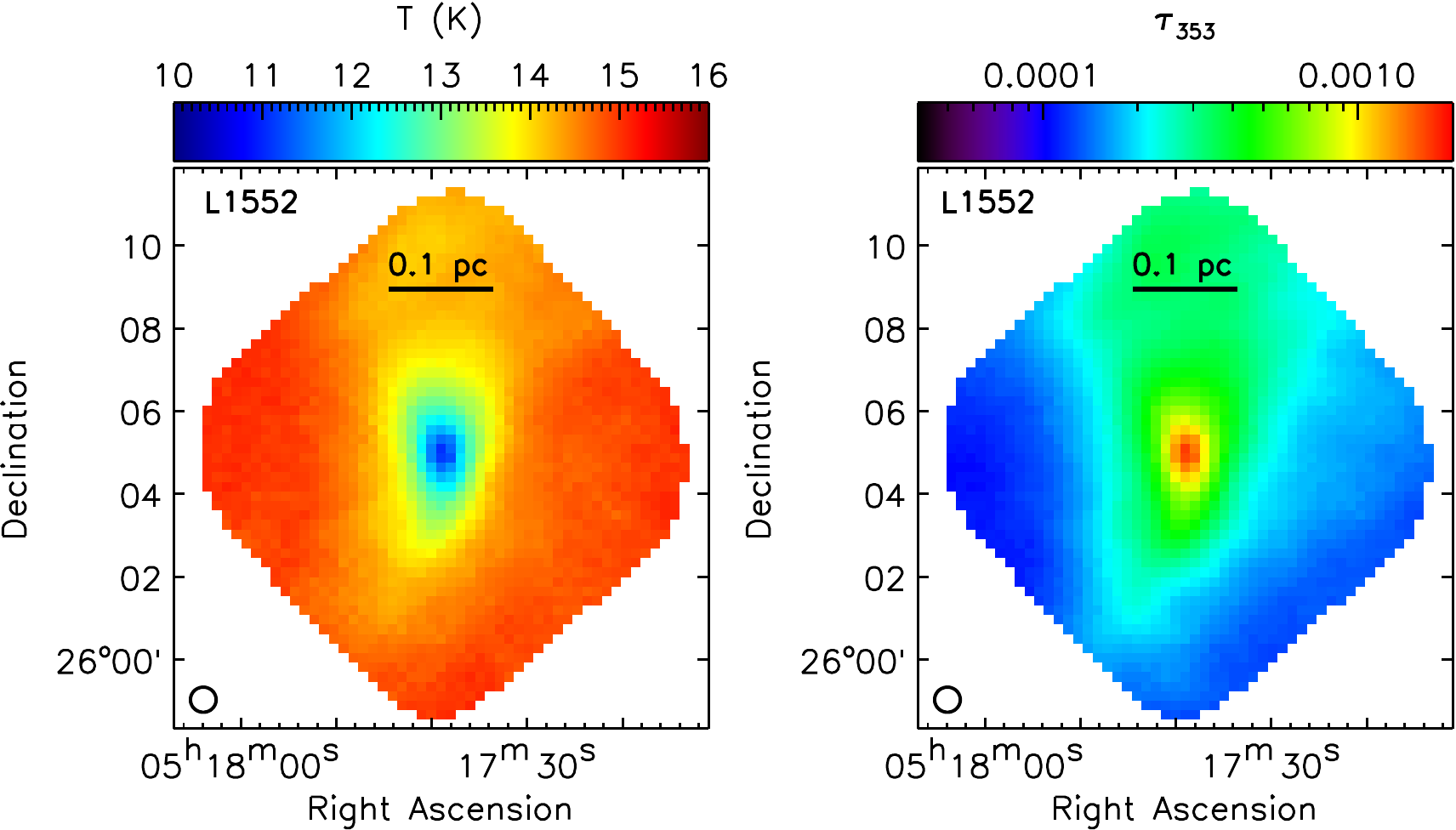}
\qquad
\includegraphics[scale=0.575]{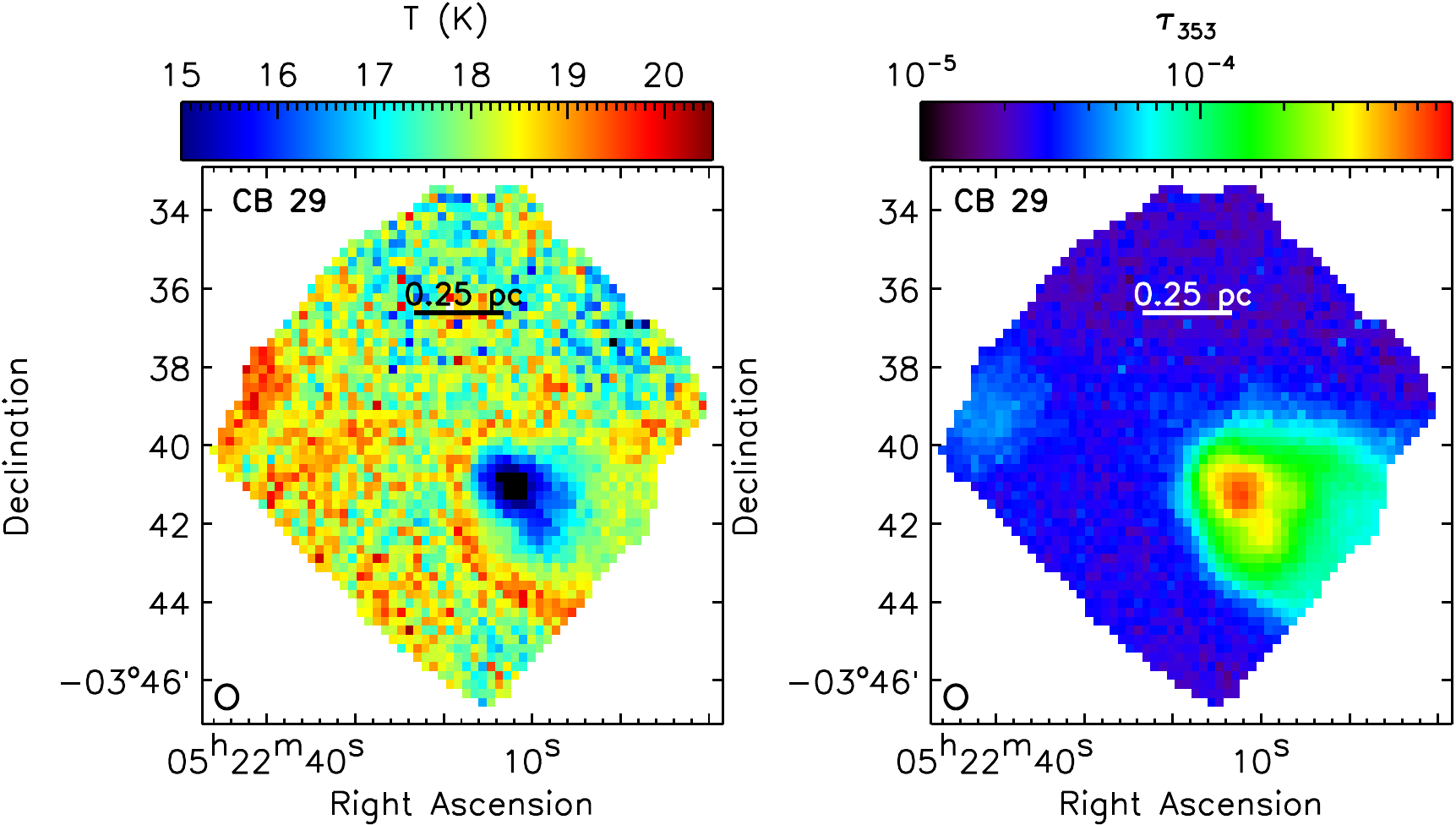}
\qquad
\includegraphics[scale=0.575]{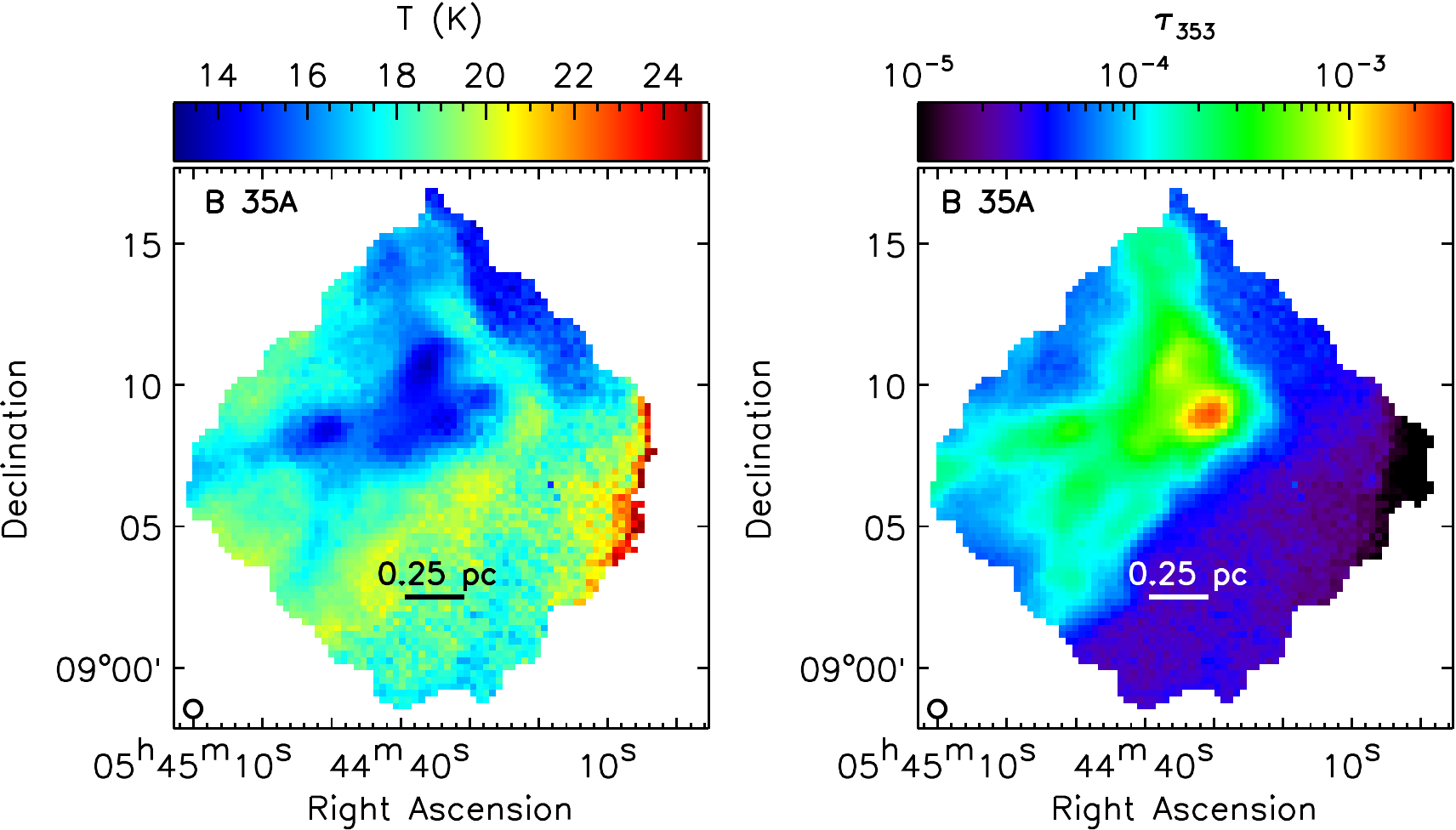}
\qquad
\includegraphics[scale=0.575]{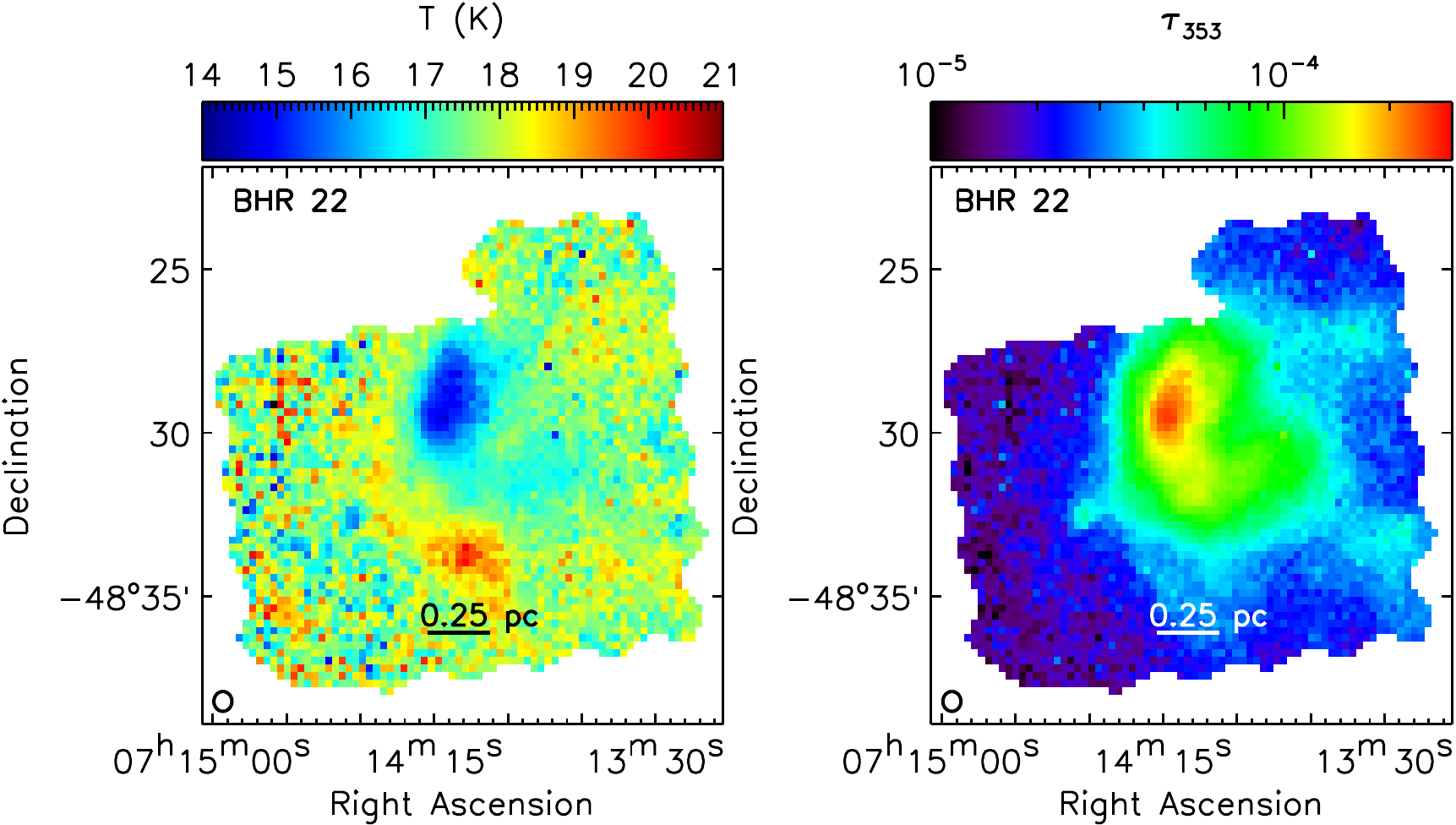}
\caption{Continued -  For L1552, CB 29, B 35A, and BHR 22}
\end{figure*}
\begin{figure*}
\ContinuedFloat
\centering
\includegraphics[scale=0.575]{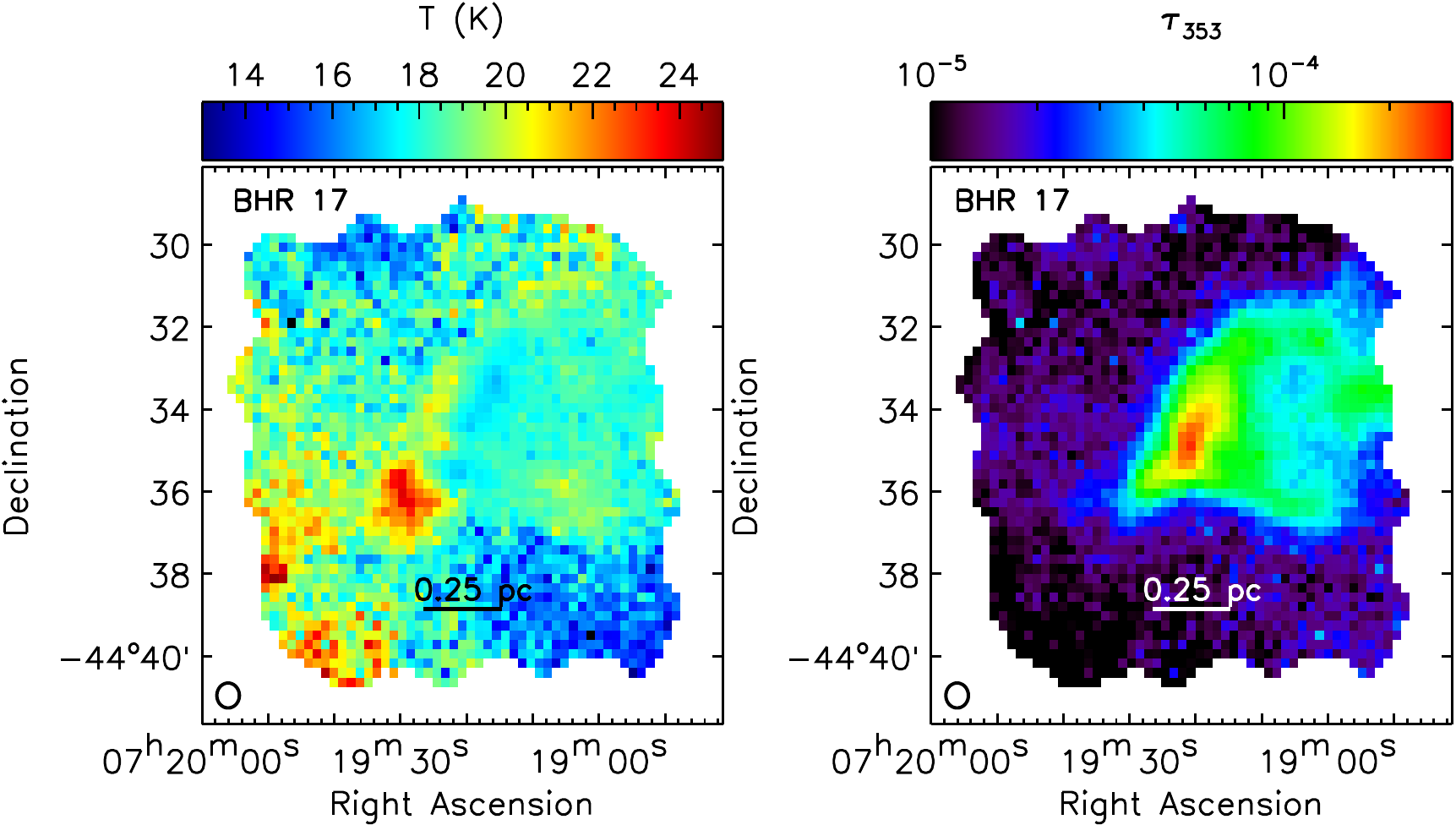}
\qquad
\includegraphics[scale=0.575]{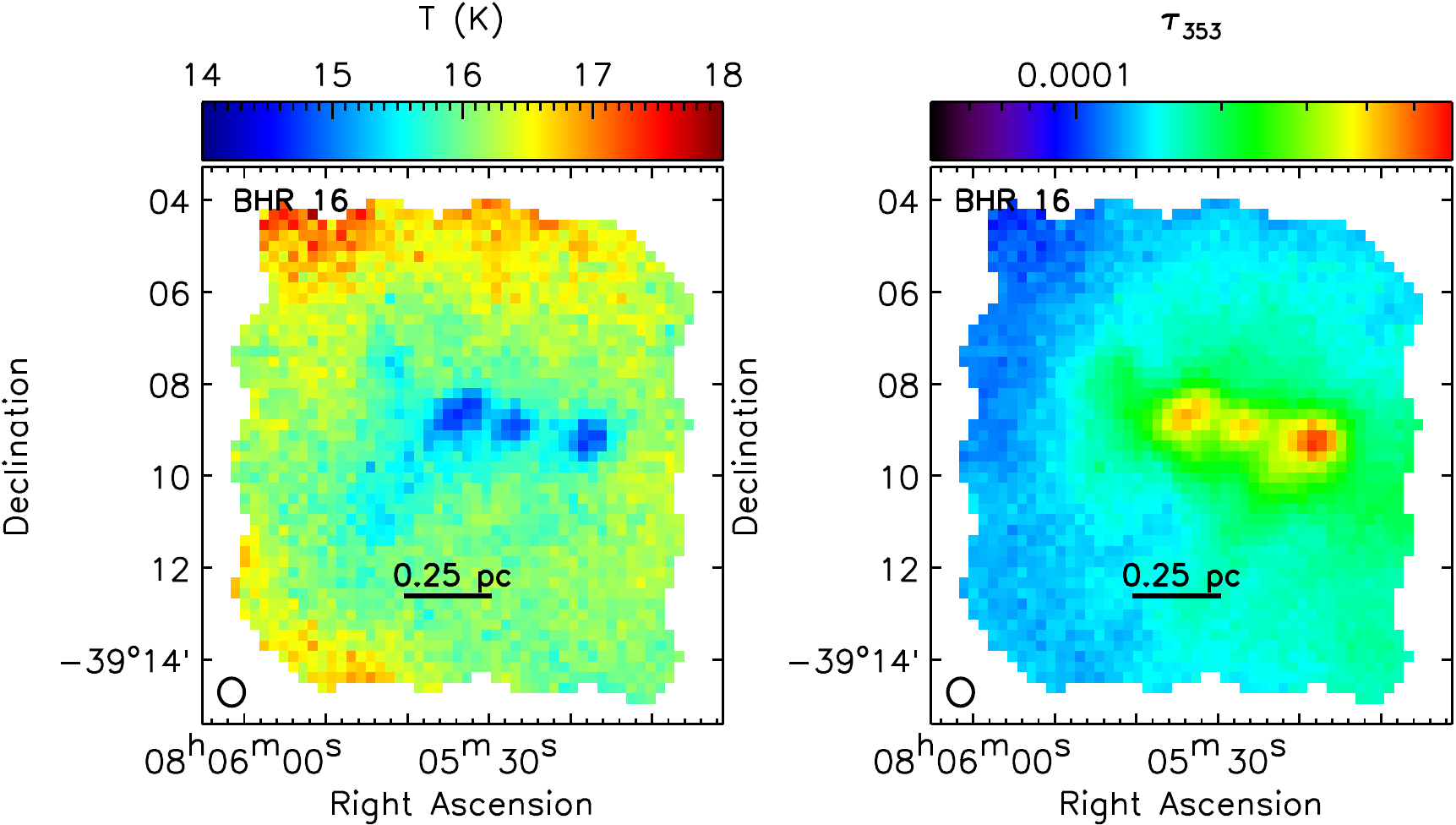}
\qquad
\includegraphics[scale=0.575]{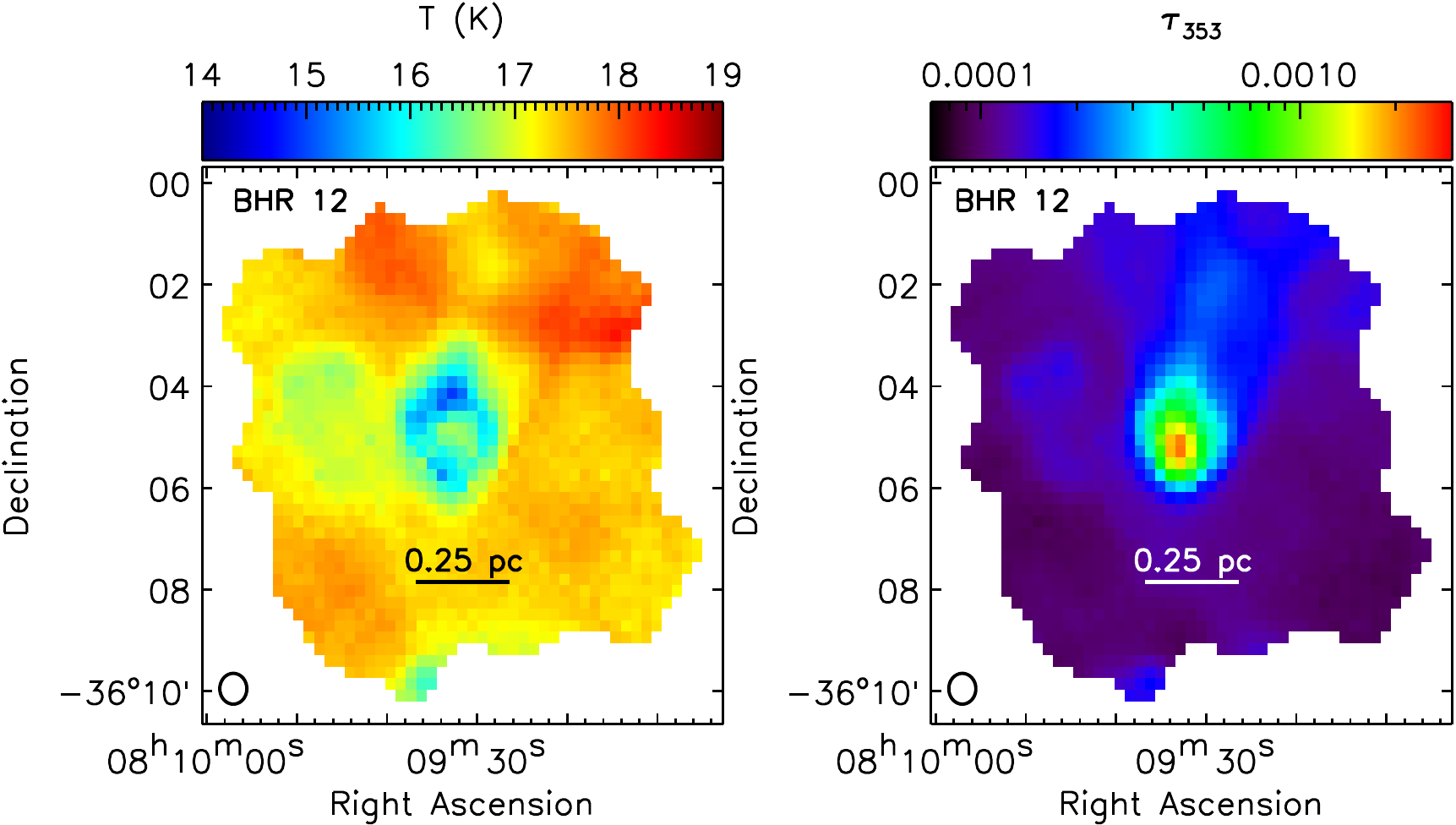}
\qquad
\includegraphics[scale=0.575]{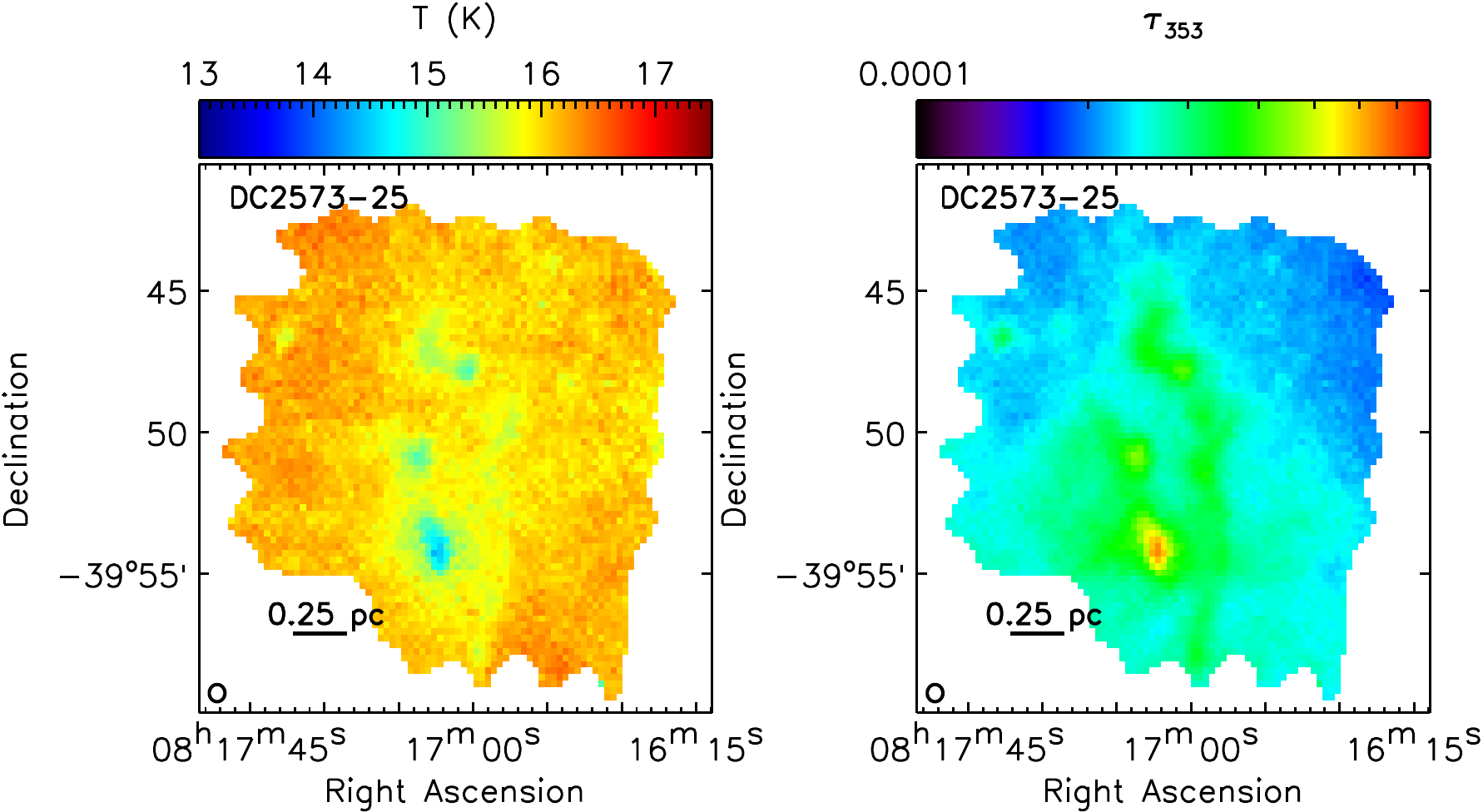}
\caption{Continued -  For BHR 17, BHR 16, BHR 12, DC2573-25}
\end{figure*}
\begin{figure*}
\ContinuedFloat
\centering
\includegraphics[scale=0.575]{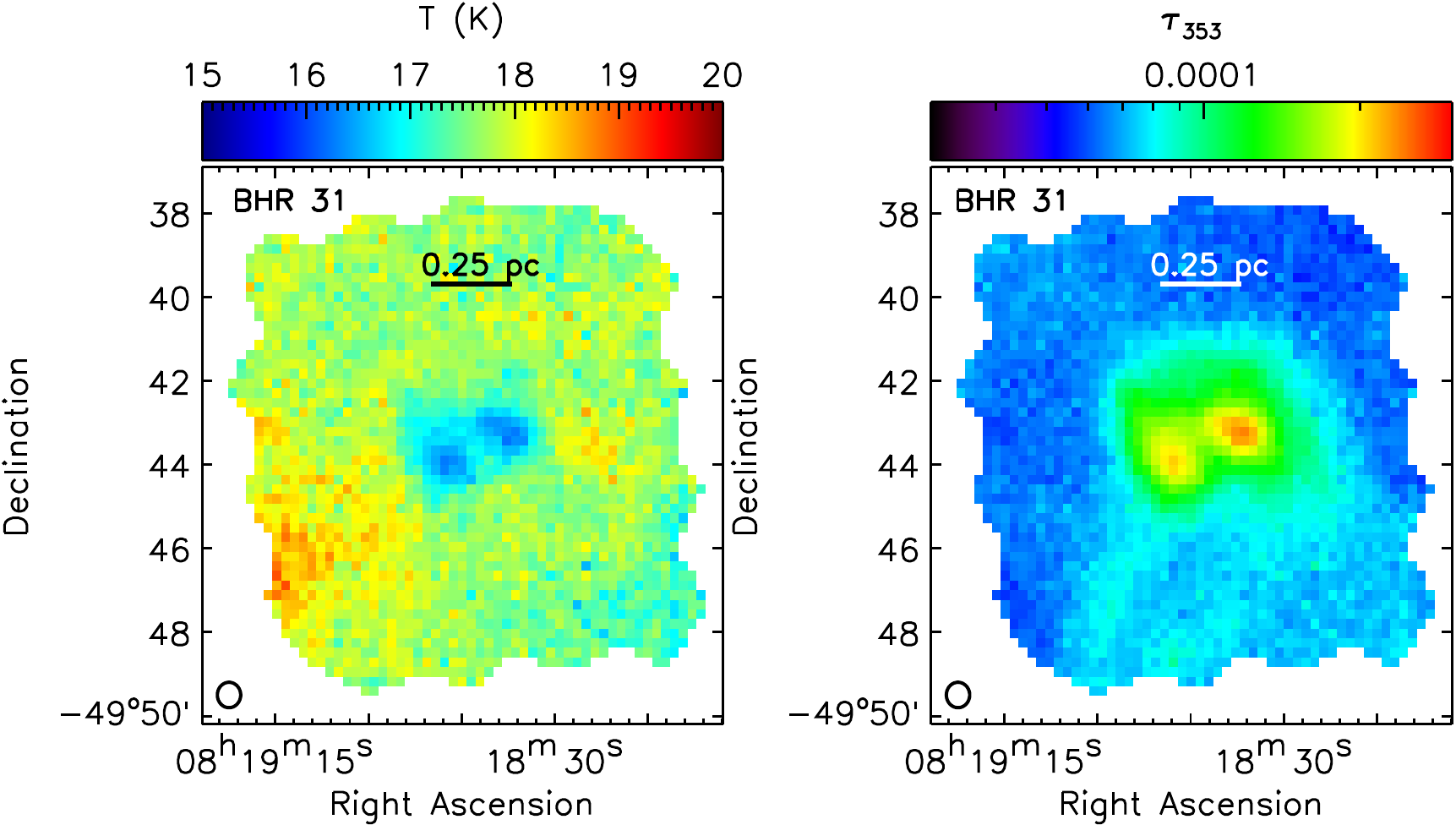}
\qquad
\includegraphics[scale=0.575]{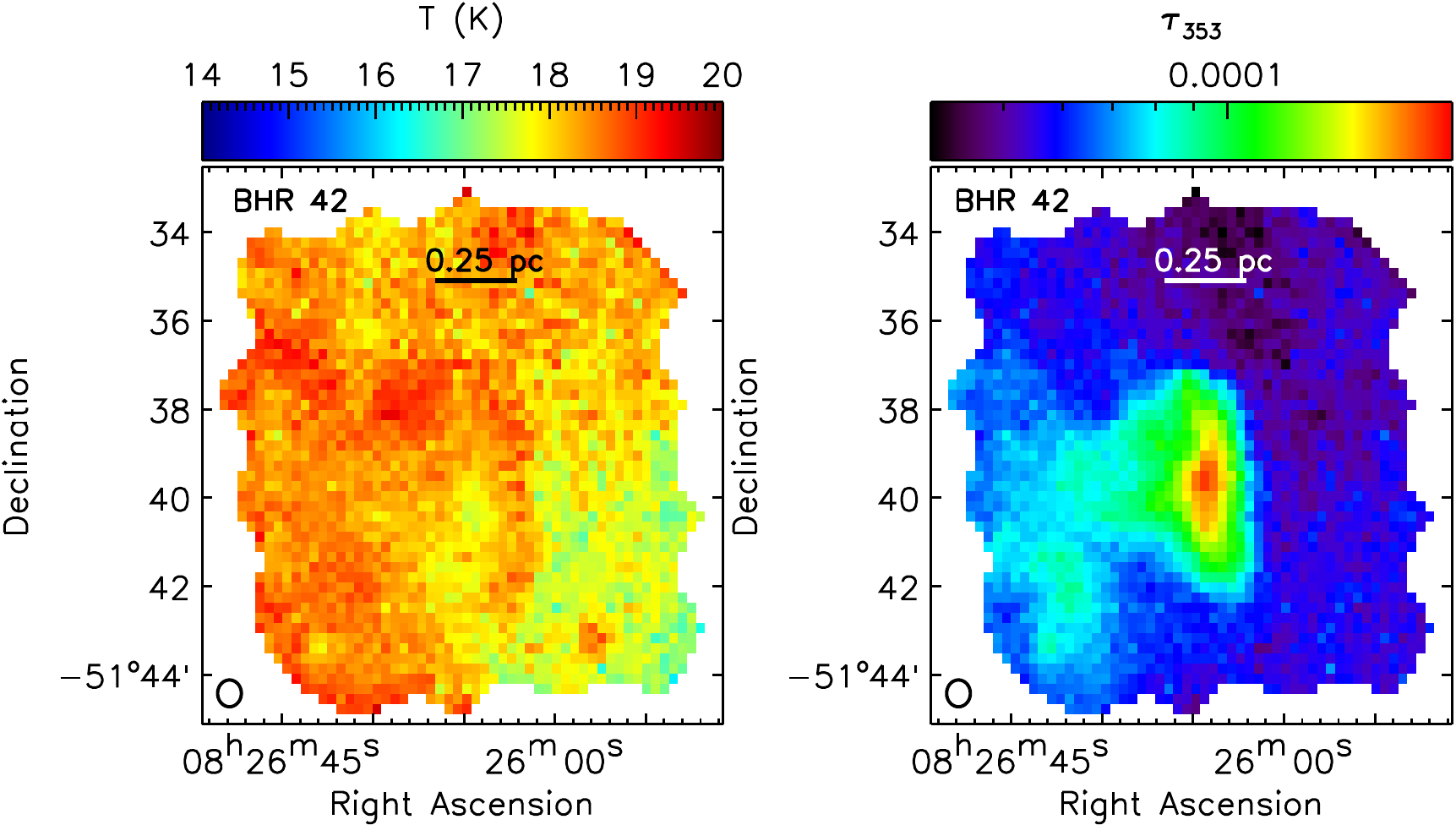}
\qquad
\includegraphics[scale=0.575]{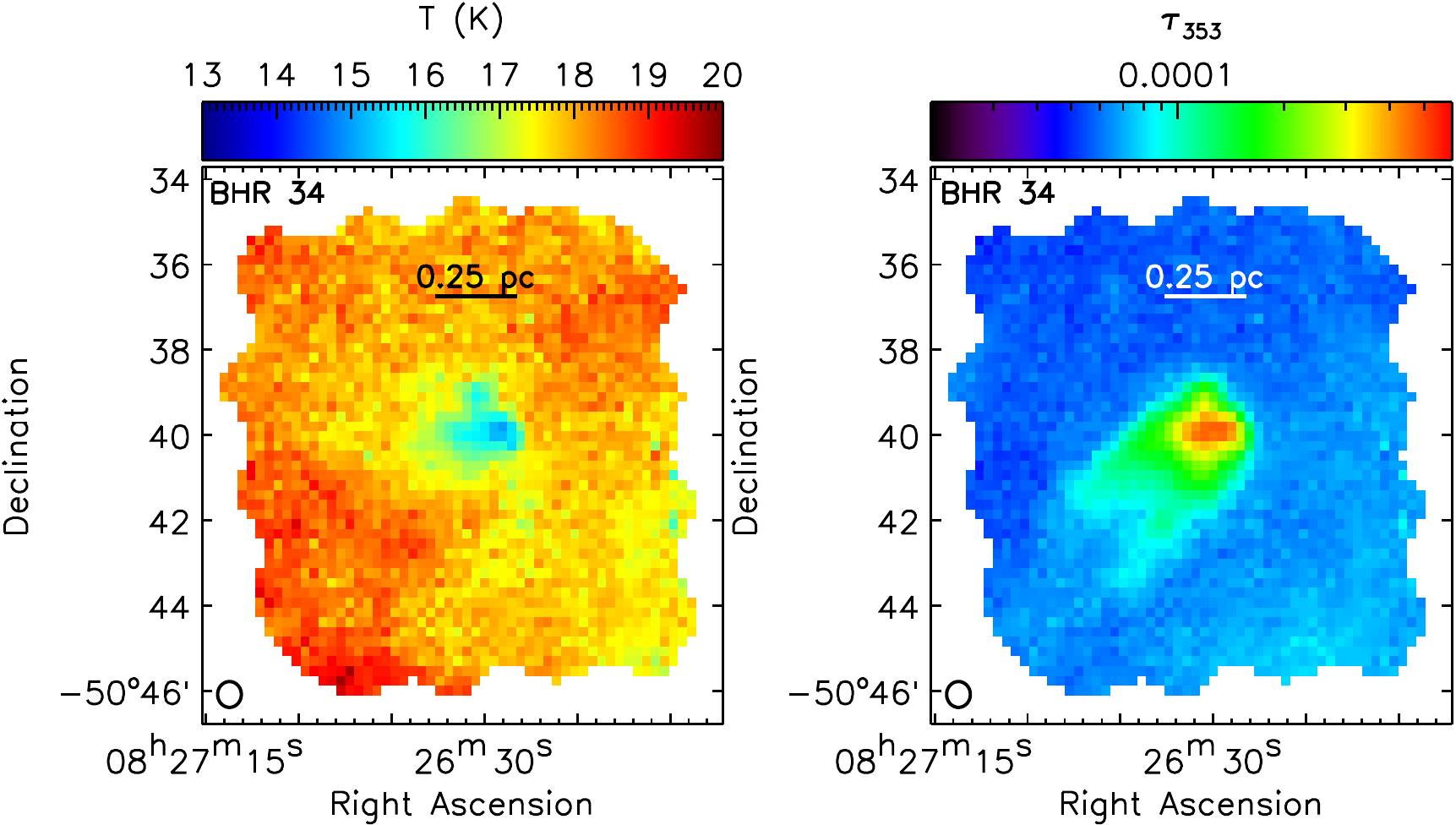}
\qquad
\includegraphics[scale=0.575]{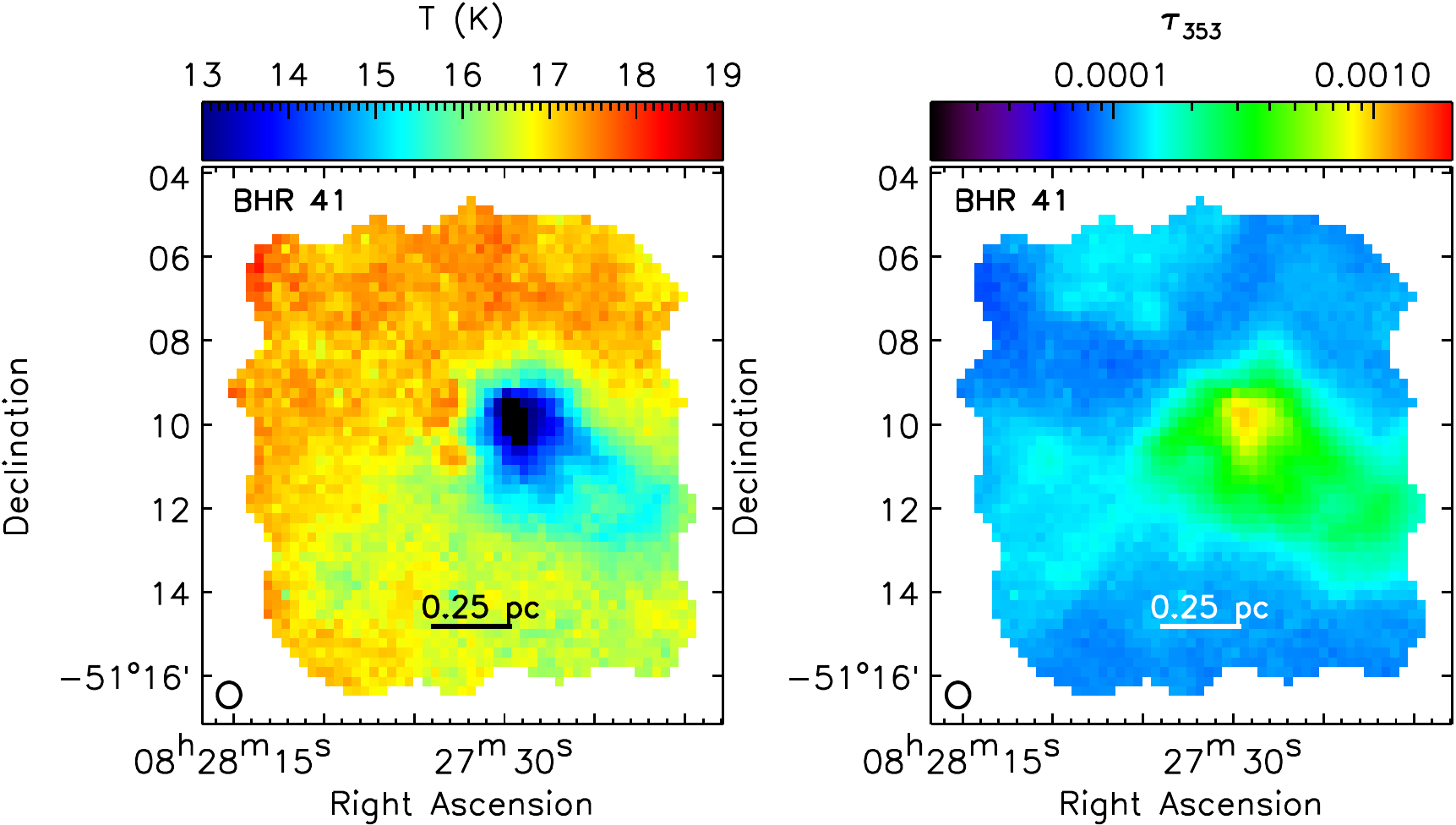}
\caption{Continued -  For BHR 31, BHR 42, BHR 34, and BHR 41}
\end{figure*}
\begin{figure*}
\ContinuedFloat
\centering
\includegraphics[scale=0.575]{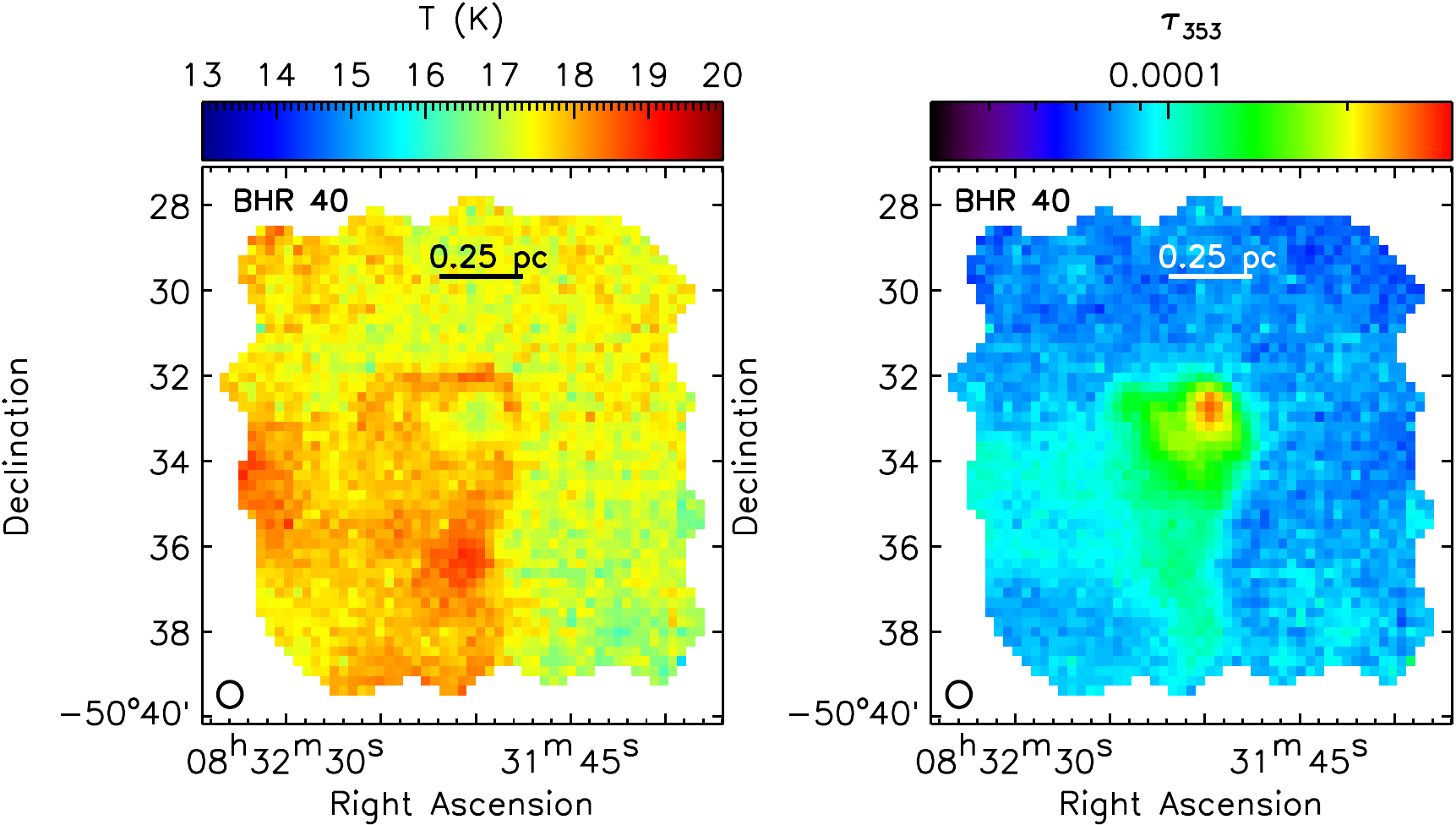}
\qquad
\includegraphics[scale=0.575]{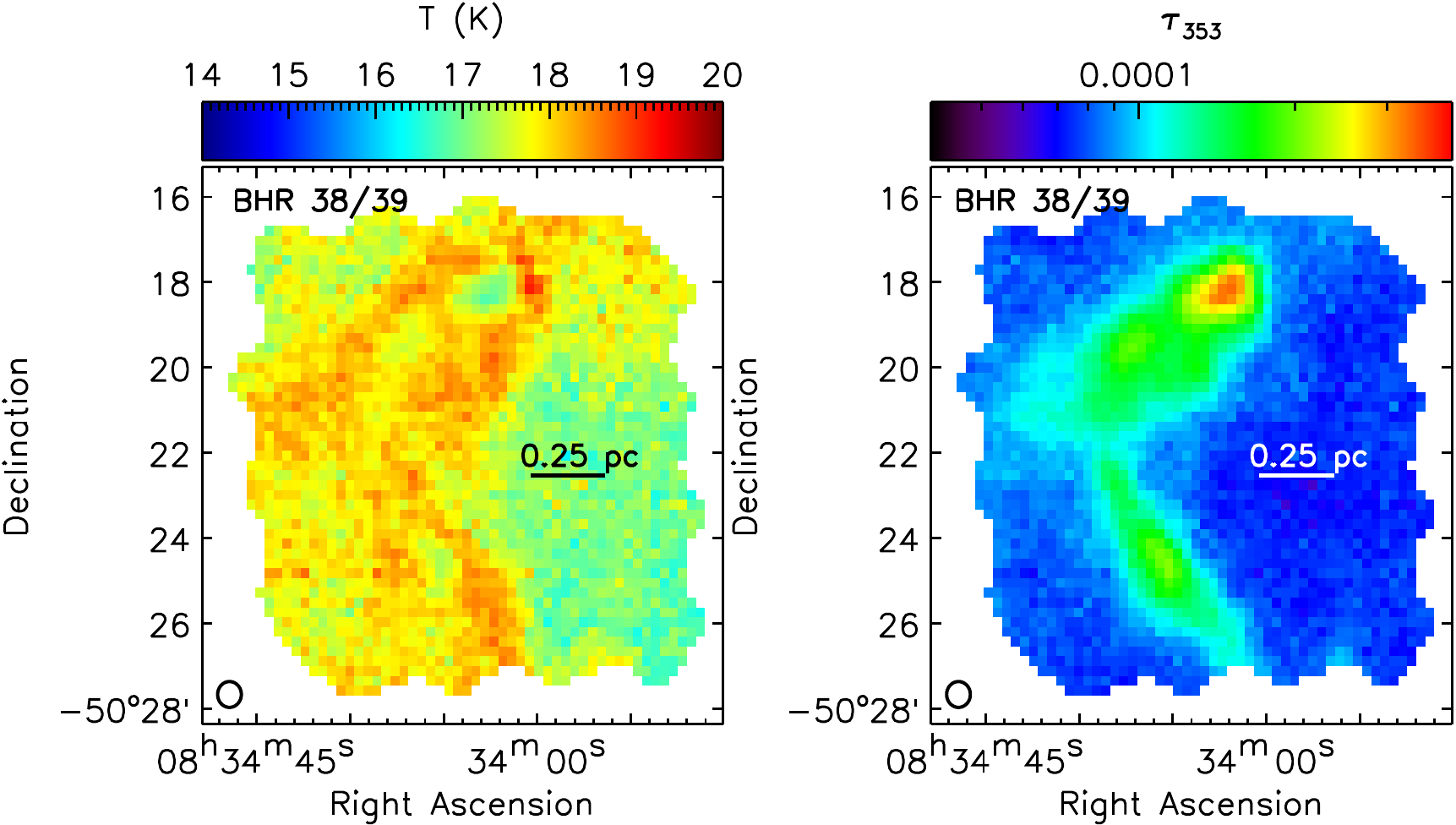}
\qquad
\includegraphics[scale=0.575]{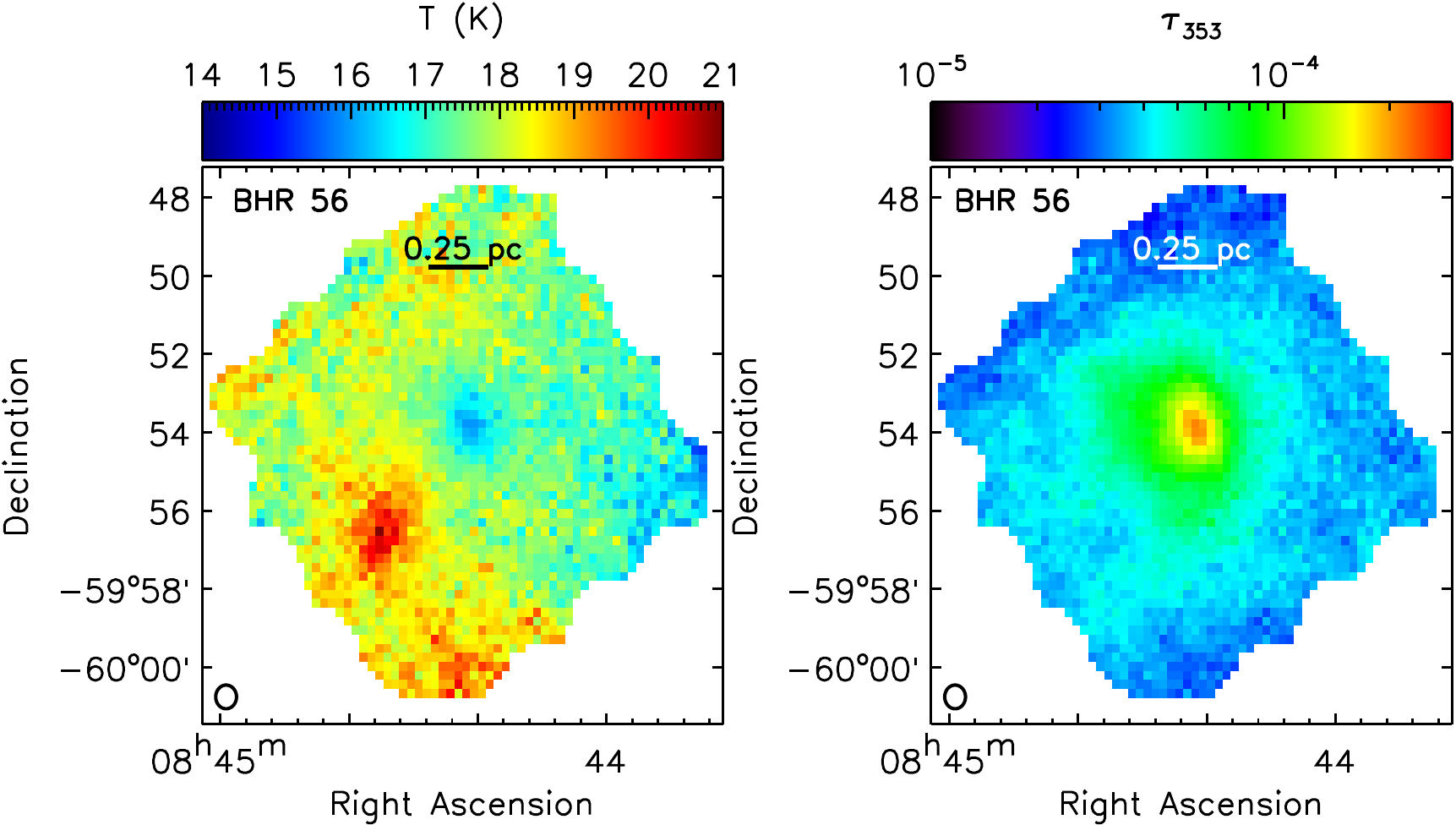}
\qquad
\includegraphics[scale=0.575]{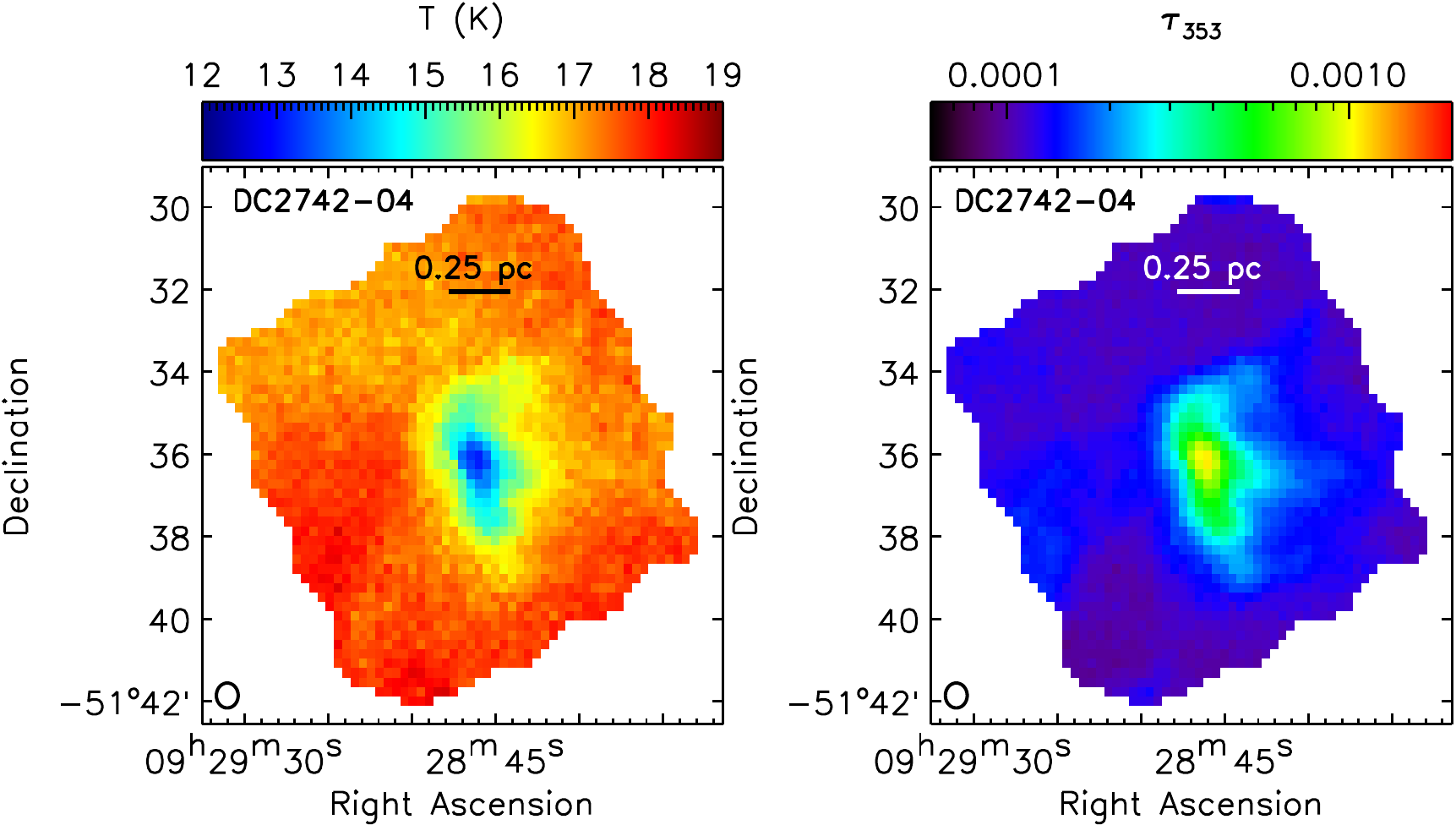}
\caption{Continued -  For BHR 40, BHR 38/39, BHR 56, and DC2742-04}
\end{figure*}
\begin{figure*}
\ContinuedFloat
\centering
\includegraphics[scale=0.575]{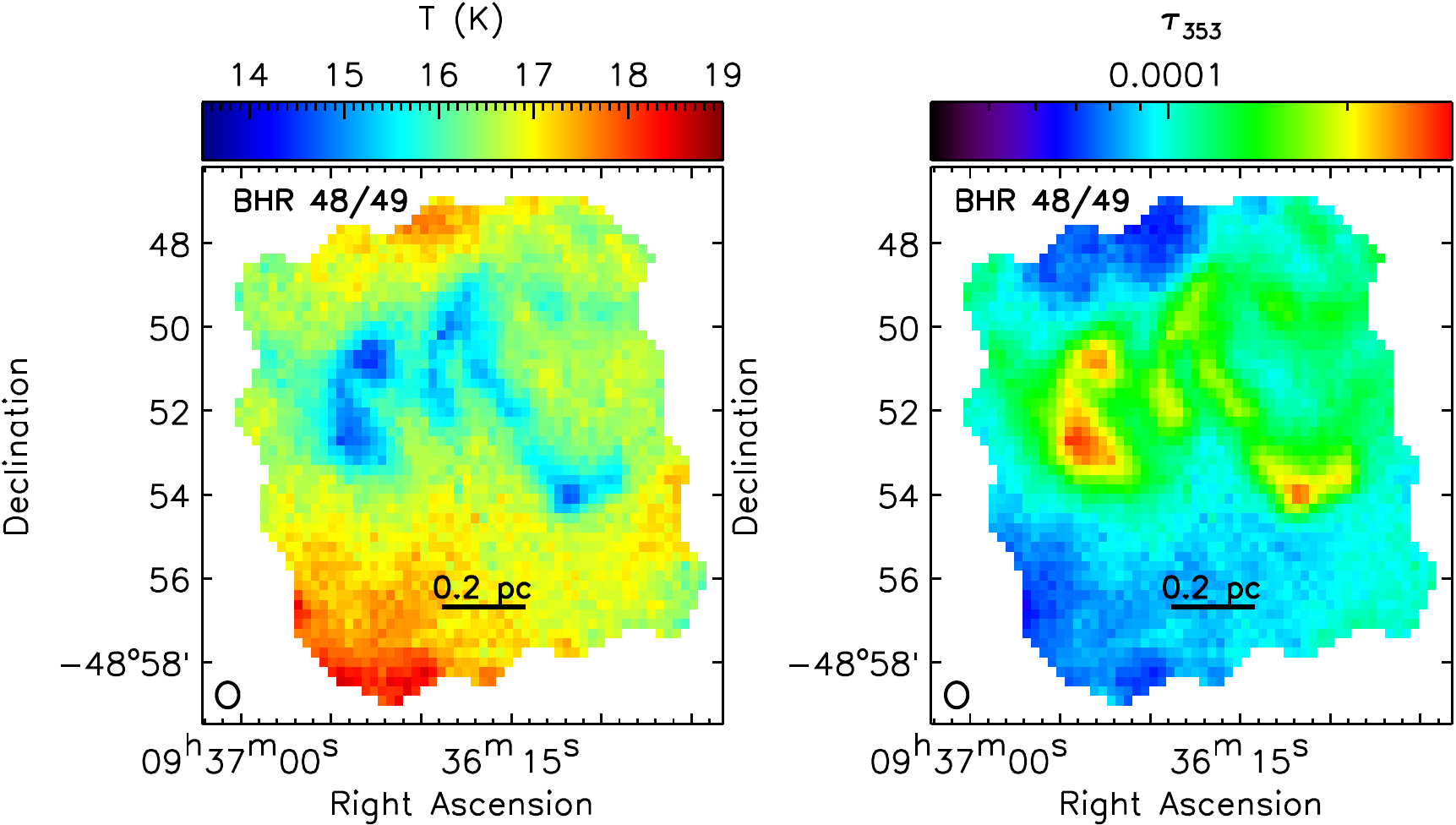}
\qquad
\includegraphics[scale=0.575]{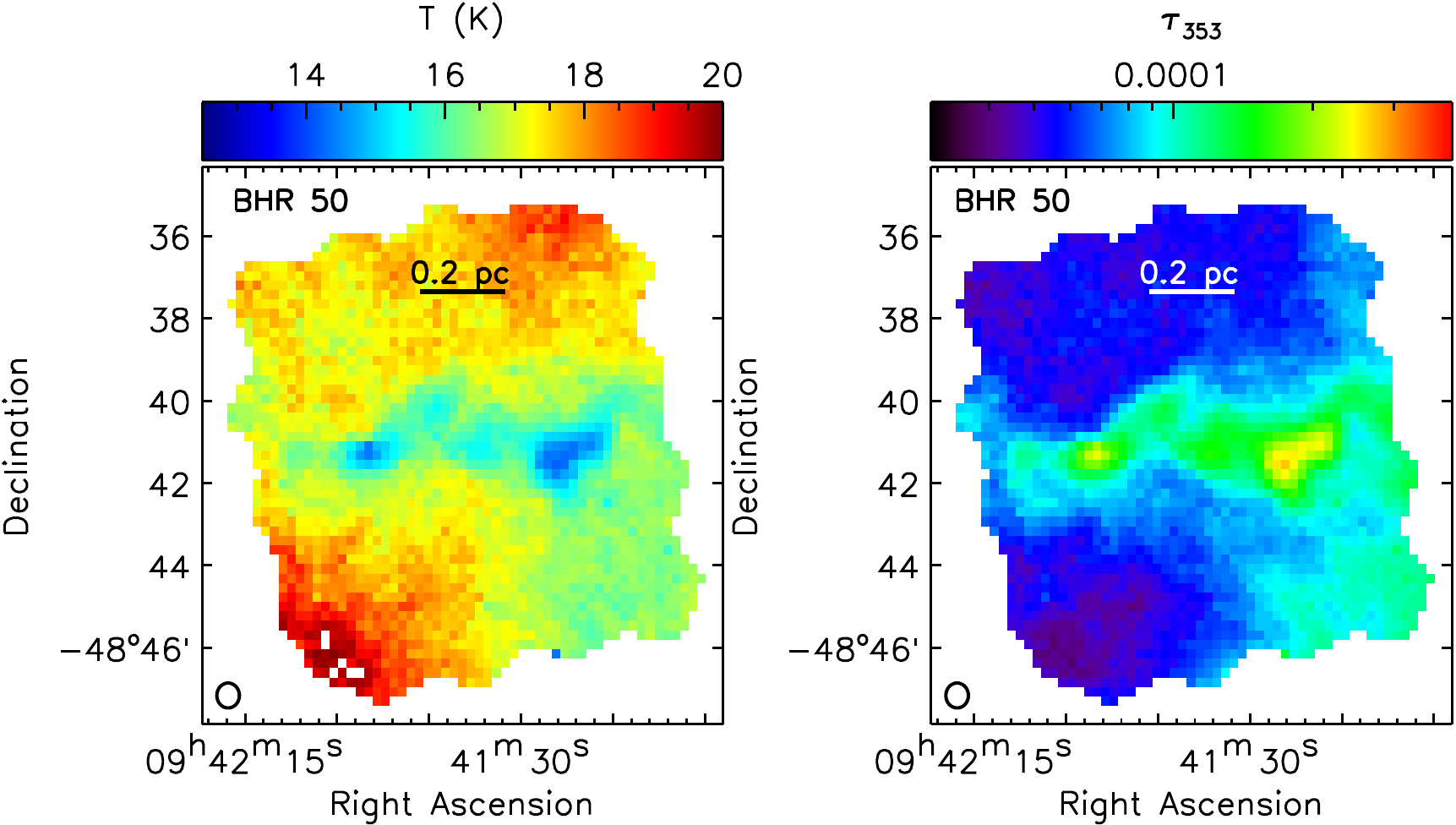}
\qquad
\includegraphics[scale=0.575]{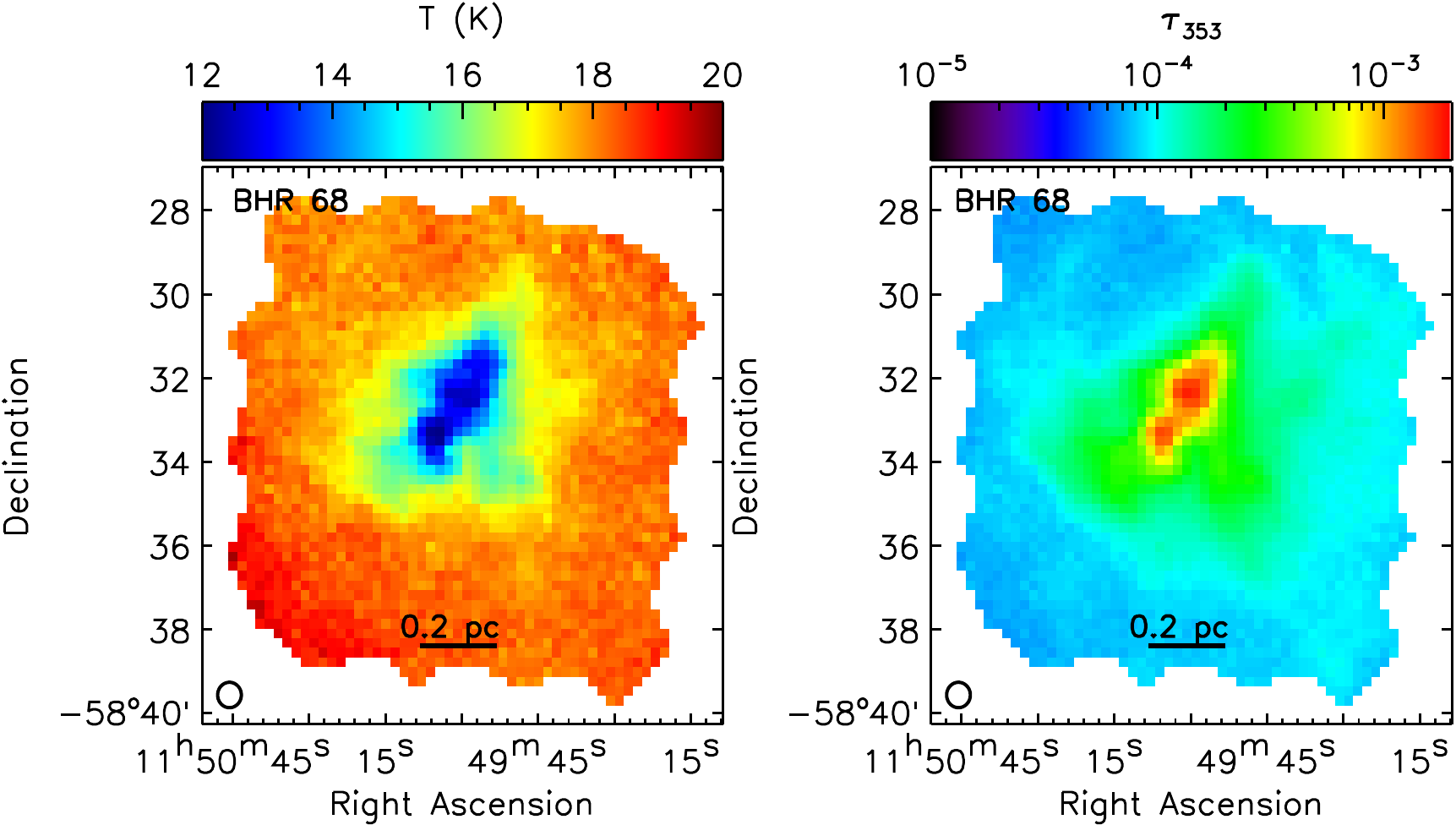}
\qquad
\includegraphics[scale=0.575]{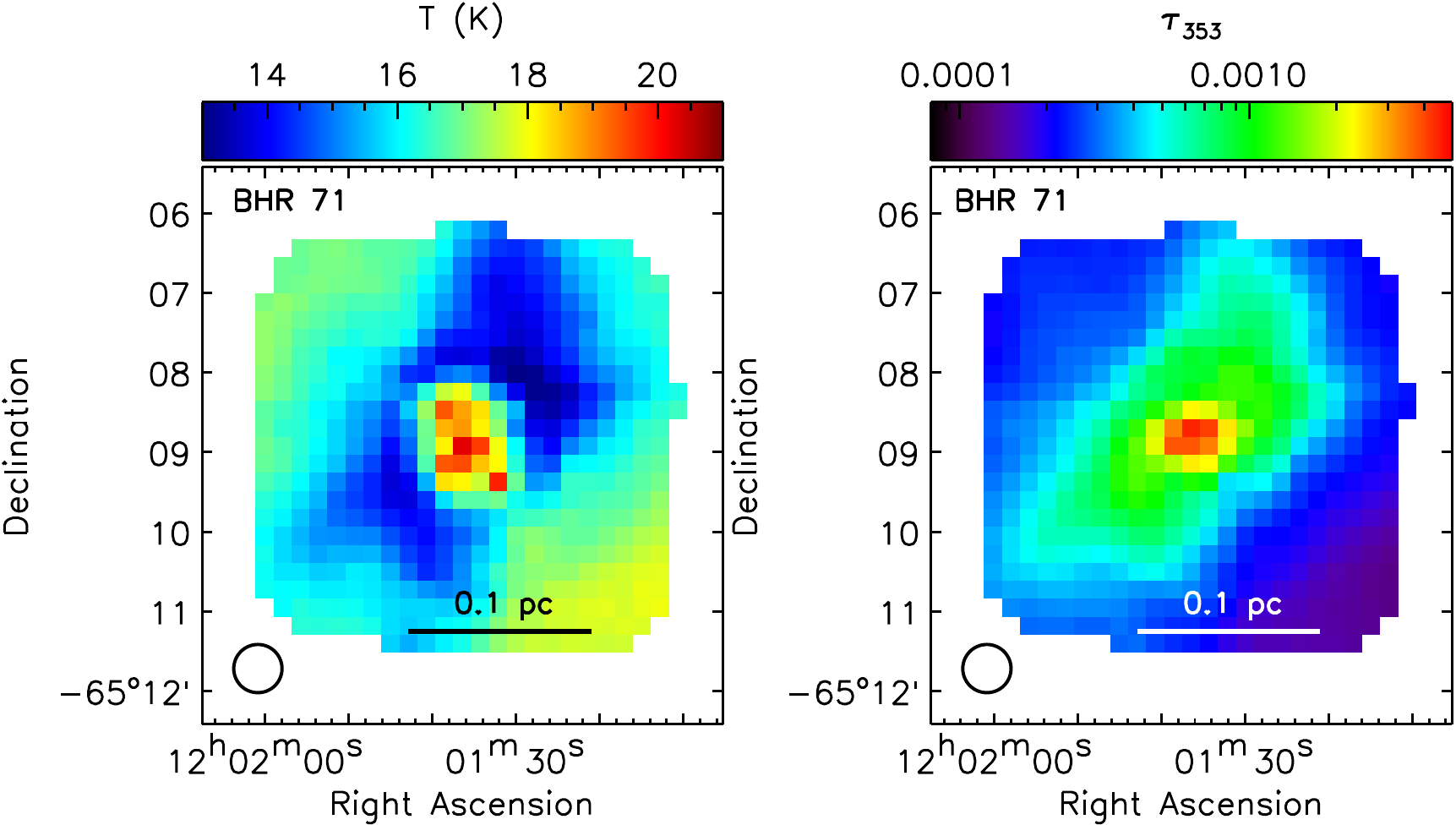}
\caption{Continued -  For BHR 48/49, BHR 50, BHR 68, and BHR 71}
\end{figure*}
\begin{figure*}
\ContinuedFloat
\centering
\includegraphics[scale=0.575]{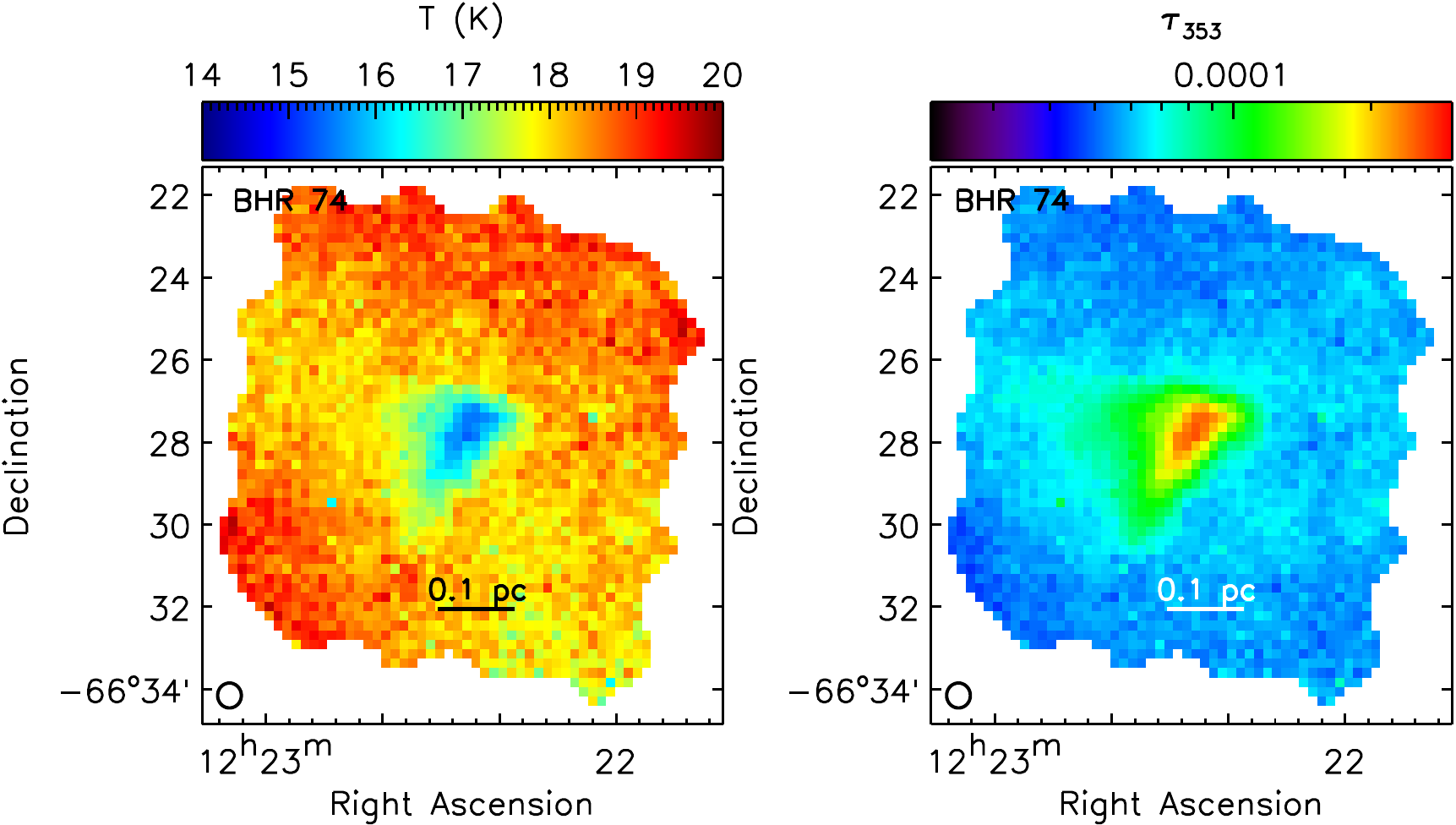}
\qquad
\includegraphics[scale=0.575]{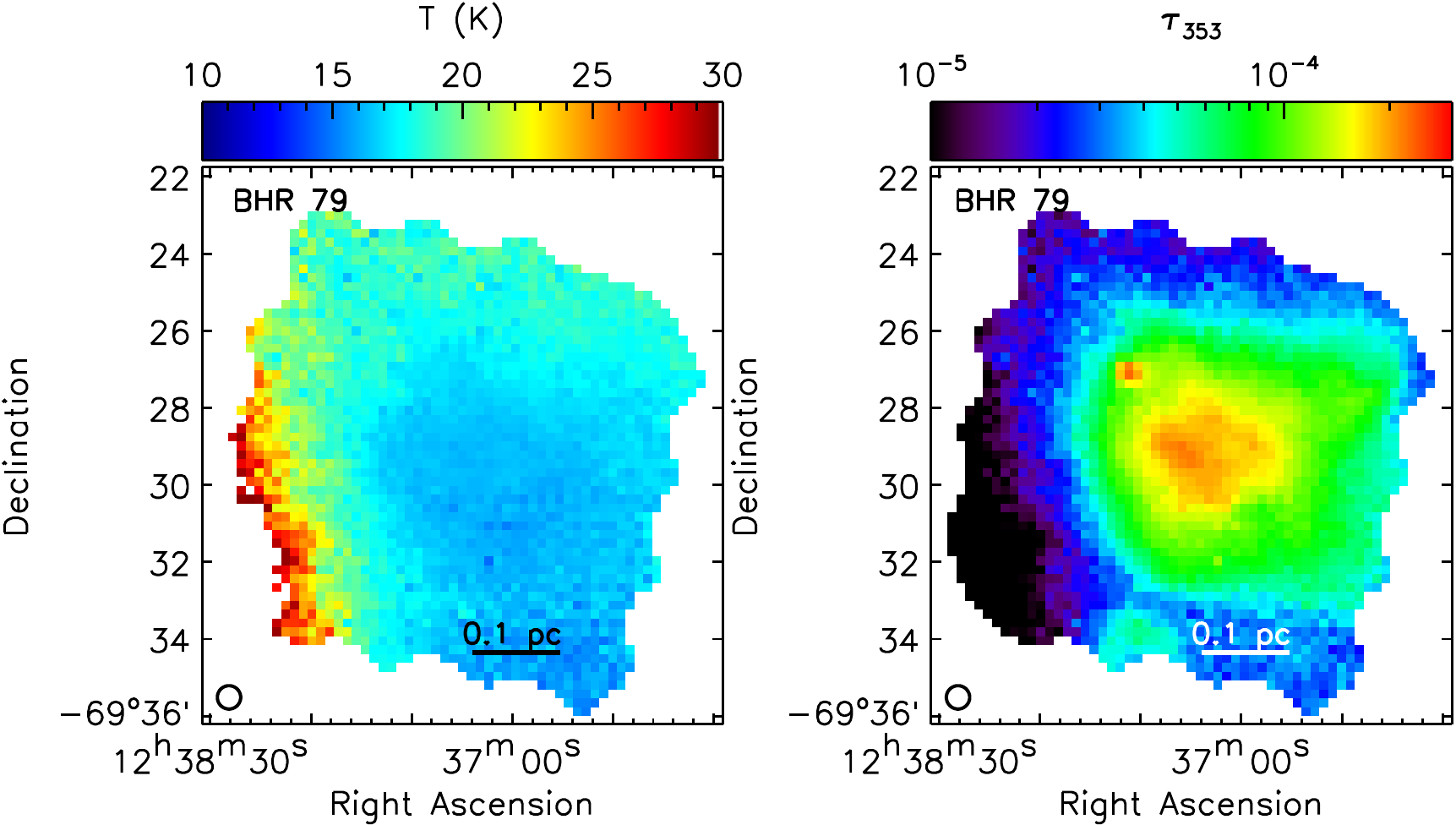}
\qquad
\includegraphics[scale=0.575]{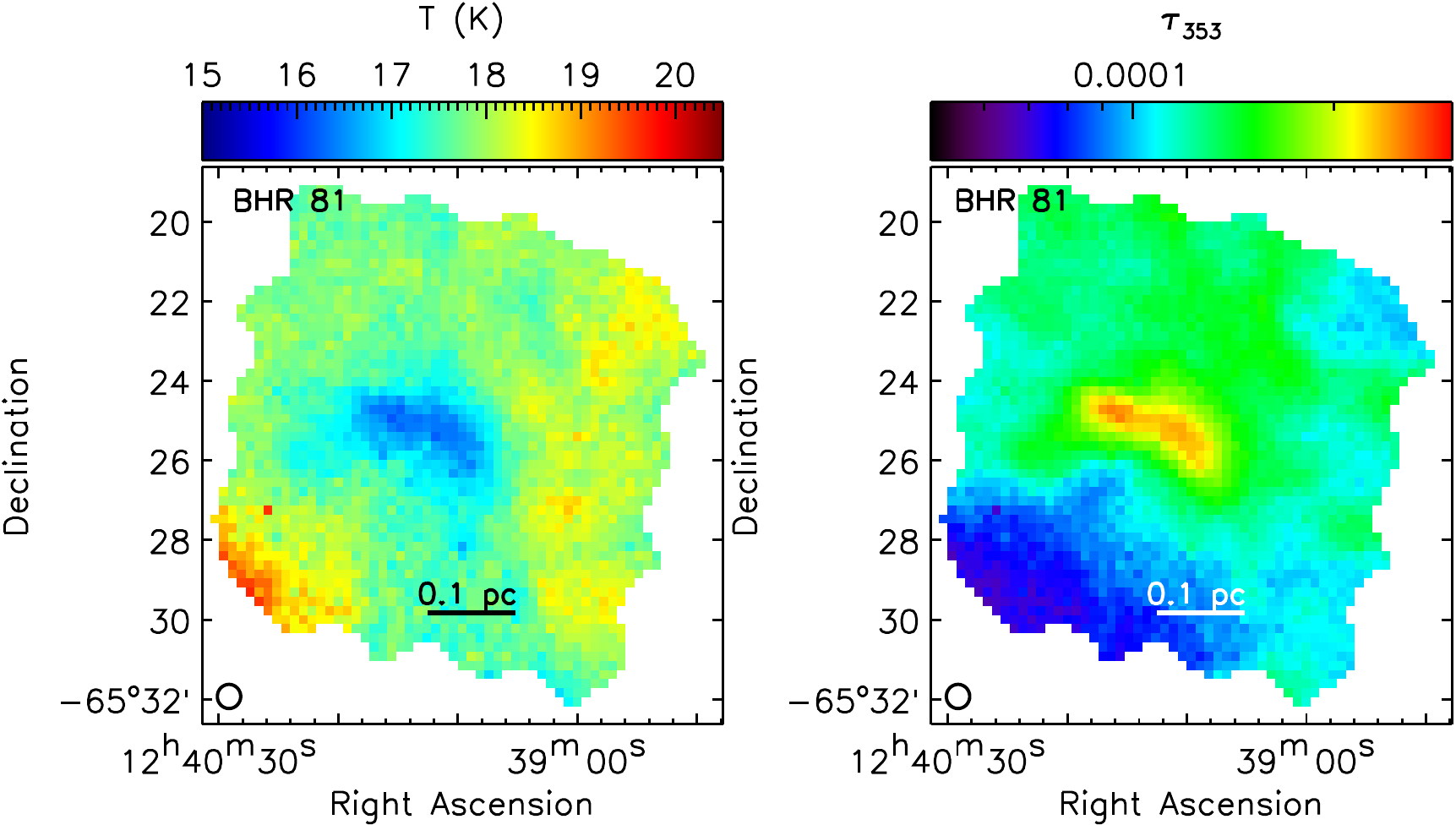}
\qquad
\includegraphics[scale=0.575]{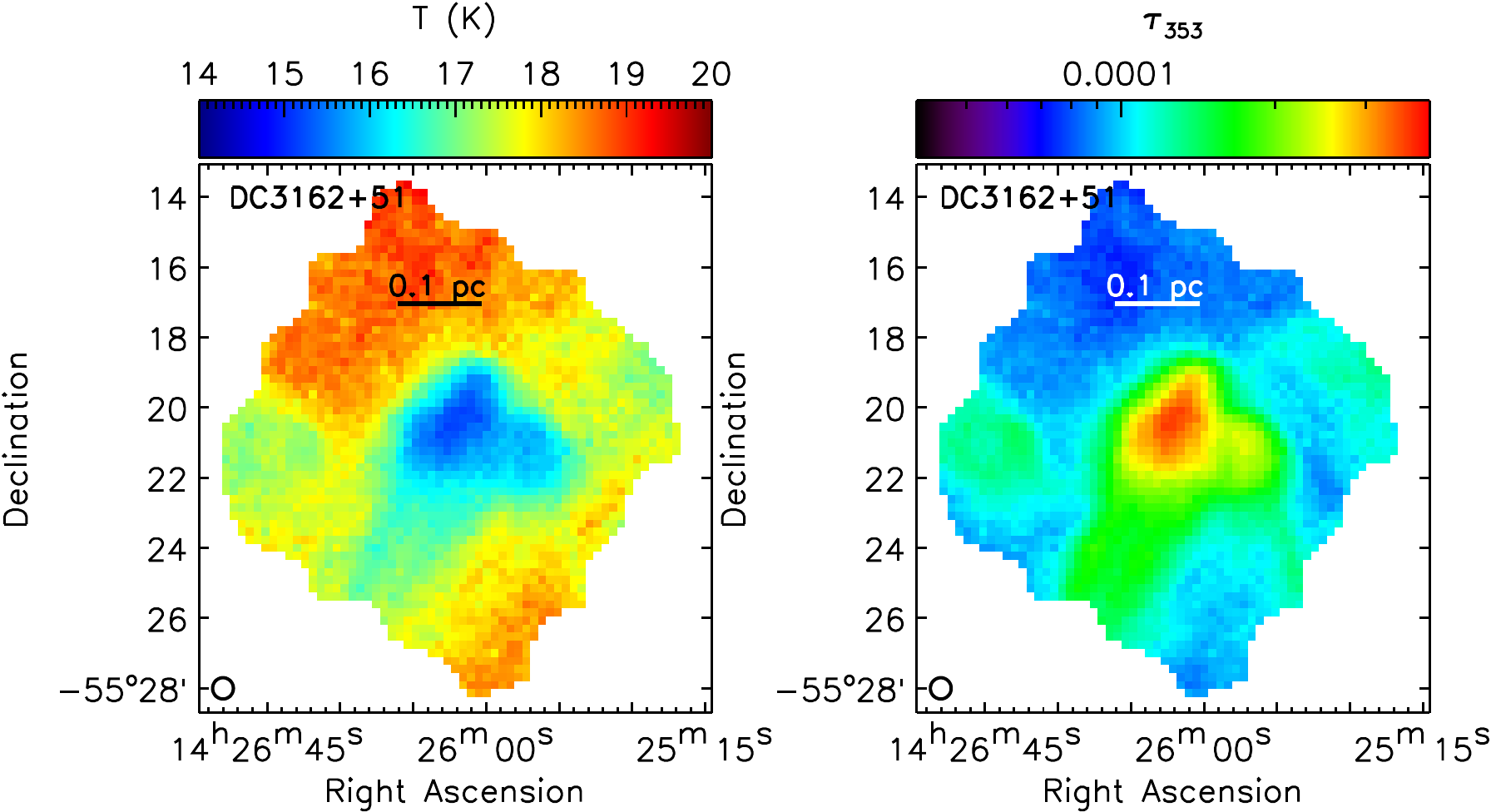}
\caption{Continued -  For BHR 74, BHR 79, BHR 81, DC3162+51}
\end{figure*}
\begin{figure*}
\ContinuedFloat
\centering
\includegraphics[scale=0.575]{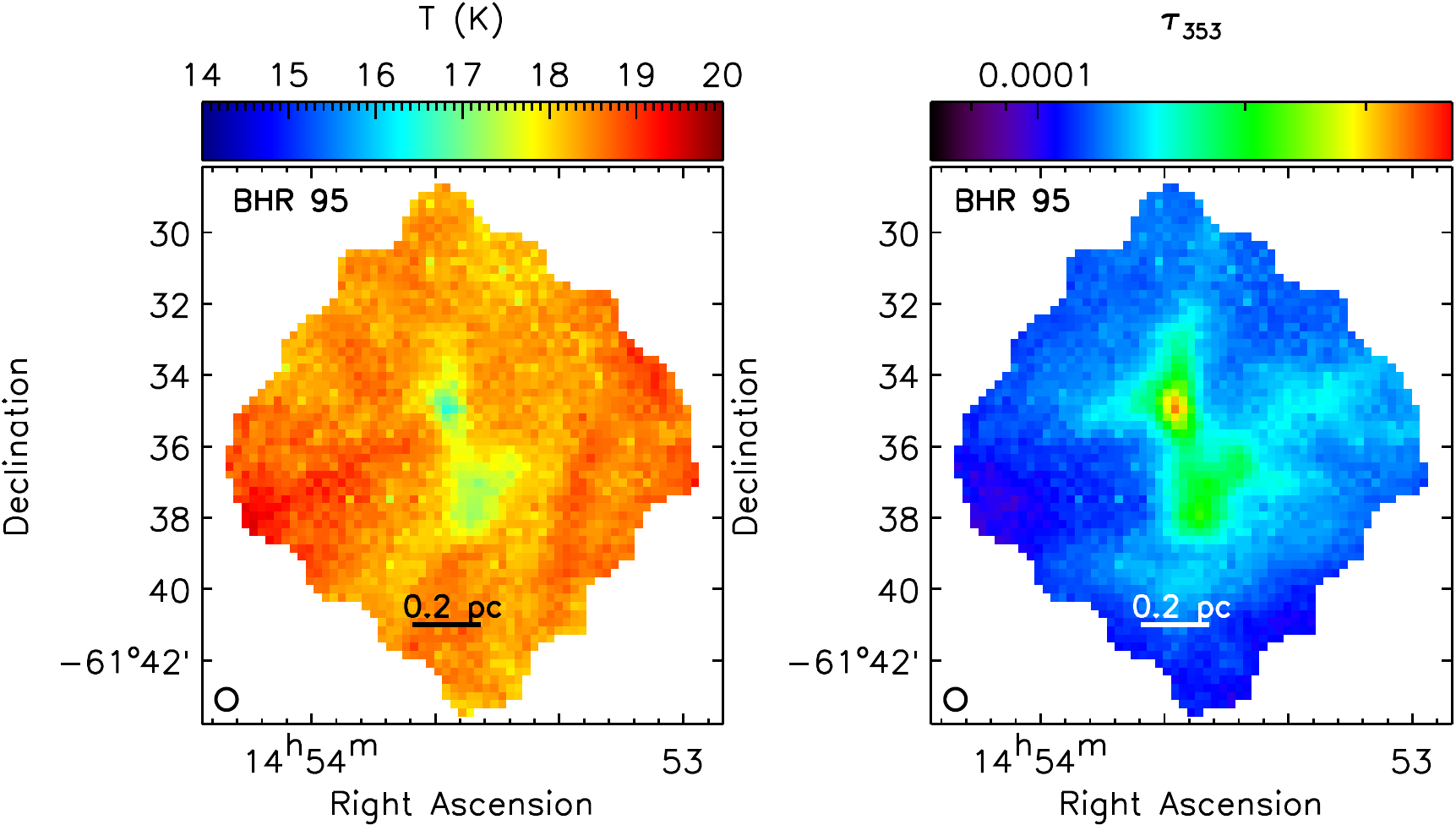}
\qquad
\includegraphics[scale=0.575]{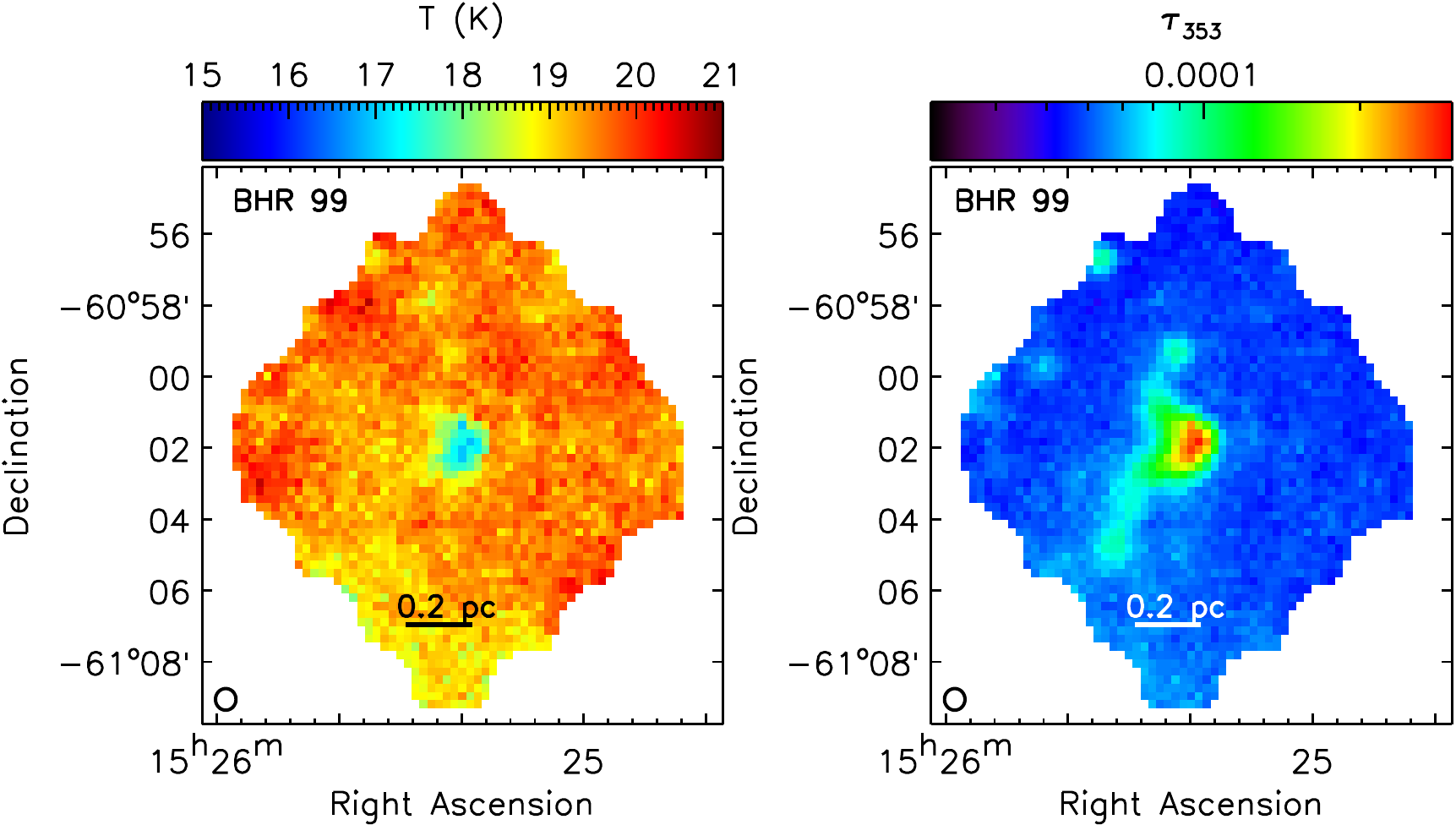}
\qquad
\includegraphics[scale=0.575]{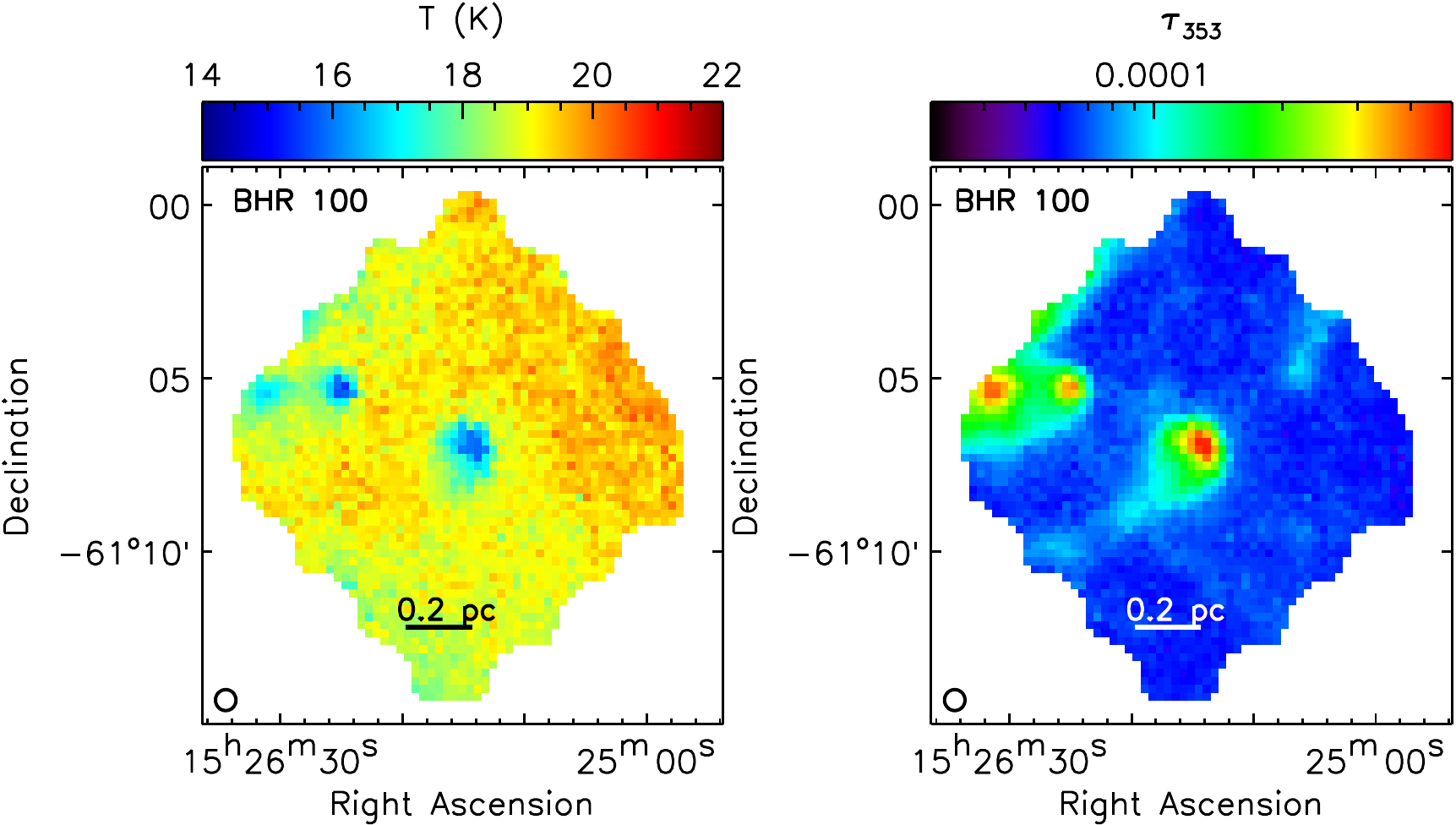}
\qquad
\includegraphics[scale=0.575]{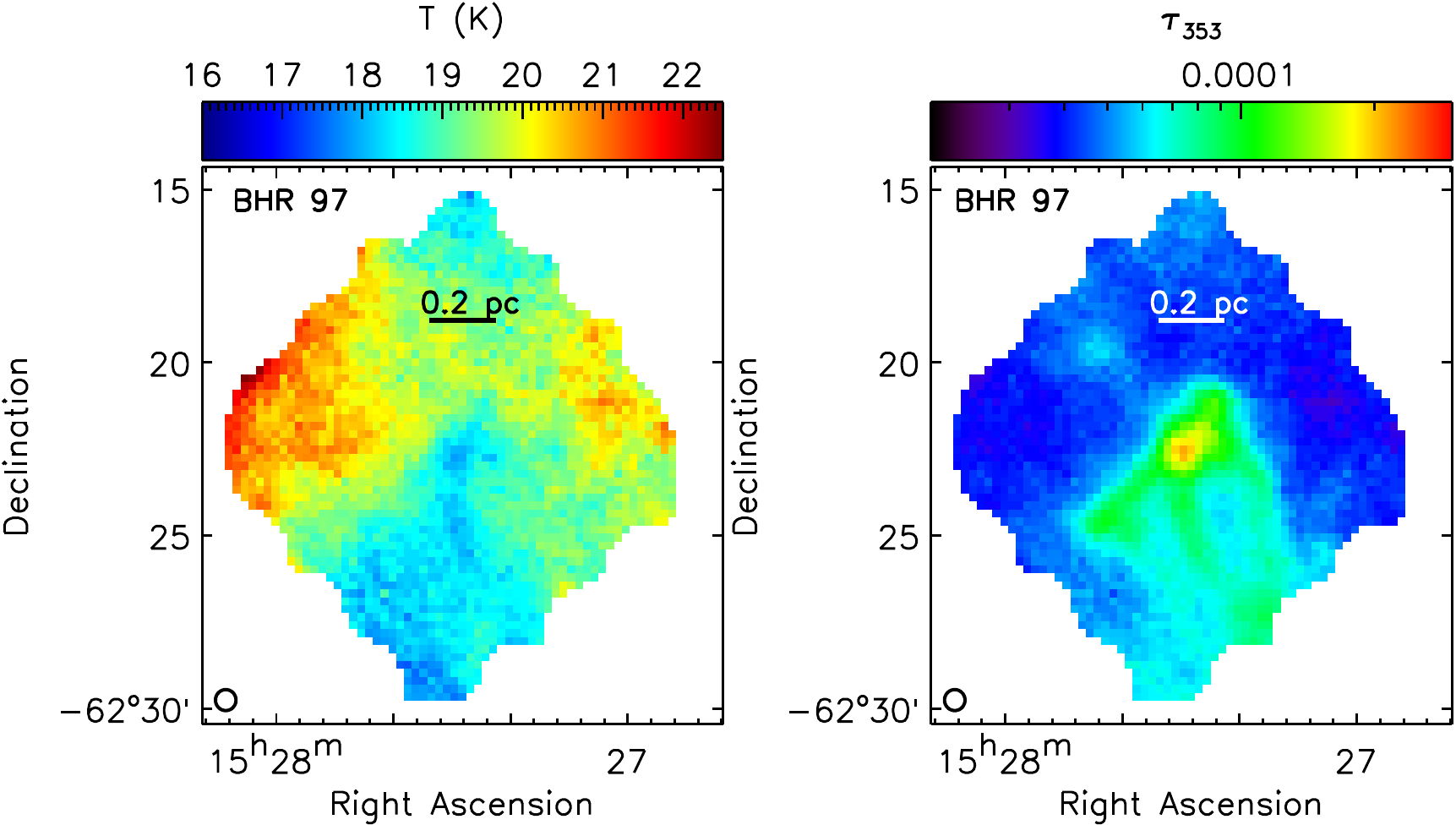}
\caption{Continued -  For BHR 95, BHR 99, BHR 100, and BHR 97}
\end{figure*}
\begin{figure*}
\ContinuedFloat
\centering
\includegraphics[scale=0.575]{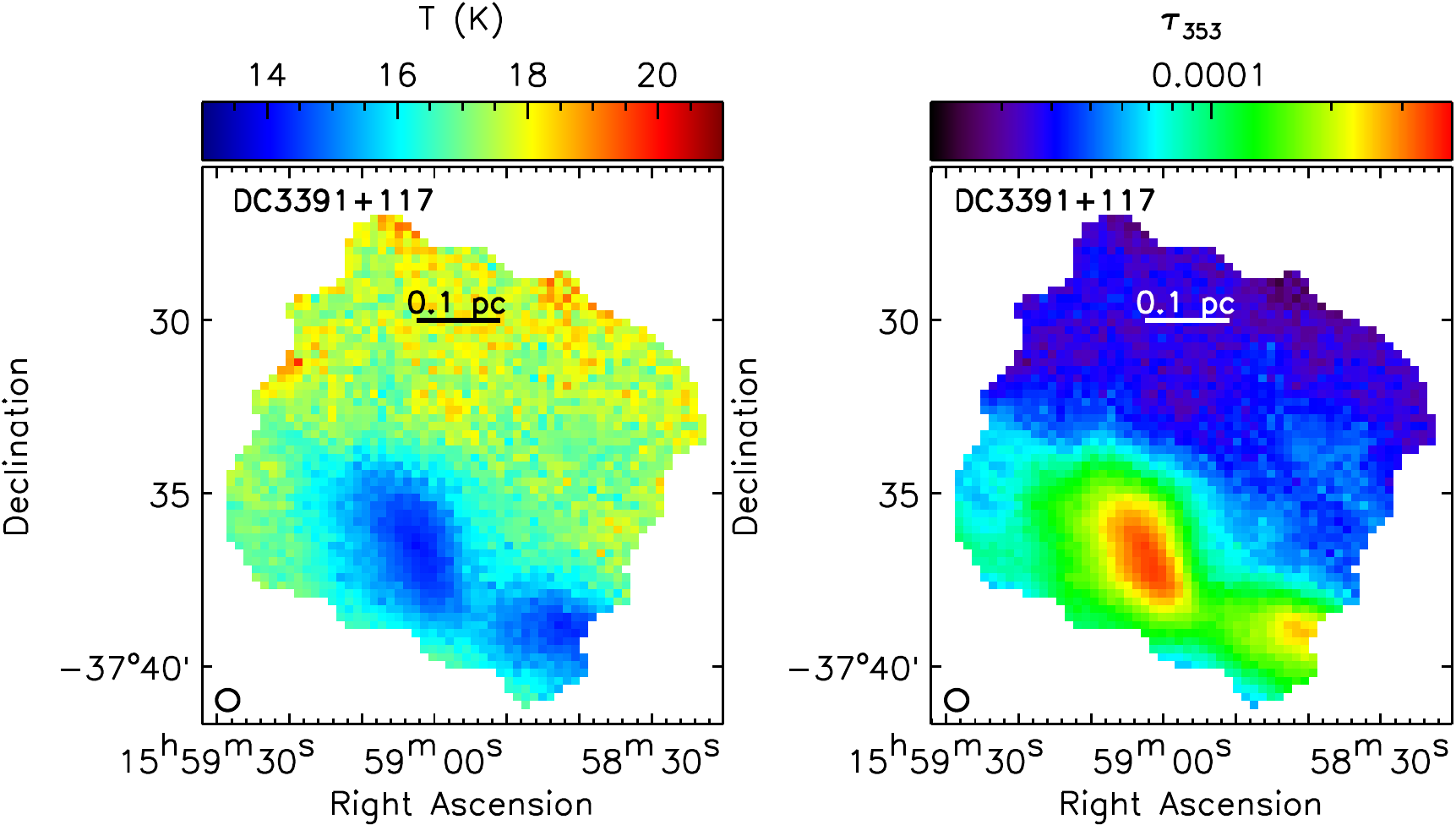}
\qquad
\includegraphics[scale=0.575]{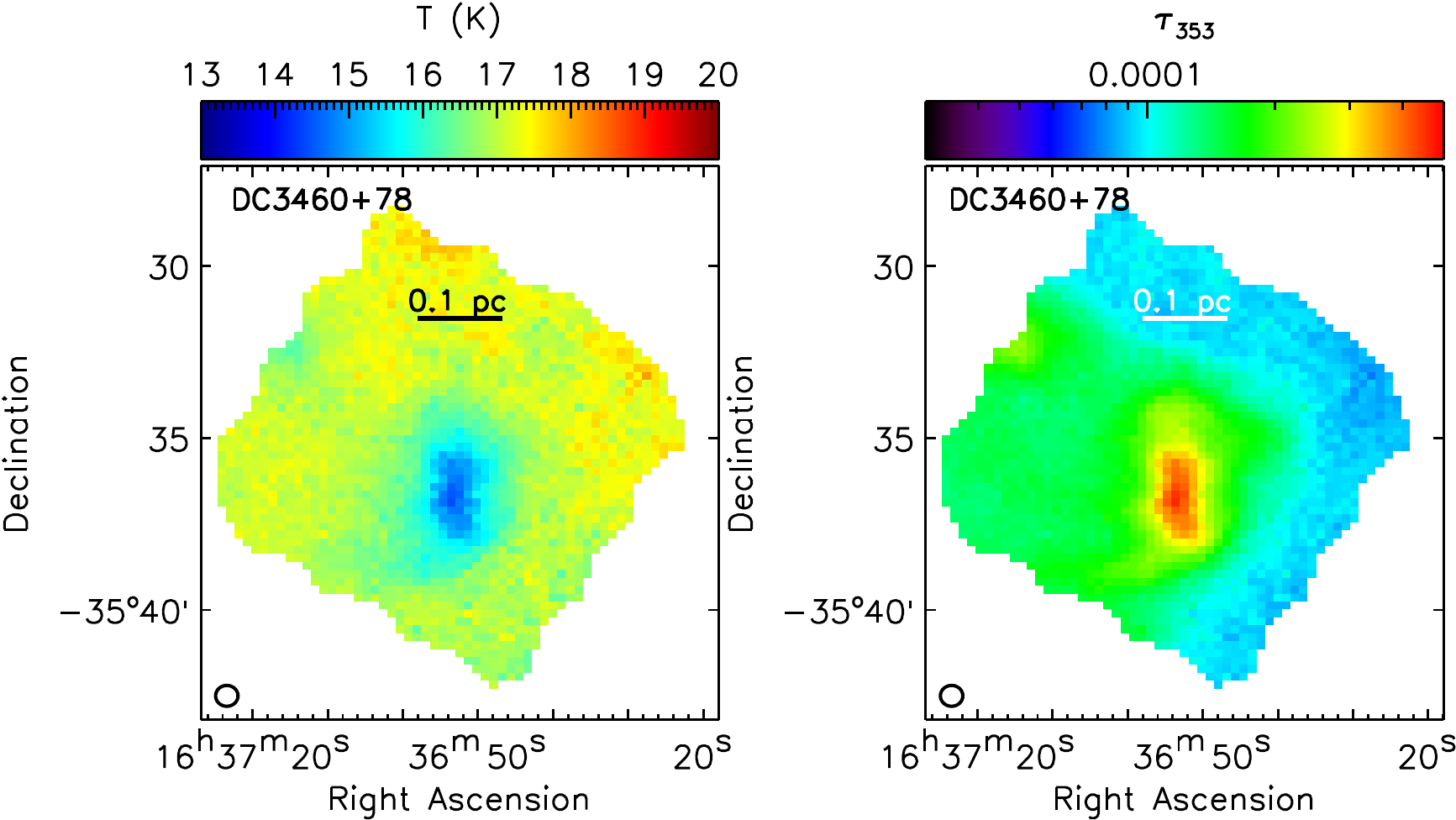}
\qquad
\includegraphics[scale=0.575]{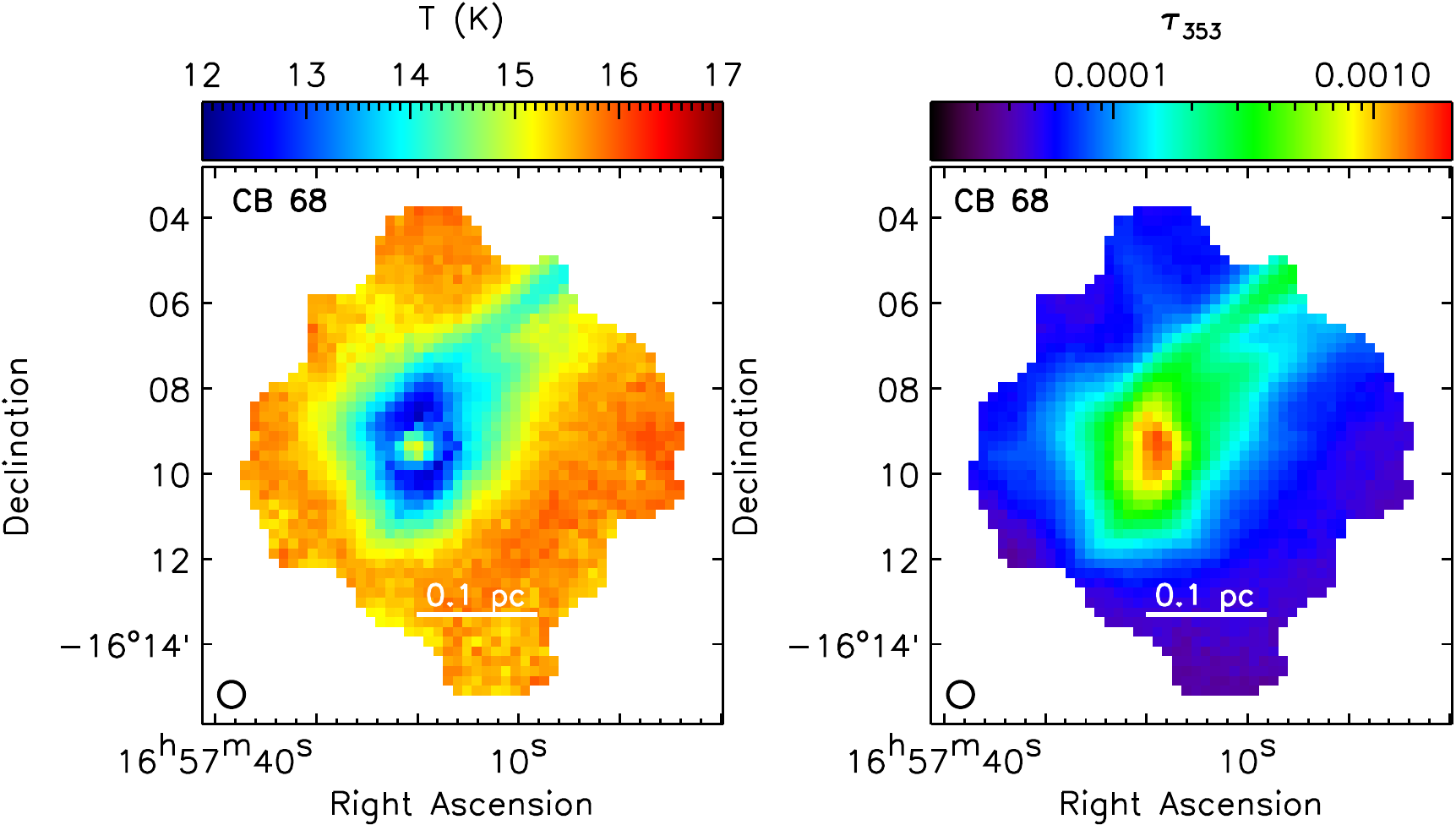}
\qquad
\includegraphics[scale=0.575]{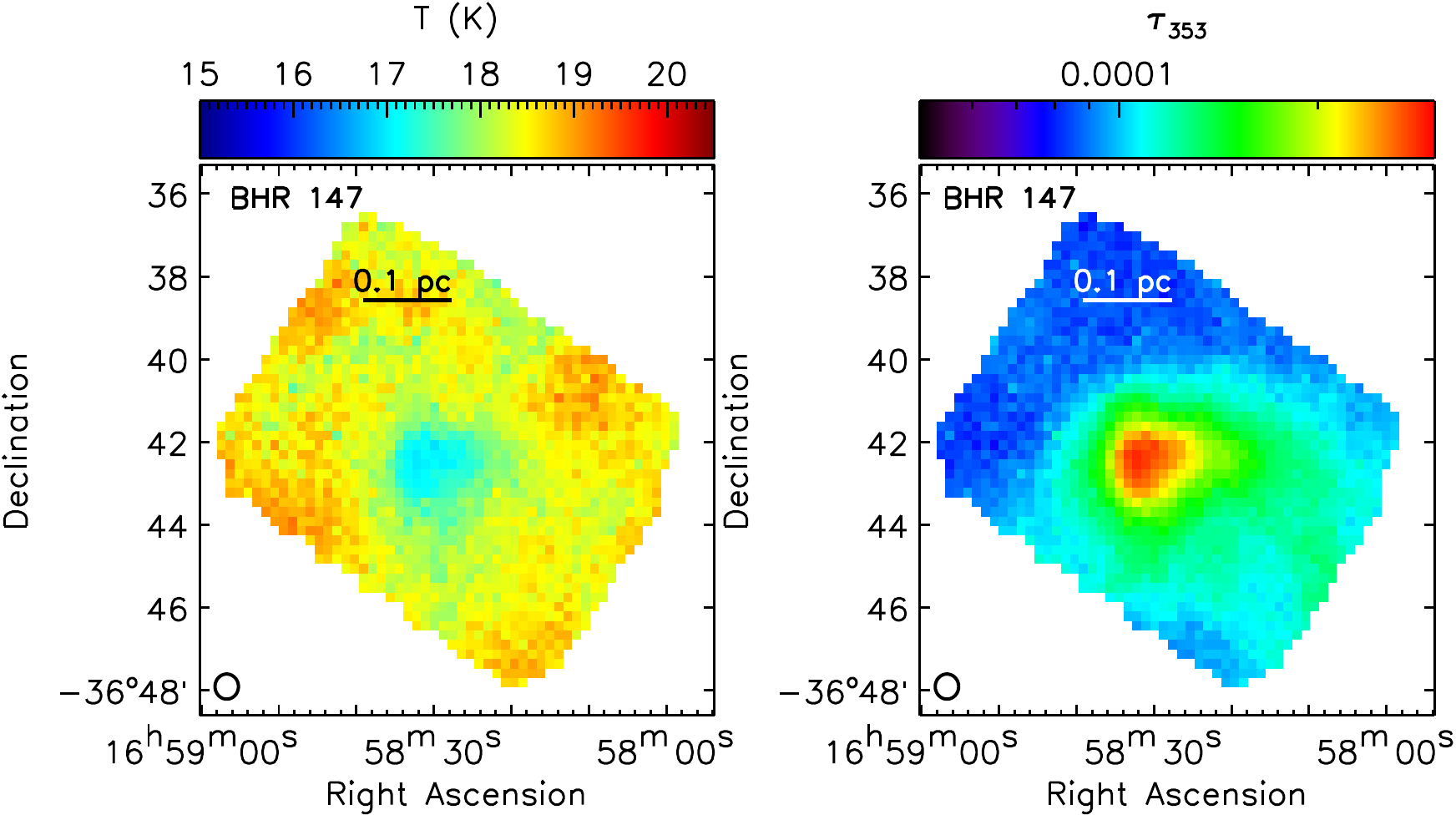}
\caption{Continued -  For DC3391+117, DC3460+78, CB 68, BHR 147}
\end{figure*}
\begin{figure*}
\ContinuedFloat
\centering
\includegraphics[scale=0.575]{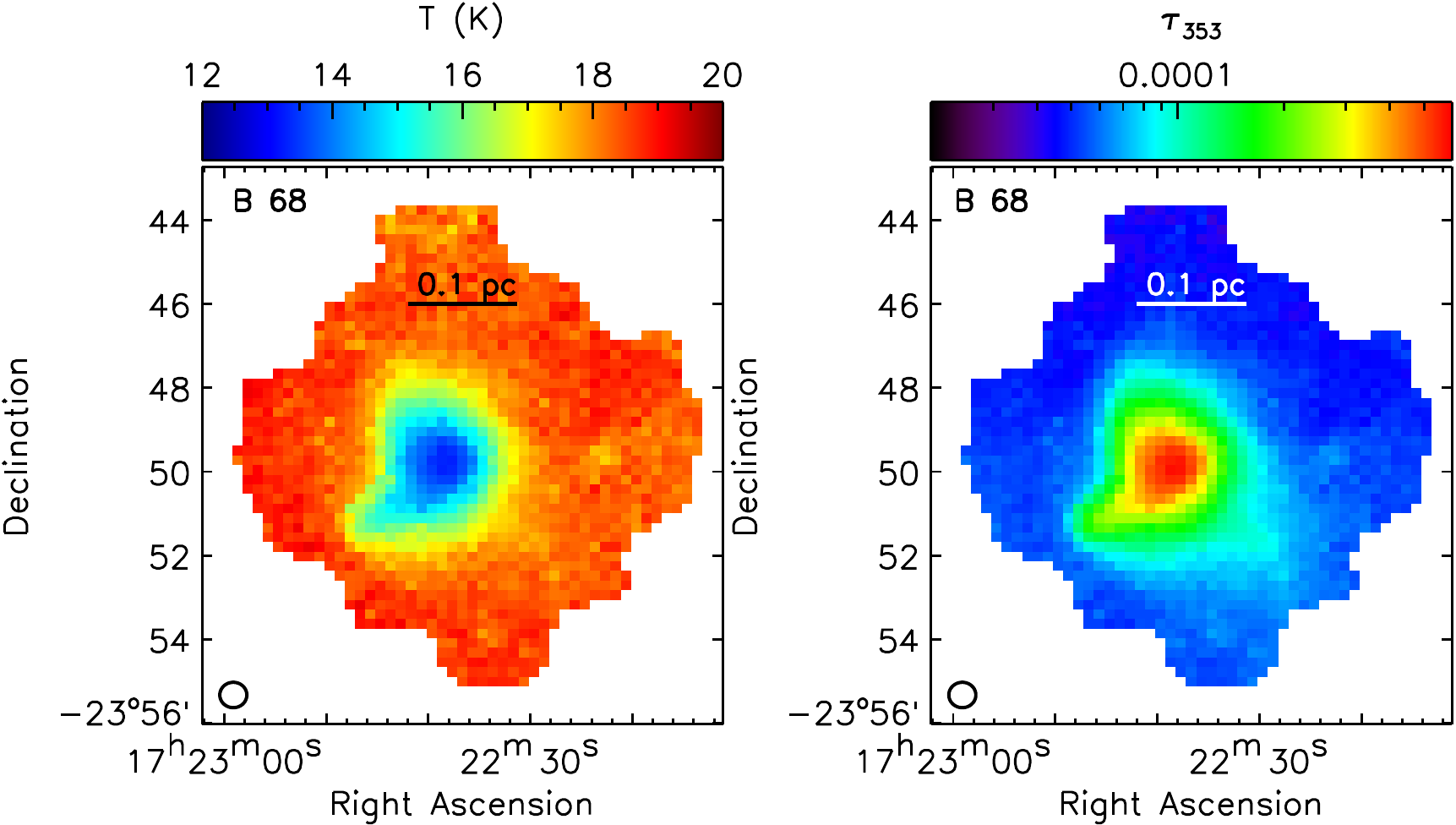}
\qquad
\includegraphics[scale=0.575]{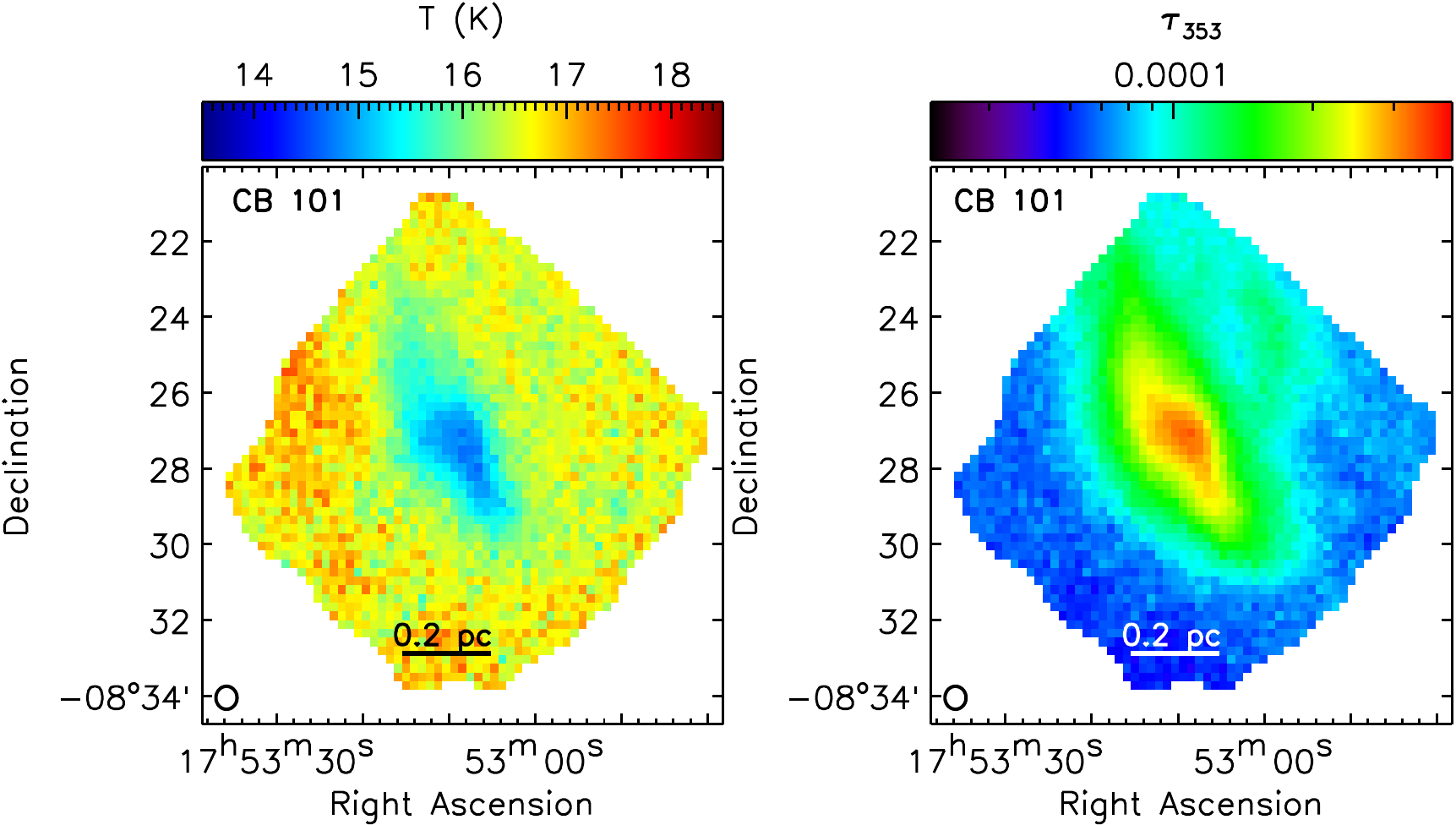}
\qquad
\includegraphics[scale=0.575]{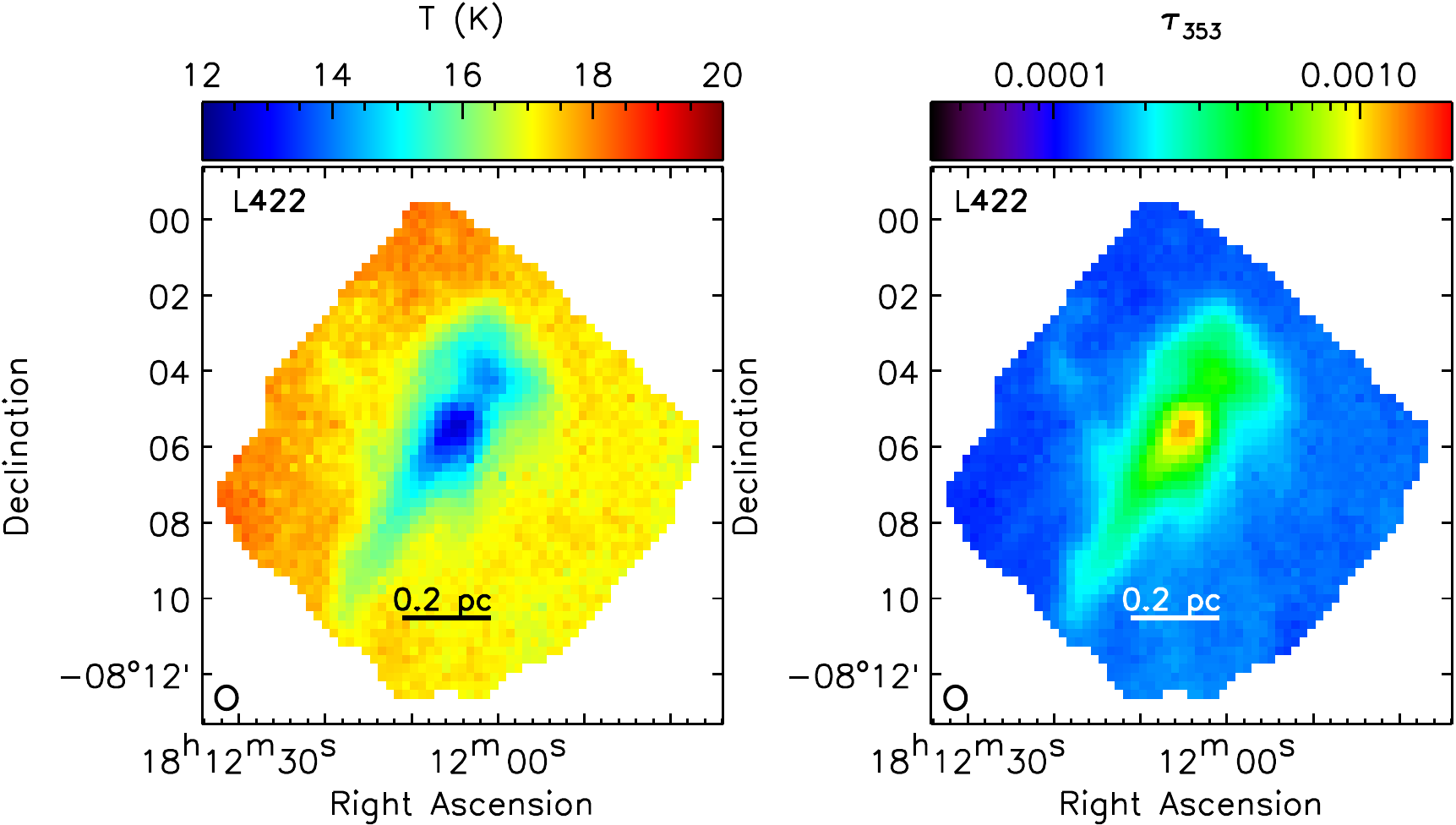}
\qquad
\includegraphics[scale=0.575]{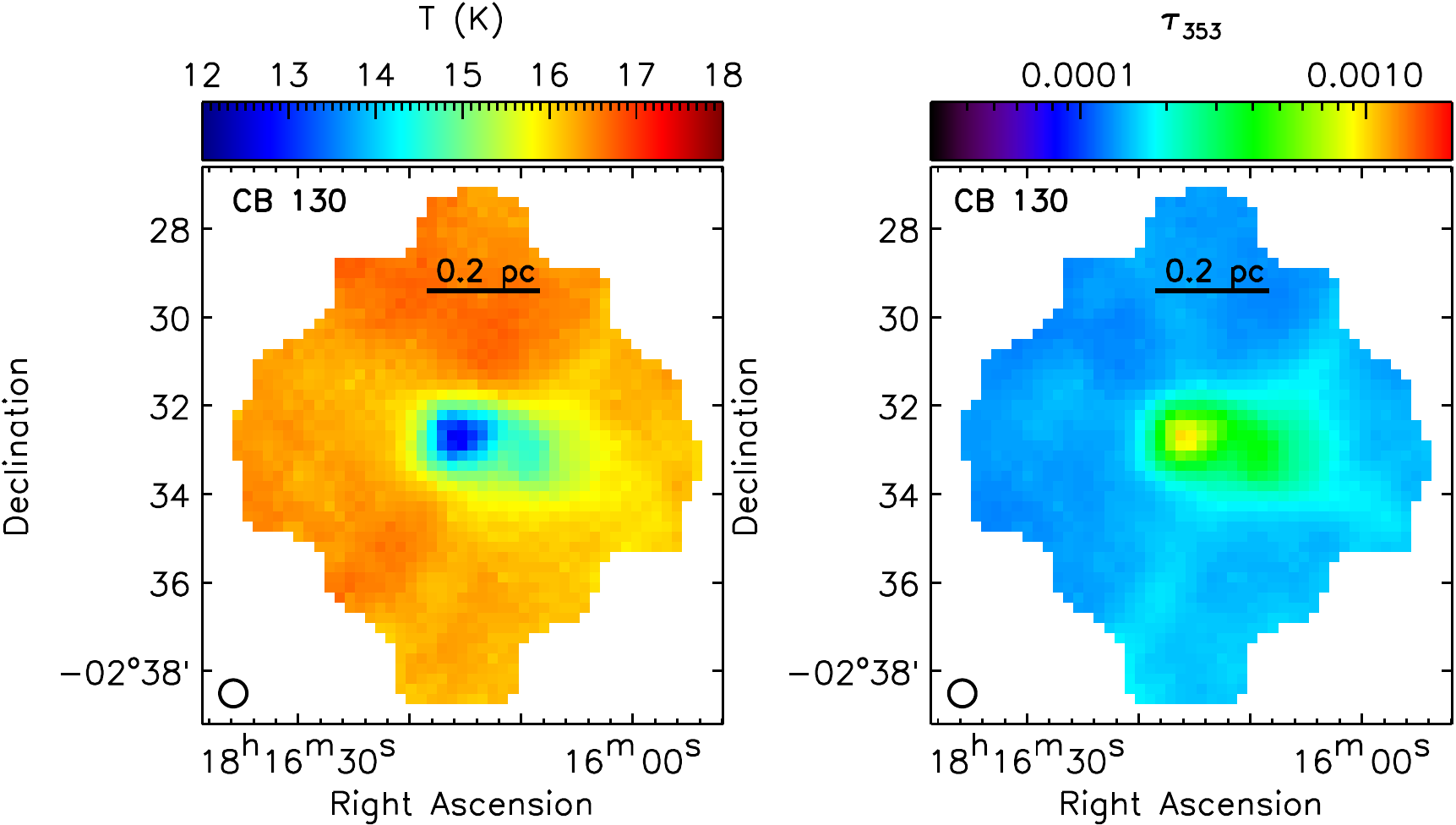}
\caption{Continued -  For B 68, CB 101, L422, and CB 130}
\end{figure*}
\begin{figure*}
\ContinuedFloat
\centering
\includegraphics[scale=0.575]{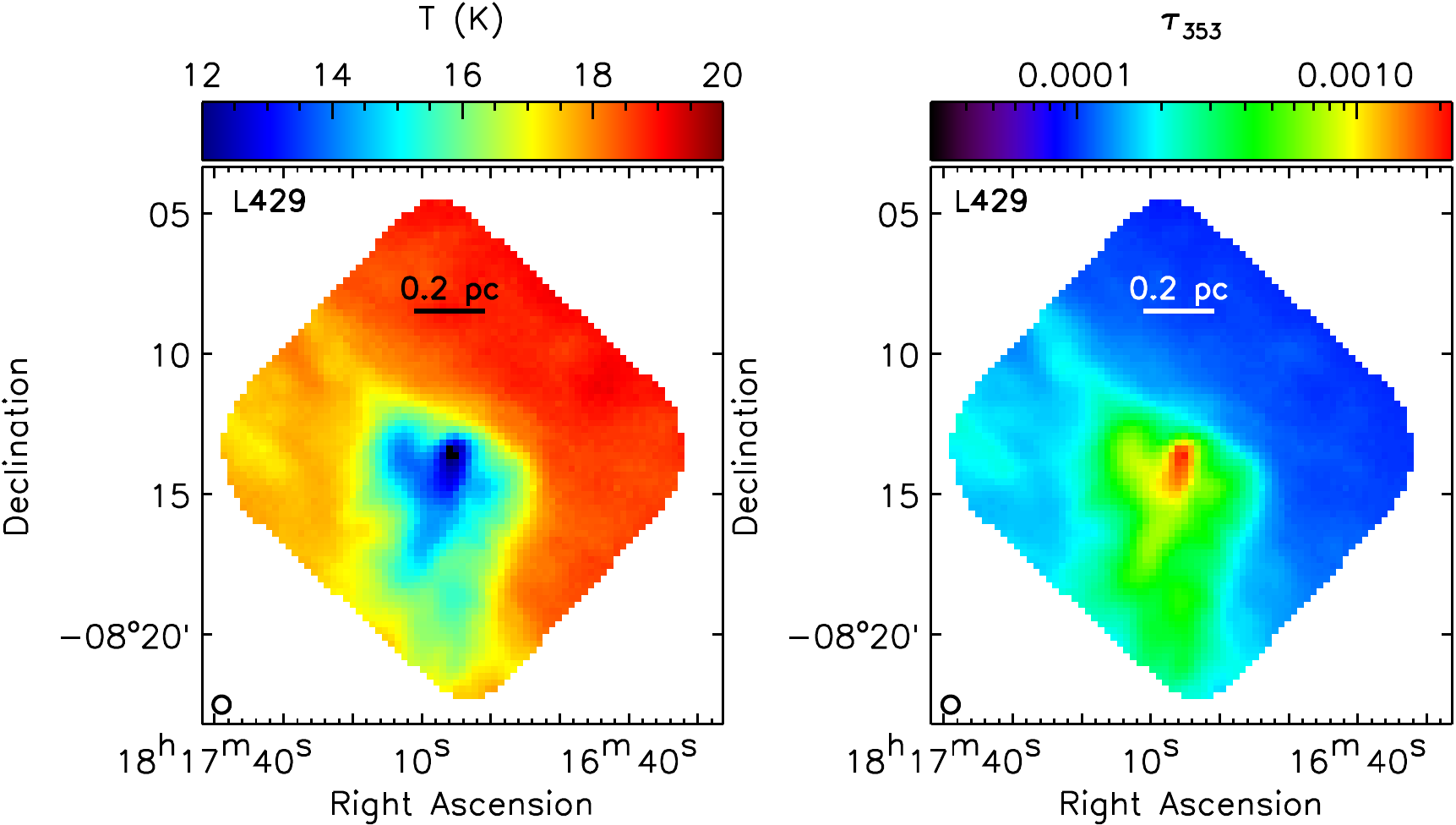}
\qquad
\includegraphics[scale=0.575]{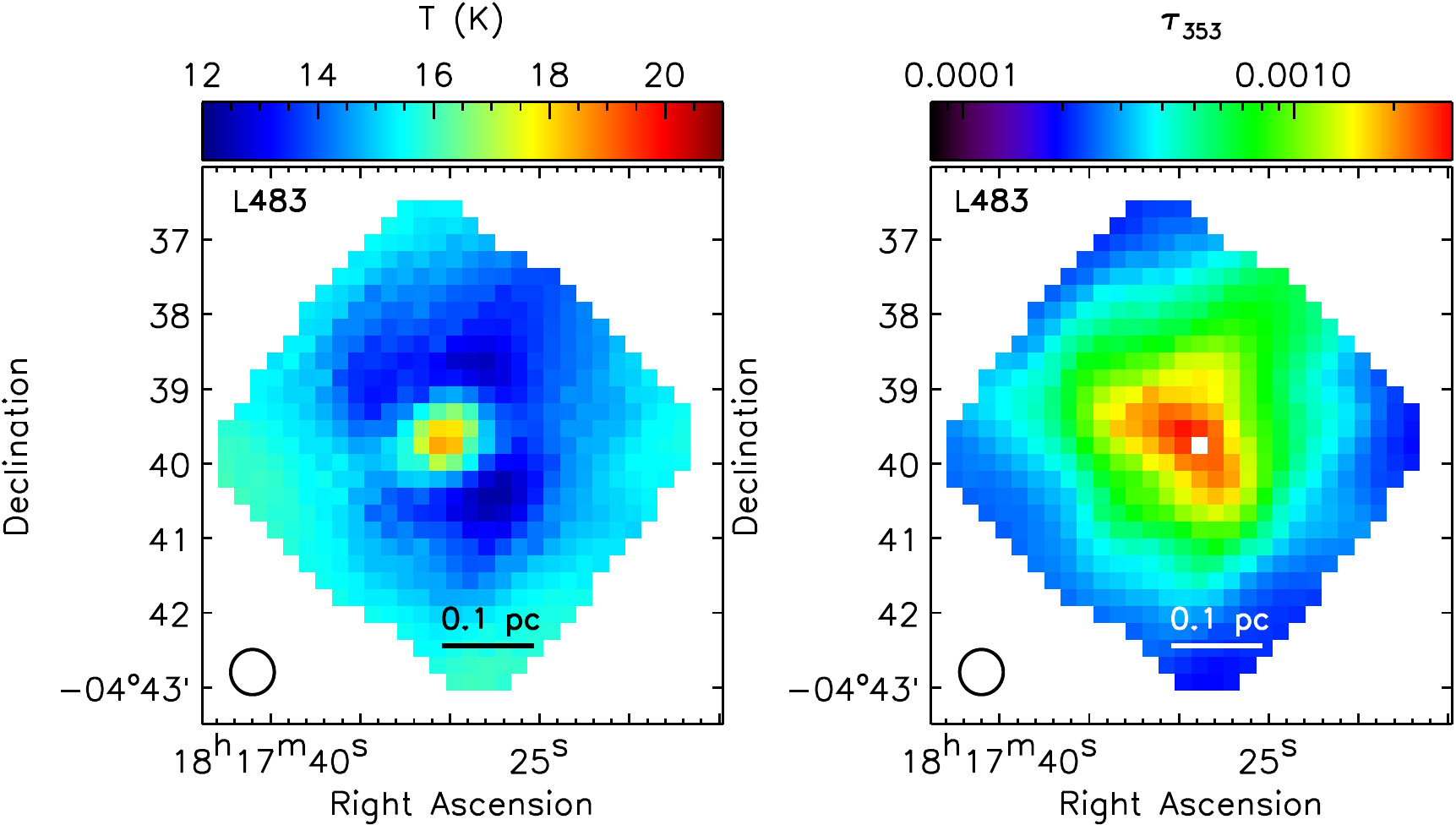}
\qquad
\includegraphics[scale=0.575]{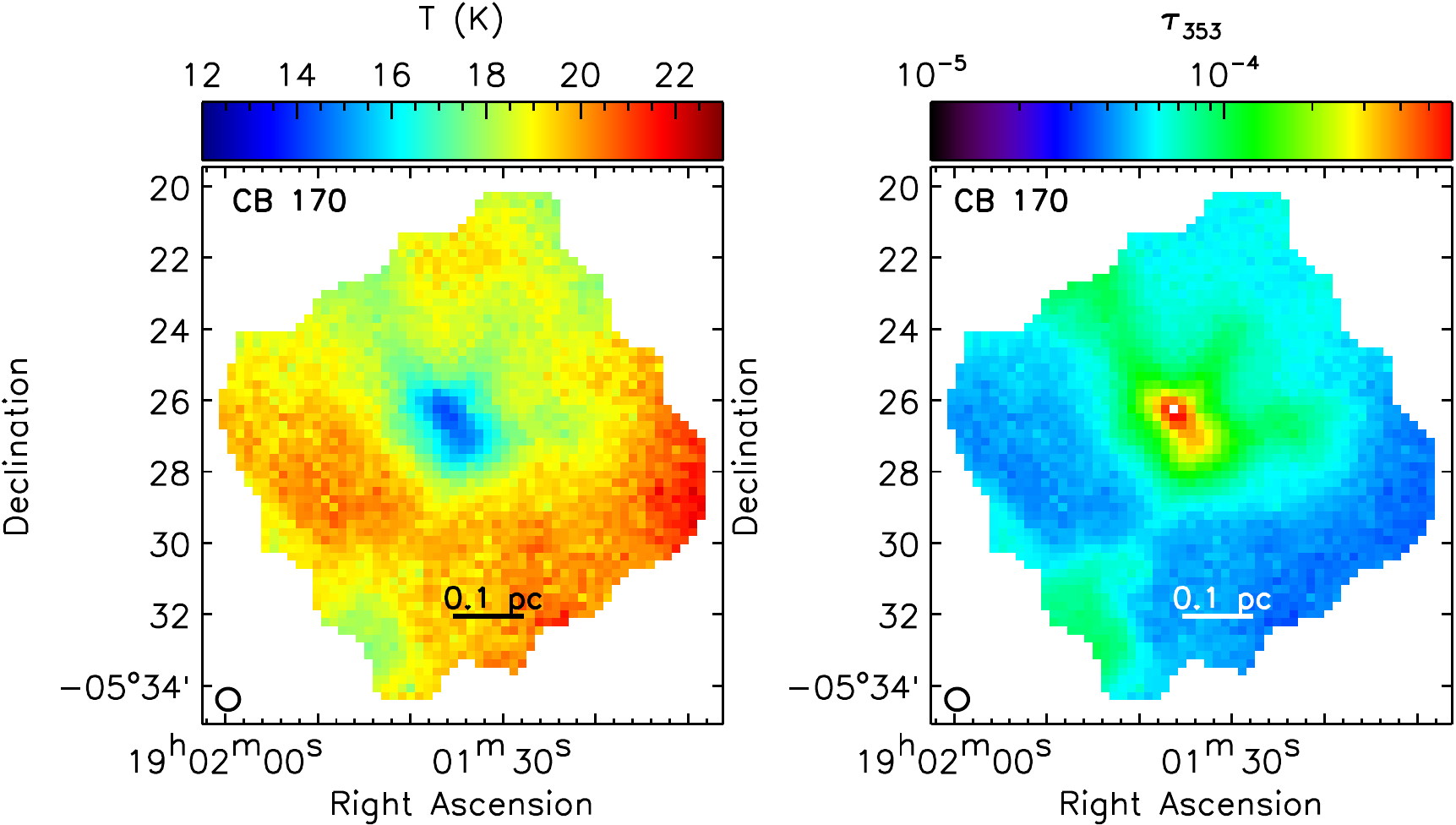}
\qquad
\includegraphics[scale=0.575]{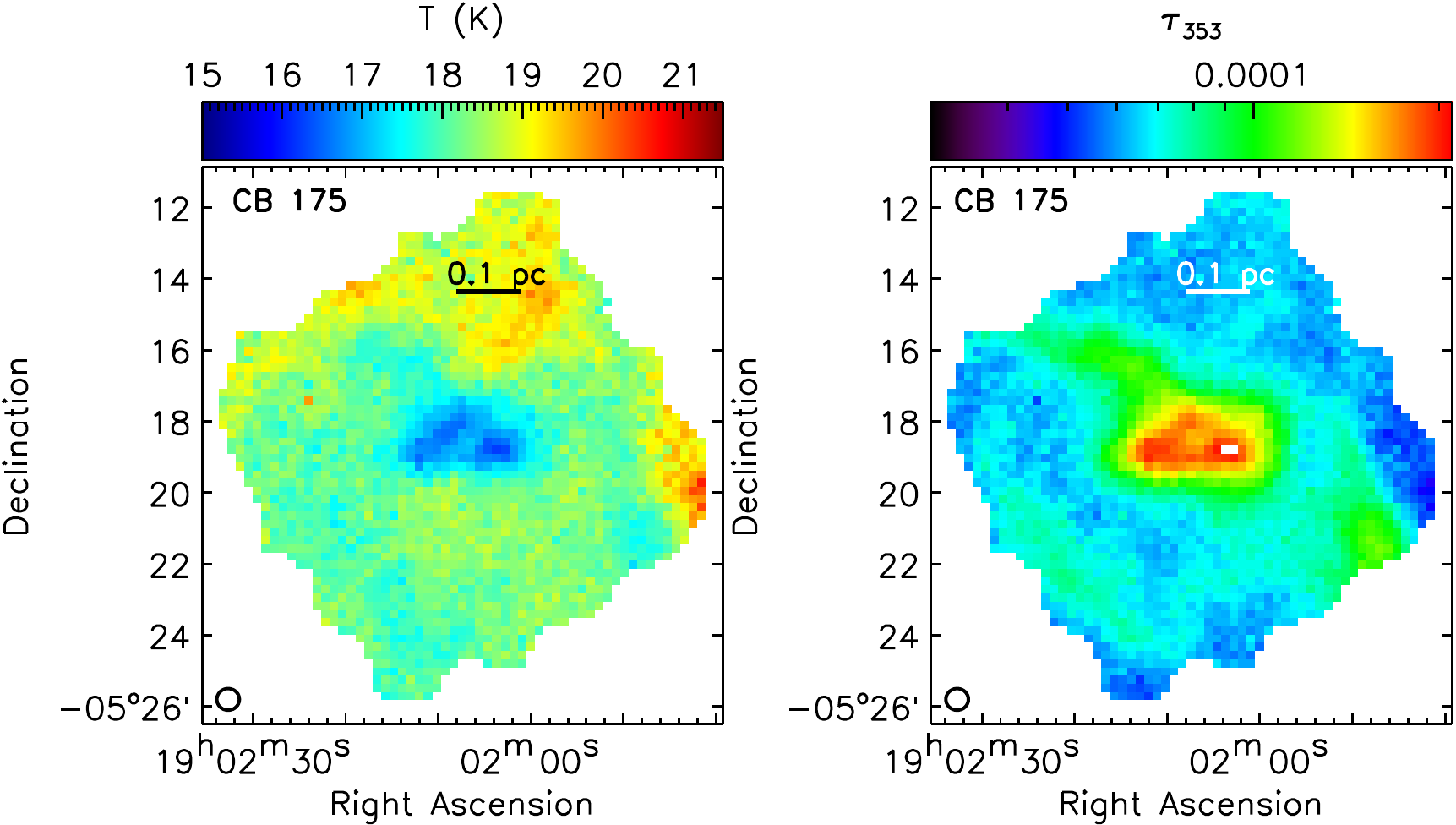}
\caption{Continued -  For L429, L483, CB 170, and CB 175}
\end{figure*}
\begin{figure*}
\ContinuedFloat
\centering
\includegraphics[scale=0.575]{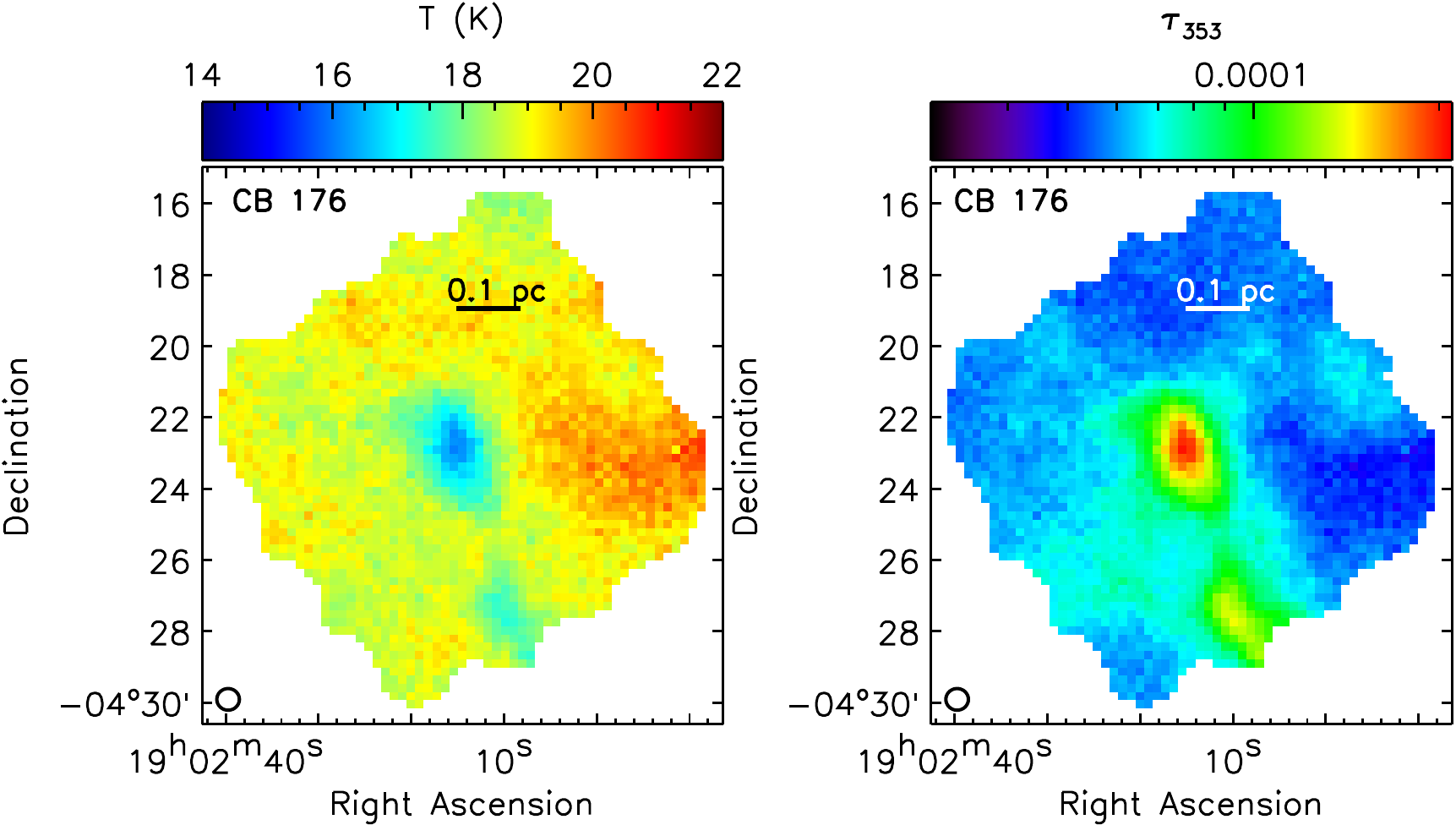}
\qquad
\includegraphics[scale=0.575]{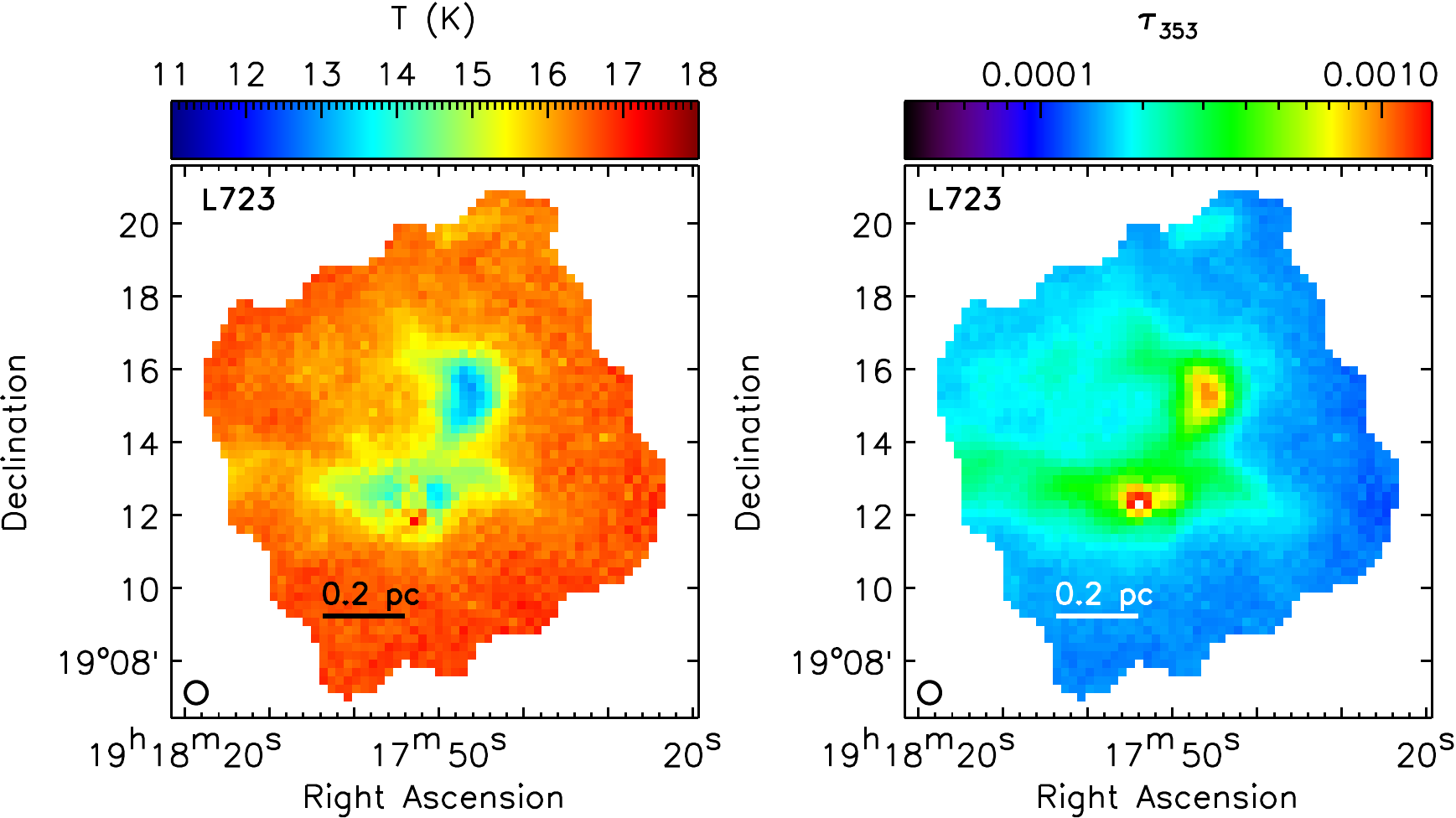}
\qquad
\includegraphics[scale=0.575]{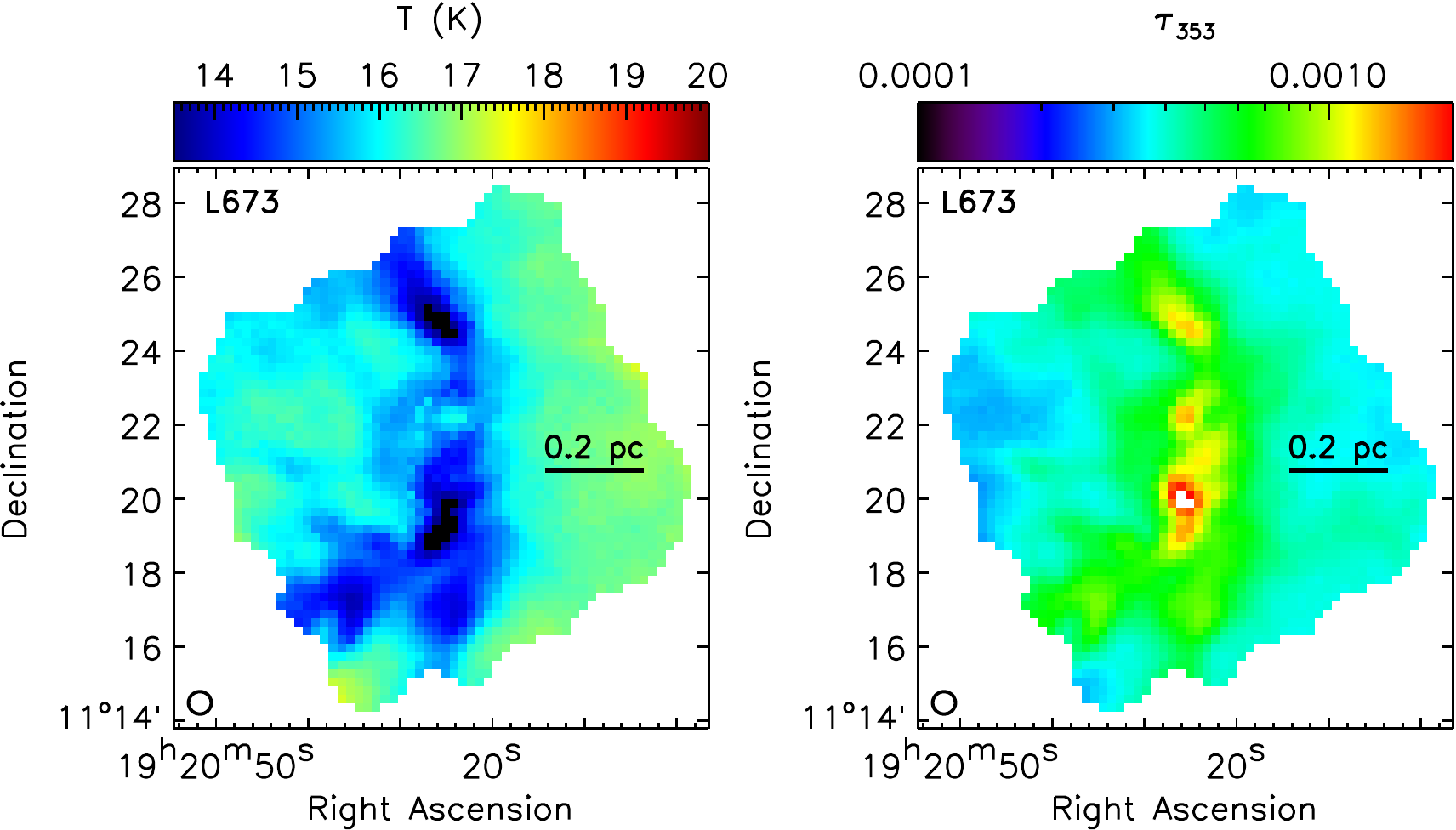}
\qquad
\includegraphics[scale=0.575]{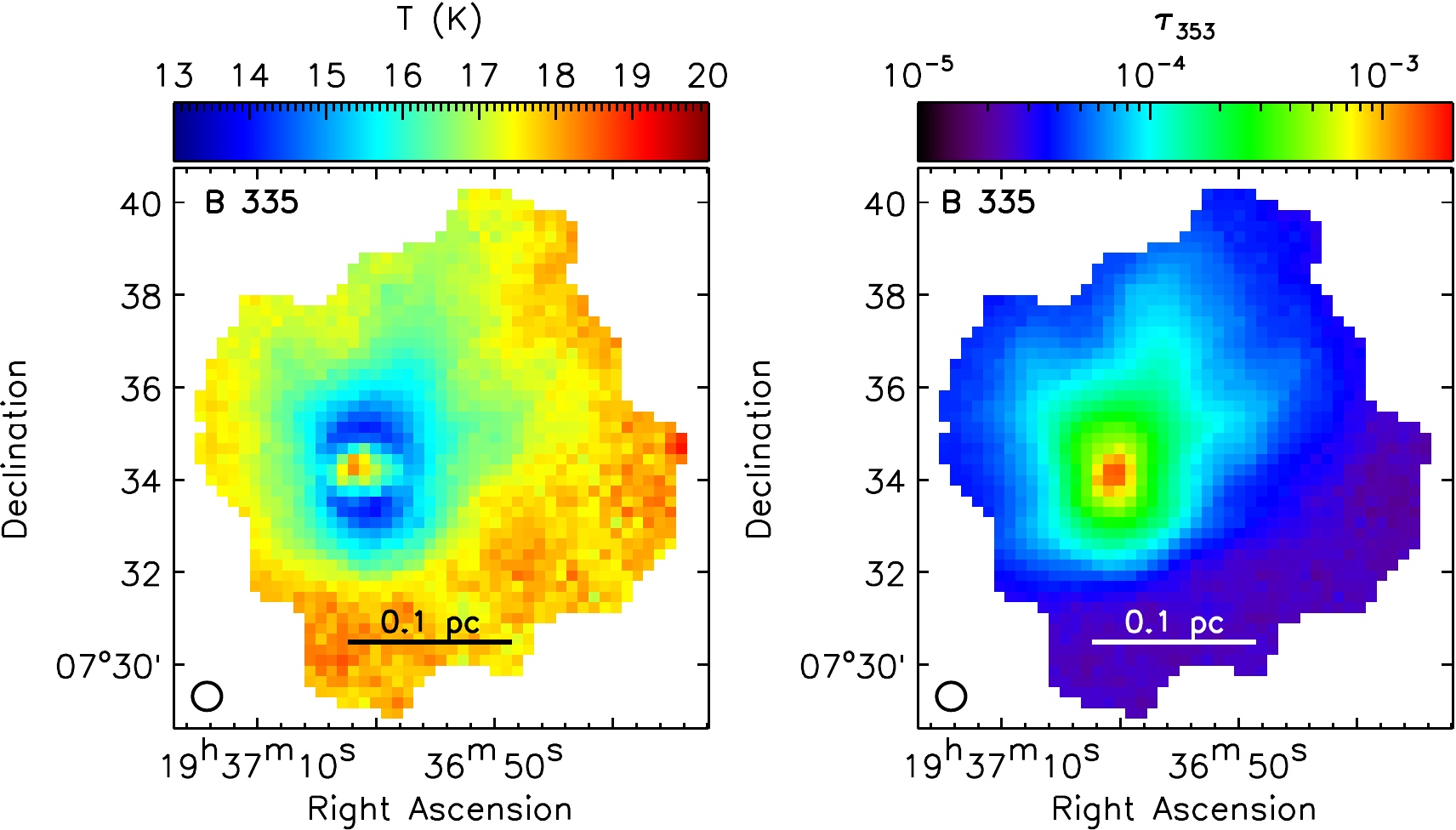}
\caption{Continued -  For CB 176, L723, L673, and B 335.}
\end{figure*}
\begin{figure*}
\ContinuedFloat
\centering
\includegraphics[scale=0.575]{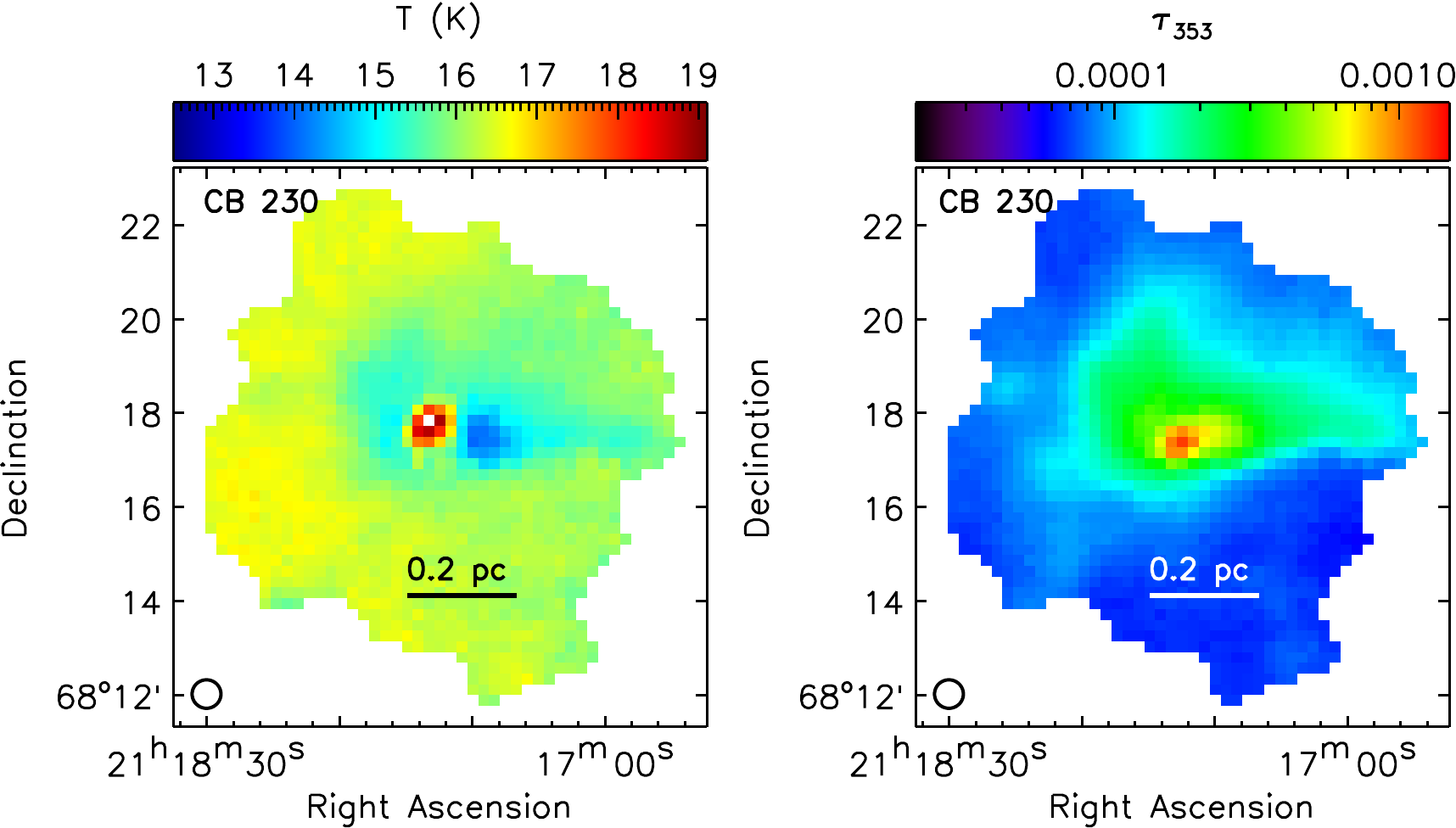}
\qquad
\includegraphics[scale=0.575]{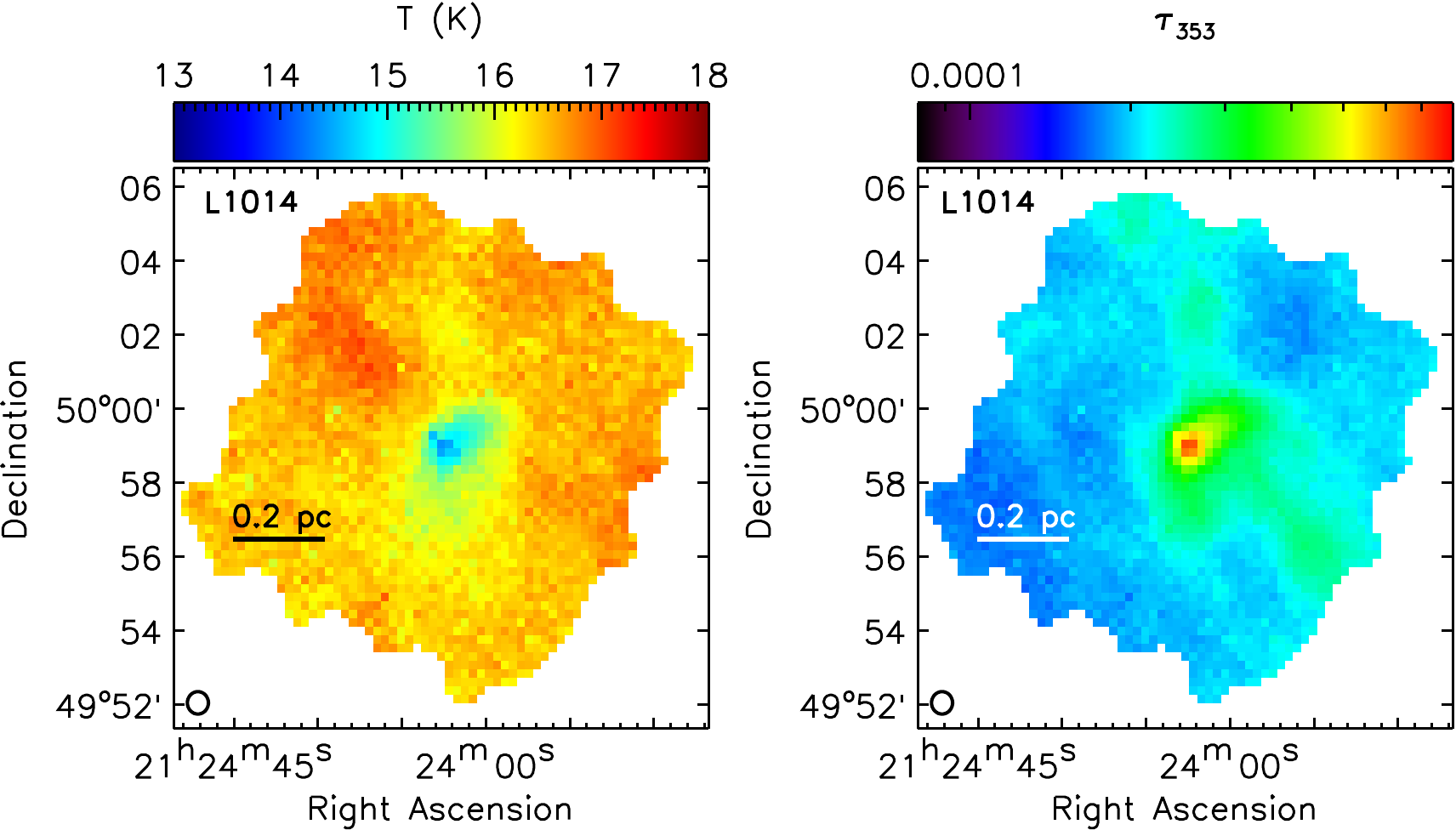}
\qquad
\includegraphics[scale=0.575]{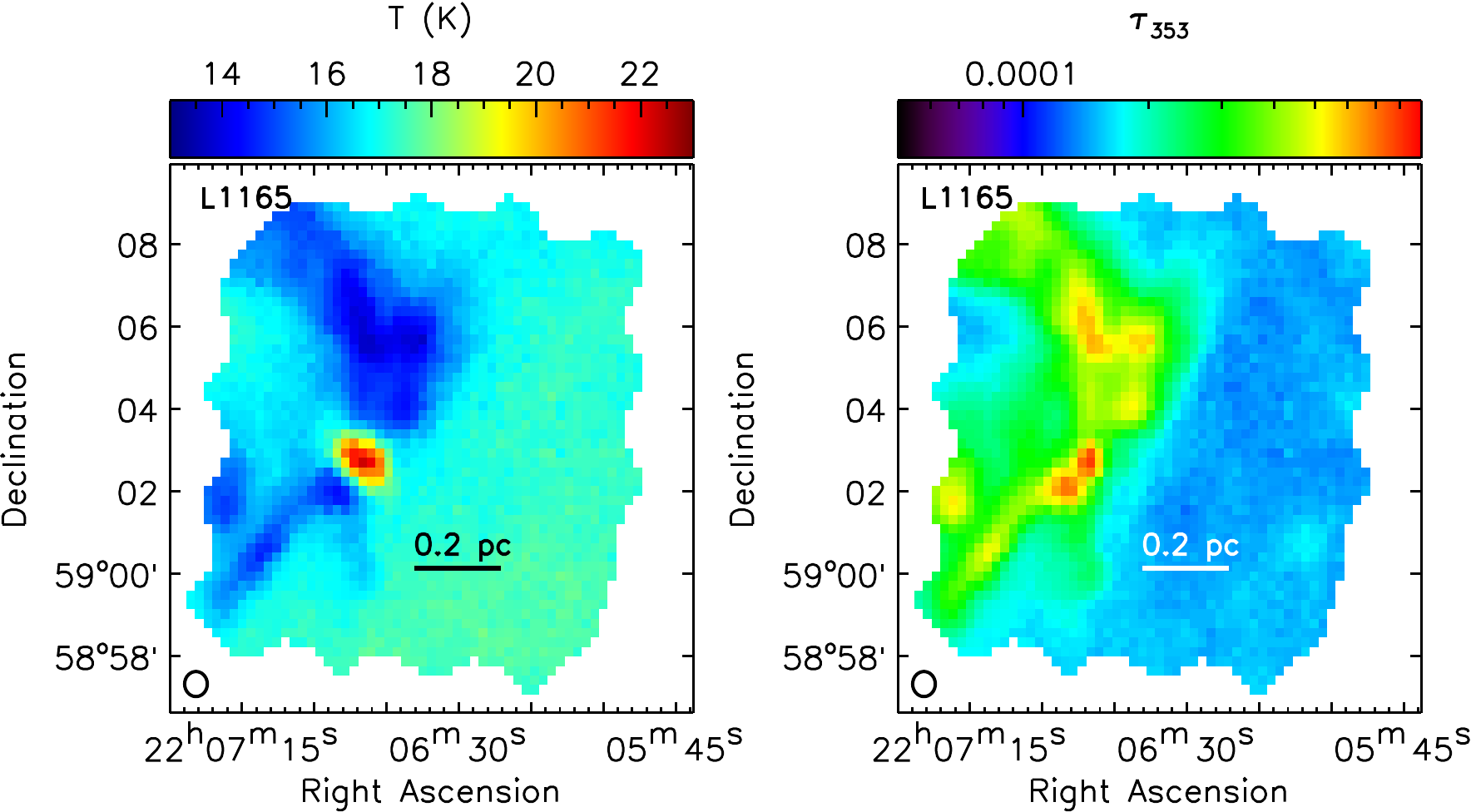}
\qquad
\includegraphics[scale=0.575]{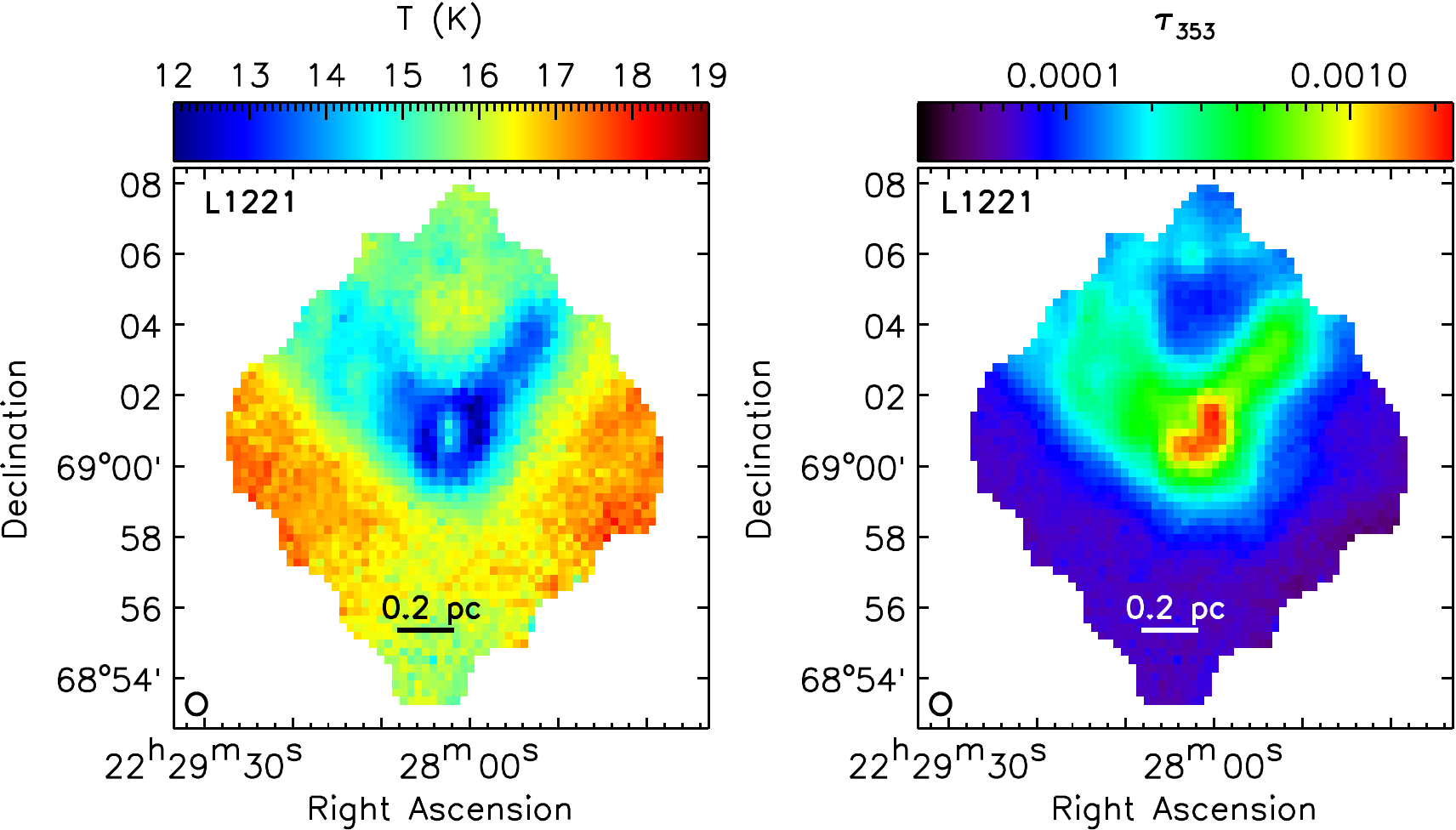}
\caption{Continued -  For CB 230, L1014, L1165, and L1221}
\end{figure*}
\begin{figure*}
\ContinuedFloat
\centering
\includegraphics[scale=0.575]{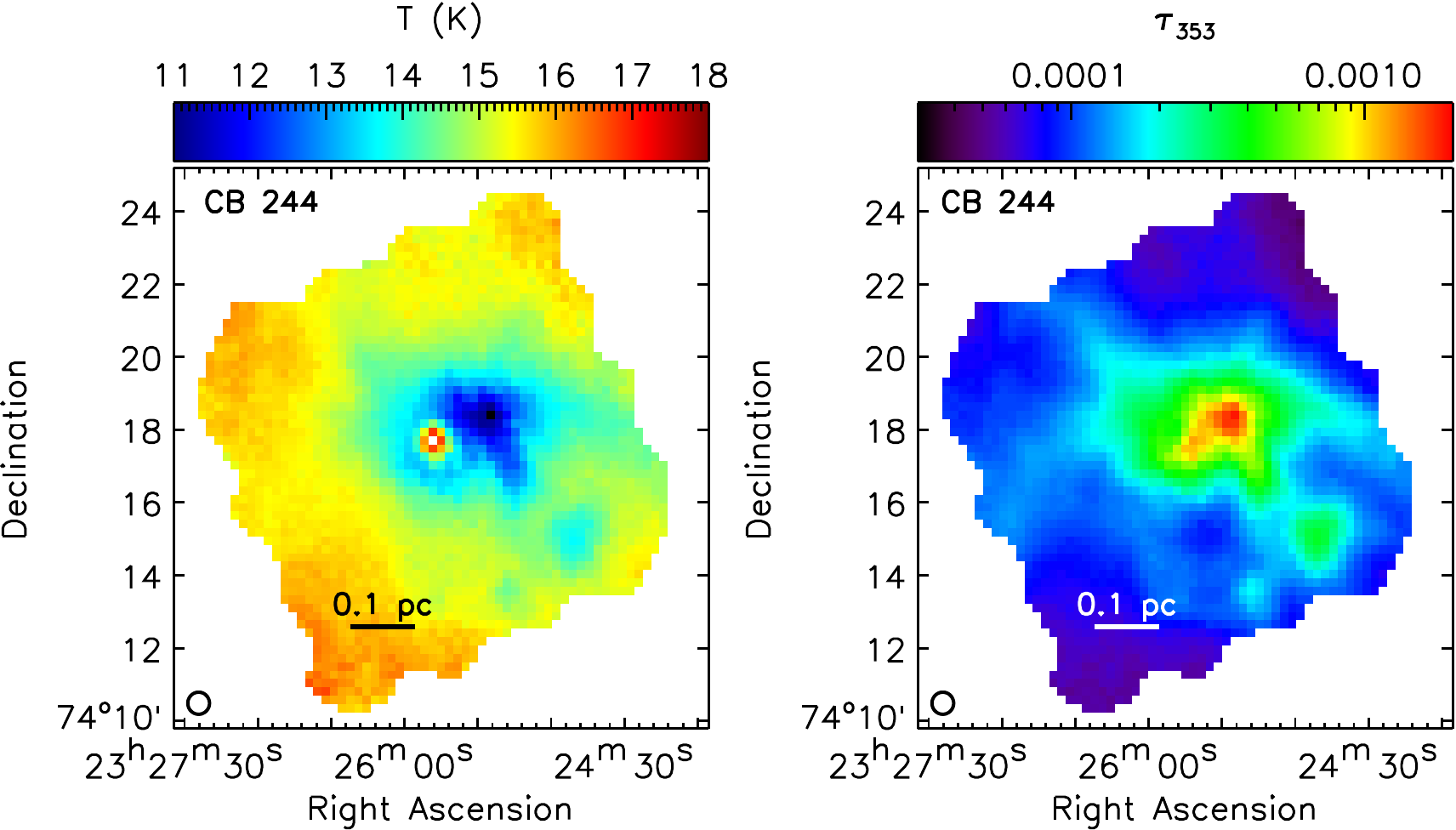}
\caption{Continued -  For CB 244}
\end{figure*}

\begin{deluxetable*}{lccc}
\tablecaption{Summary of the \emph{Herschel} Corrections Groups \label{table_sum}}
\tablehead{
\multirow{2}{*}{Globule } & \colhead{PACS} & \colhead{PACS} & \multirow{2}{*}{SPIRE}   \\
\colhead{ } & \colhead{100 \um} & \colhead{160 \um} & \colhead{ } 
}
\startdata
CB 4	&	A	&	A	&	A	\\
CB 6	&	A	&	A	&	A	\\
CB 17	&	A	&	A	&	A	\\
L1521F	&	C	&	C	&	B	\\
CB 26	&	A	&	C	&	A	\\
L1544	&	A	&	B	&	A	\\
CB 27	&	C	&	C	&	A	\\
L1552	&	B	&	B	&	A	\\
CB 29	&	C	&	C	&	A	\\
B 35A	&	C	&	C	&	A	\\
BHR 22	&	B	&	B	&	A	\\
BHR 17	&	C	&	A	&	A	\\
BHR 16	&	C	&	B	&	B	\\
BHR 12	&	B	&	B	&	C	\\
DC2573-25	&	C	&	B	&	B	\\
BHR 31	&	A	&	A	&	A	\\
BHR 42	&	B	&	B	&	A	\\
BHR 34	&	B	&	B	&	A	\\
BHR 41	&	A	&	B	&	A	\\
BHR 40	&	C	&	C	&	B	\\
BHR 38/39	&	C	&	C	&	A	\\
BHR 56	&	C	&	B	&	B	\\
DC2742-04	&	B	&	B	&	A	\\
BHR 48/49	&	B	&	B	&	C	\\
BHR 50	&	A	&	B	&	C	\\
BHR 68	&	A	&	B	&	A	\\
BHR 71	&	B	&	C	&	B	\\
BHR 74	&	A	&	A	&	A	\\
BHR 79	&	B	&	B	&	A	\\
BHR 81	&	B	&	B	&	B	\\
DC3162+51	&	C	&	C	&	C	\\
BHR 95	&	B	&	A	&	B	\\
BHR 99	&	B	&	B	&	A	\\
BHR 100	&	A	&	C	&	A	\\
BHR 97	&	B	&	B	&	C	\\
DC3391+117	&	B	&	B	&	A	\\
DC3460+78	&	C	&	C	&	B	\\
CB 68	&	A	&	A	&	A	\\
BHR 147	&	B	&	B	&	B	\\
B 68	&	B	&	C	&	A	\\
CB 101	&	C	&	C	&	A	\\
L422	&	A	&	B	&	B	\\
CB 130	&	A	&	A	&	A	\\
L429	&	C	&	B	&	B	\\
L483	&	C	&	C	&	A	\\
CB 170	&	C	&	B	&	A	\\
CB 175	&	B	&	B	&	A	\\
CB 176	&	B	&	A	&	A	\\
L723	&	B	&	C	&	A	\\
L673	&	C	&	B	&	B	\\
B 335	&	A	&	A	&	A	\\
CB 230	&	A	&	B	&	A	\\
L1014	&	A	&	C	&	B	\\
L1165	&	B	&	B	&	A	\\
L1221	&	A	&	A	&	A	\\
CB 244	&	A	&	B	&	B	
\enddata
\end{deluxetable*} 

\end{appendix}

\end{document}